\documentclass[preprint]{JASA}

\usepackage{algpseudocode}
\usepackage{graphicx}
\usepackage{amsmath}
\usepackage{commath}
\usepackage{multirow}
\usepackage[flushleft]{threeparttable}
\usepackage[linesnumbered,ruled,vlined]{algorithm2e}
\DeclareMathOperator*{\argmin}{arg\,min}

\def\bea{\begin{eqnarray}}
\def\eea{\end{eqnarray}}

\begin{document}

\title{A deep-learning based generalized reduced-order model of glottal flow during normal phonation}
\author{Yang Zhang}
\author{Weili Jiang}
\affiliation{Department of Mechanical Engineering, University of Maine, Orono, ME, 04469, USA}
\author{Luning Sun}
\author{Jianxun Wang}		
\affiliation{Department of Aerospace and Mechanical Engineering, University of Notre Dame, Notre Dame, IN, 46556, USA}

\author{Simeon Smith}
\author{Ingo R. Titze}
\affiliation{The National Center for Voice and Speech, University of Utah, Salt Lake City, UT, 84101, USA}

\author{Xudong Zheng}
\email{xudong.zheng@maine.edu}
\author{Qian Xue}
\email{qian.xue@maine.edu}

\affiliation{Department of Mechanical Engineering, University of Maine, Orono, ME, 04469, USA}


\date{\today} 

\begin{abstract}
This paper proposes a deep-learning based generalized reduced-order model (ROM) that can provide a fast and accurate prediction of the glottal flow during normal phonation. The approach is based on the assumption that the vibration of the vocal folds can be represented by a universal kinematics equation (UKE), which is used to generate a glottal shape library. For each shape in the library, the ground truth values of the flow rate and pressure distribution are obtained from the high-fidelity Navier-Stokes (N-S) solution. A fully-connected deep neural network (DNN)is then trained to build the empirical mapping between the shapes and the flow rate and pressure distributions. The obtained DNN based reduced-order flow solver is coupled with a finite-element method (FEM) based solid dynamics solver for FSI simulation of phonation. The reduced-order model is evaluated by comparing to the Navier-Stokes solutions in both statics glottal shaps and FSI simulations. The results demonstrate a good prediction performance in accuracy and efficiency.  
\end{abstract}


\maketitle


\section{\label{sec:1} Introduction}
Voiced sound production in the human larynx is a complex fluid-structure interaction (FSI) process in which the forced air from the lungs interacts with vocal fold tissues to initiate sustained vibrations that modulate the glottal airflow \cite{ingo1994principles}. An accurate prediction of the vocal fold vibration and sound source relies on an accurate prediction of intraglottal pressure and glottal flow rate. In the past, the most commonly used glottal flow model for simulating FSI is the Bernoulli equation which simplifies the flow as a 1D inviscid flow \cite{ruty2007vitro,wurzbacher2006model,zanartu2007influence}. By coupling with lumped-mass or continuum vocal fold models, the model has provided important understandings of the dynamics of FSI during voice production \cite{ishizaka1972synthesis,titze1988physics,story1995voice,steinecke1995bifurcations,jiang2002chaotic,zhang2008nonlinear,tao2008chaotic,erath2011nonlinear,alipour2000finite}. Yet, the inviscid assumption has made the model inaccurate in predicting the glottal flow rate and intraglottal pressures, especially during glottal closing when the glottis is typically in a divergent shape in which rich viscous effects occur such as flow separation, shear layer instability and intraglottal vortices \cite{scherer1983pressure,pelorson1994theoretical,deverge2003influence}. To improve the accuracy,  research efforts have been made to incorporate various viscous loss terms into the Bernoulli equation \cite{van1957air,ishizaka1972synthesis,deverge2003influence,zhang2016evaluation}. While the results showed improvement over the original Bernoulli equation, the modified model is largely based on assumptions of simple glottal shapes. On the other hand,  the quick advancement of the continuum vocal fold model from simple 2D configurations to complex 3D subject-specific configurations increasingly requires a more sophisticated glottal flow model that can represent glottal flow dynamics in complex glottal shapes. The Navier-Stokes (N-S) equation based model, i.e., the full-order model (FOM) can satisfy the requirement \cite{luo2008immersed,mittal2011toward,zheng2010coupled,xue2014subject}, but the very high computational cost limits its use in statistical studies.  Therefore, there is a need and interest in developing a glottal flow model that can provide accurate and fast solution of glottal flow dynamics in complex glottal shapes. 

It has been shown that self-sustained oscillation of vocal folds is dominated by a few modes of vibration, even when the motion is abnormal \cite{berry1994interpretation,berry2001mechanisms,dollinger2005medial}. This high predictability of the vibratory pattern of the vocal folds makes it feasible to model the glottal flow dynamics based on the glottal shapes using deep-learning approach. Nevertheless, related research focusing on this area is still rare. A deep-learning based reduced-order model (ROM) model for glottal flow was proposed in our previous study \cite{yang2020DNN}. The model was based on the Bernoulli equation with a viscous loss term predicted by a deep neural network (DNN) model. With the trained DNN-Bernoulli model, the flow resistance coefficient as well as the flow rate and pressure distribution of a given glottal shape can be predicted. However, the DNN-Bernoulli model was developed under certain initial and geometry conditions and the generalization ability of the model may be limited. In this paper, a deep-learning based generalized ROM of the glottal flow during normal phonations is proposed. The underlying assumption of the approach is that the vocal fold kinematics can be approximated by a few vibration modes described by the surface-wave approach \cite{smith2018vocal}. A number of past studies showed that the vocal fold vibration in normal phonation is dominated by two modes \cite{berry1994interpretation,berry2001mechanisms,dollinger2005medial,smith2018vocal}. Therefore, in the present work, we assume that the vibration of the vocal folds is approximated by a linear combination of the modal displacement of the two dominant modes, and then a universal kinematics equation (UKE) can be obtained. The UKE is examined by generating a large number of glottal shapes from FSI simulations with various vocal fold material properties and subglottal pressures and fitting the glottal shapes with the UKE using the genetic algorithm (GA) \cite{goldberg2006genetic,mitchell1998introduction,forrest1996genetic}. The probability distribution function (PDF) of each fitting parameter is then obtained and used to construct a generalized glottal shape library by appropriately resampling the PDF of the fitting parameters. For each shape in the library, the ground truth value of the flow rate and pressure distribution are obtained from high-fidelity N-S solutions. A fully-connected DNN \cite{goodfellow2016deep} is then used to build the empirical mapping between input parameters (fitting parameters in the UKE and subglottal pressure) and output parameters (flow rate and pressure distribution). K-fold cross validation is performed to fine tune the architecture and hyperparameters and evaluate the prediction performance of the DNN. The developed reduced order glottal flow model is therefore composed of two parts: (a) glottal shape parameterization using the UKE and GA, and (b) glottal flow rate and intraglottal pressure prediction using the trained DNN. The performance of the developed flow model (ROM) is evaluated by comparing to the N-S solutions (FOM) in both static glottal shapes and FSI simulations. 

The outline of the paper is organized as follows: the three-dimensional shape of the vocal fold during vibration, including the prephonatory geometry and UKE, are introduced in Section \ref{sec:2}; the process of building up the generalized glottal shape library is elaborated in Section \ref{sec:3}; details about the implementation and evaluation of the DNN model are discussed in Section \ref{sec:4}; implementation and evaluation of the performance of the present ROM for FSI Simulation are discussed in Section \ref{sec:5}; finally, the conclusions are summarized in Section \ref{sec:6}.

\section{\label{sec:2} Three-dimensional Shape of Vocal Fold during Vibration}

\subsection{Prephonatory Geometry}
\begin{figure}[!ht]
  \begin{center}
	\includegraphics[width=0.7\textwidth]{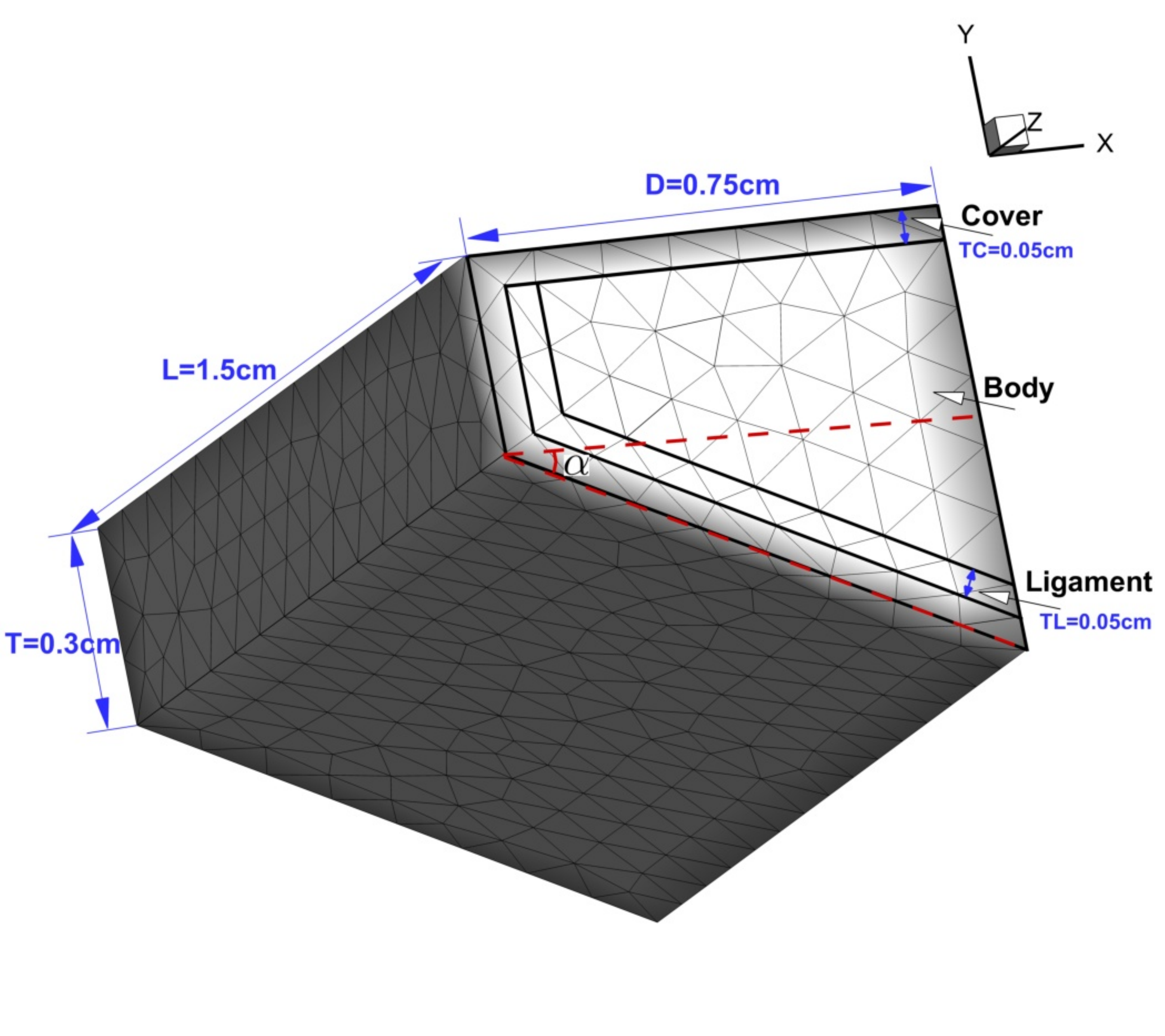}
  \end{center}
  \caption{Prephonatory geometry of the vocal fold.}
  \label{fig:3layersGeo}
\end{figure}

The prephonatory geometry of the vocal fold (right half) is shown in Figure \ref{fig:3layersGeo}. The length $L$ along the anterior-posterior direction ($z$), medial surface thickness $T$ along the inferior-superior direction ($y$) and depth $D$ along the lateral direction ($x$) are $1.5cm$, $0.3cm$ and $0.75cm$, respectively. The subglottal angle $\alpha$ equals to $\arctan0.5$. An initial gap $\Delta x=0.002cm$ along the lateral direction ($x$) exists between the left and right counterpart. The vocal fold is divided into three layers including the cover, ligament, and body. The thickness of the cover ($T_C$) and ligament ($T_L$) layers are both $0.05cm$. Each layer is assumed to be invariant in the anterior-posterior direction.

\subsection{Universal Kinematics Equation (UKE)}
\begin{figure*}
\baselineskip=12pt
\figline{\fig{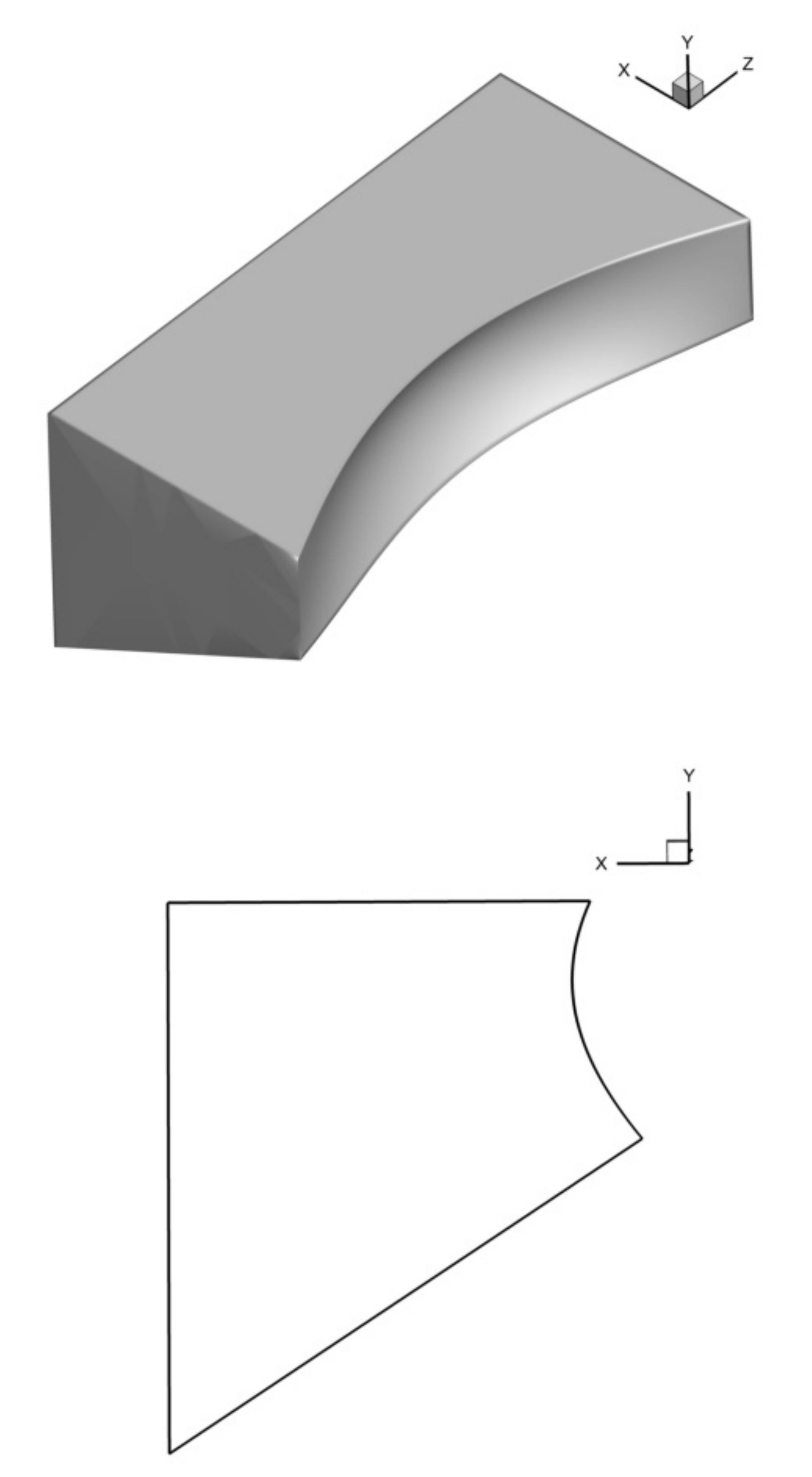}{.33\textwidth}{(a) Mode $(1,0)$}
\fig{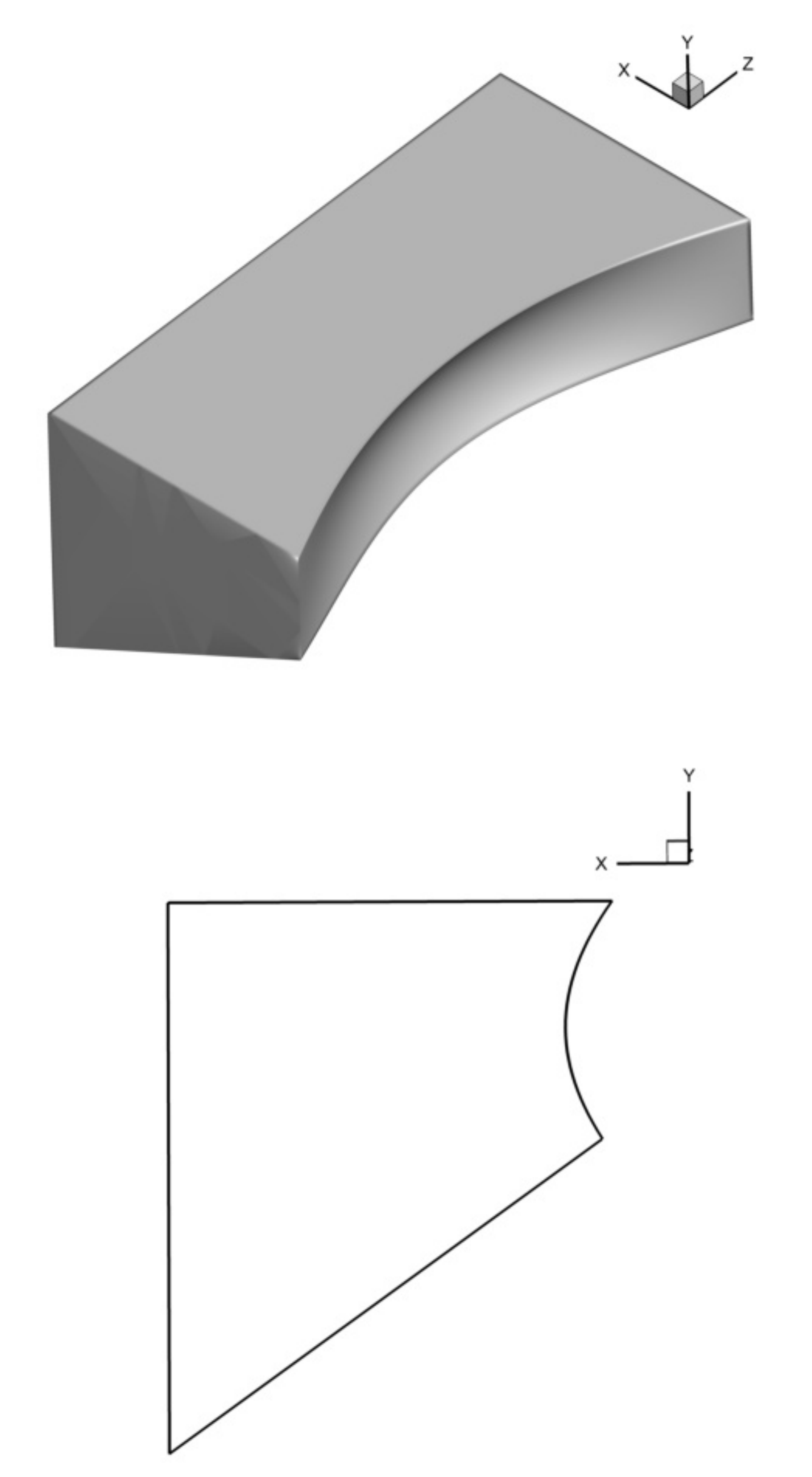}{.33\textwidth}{(b) Mode $(1,1)$}}
\figline{\fig{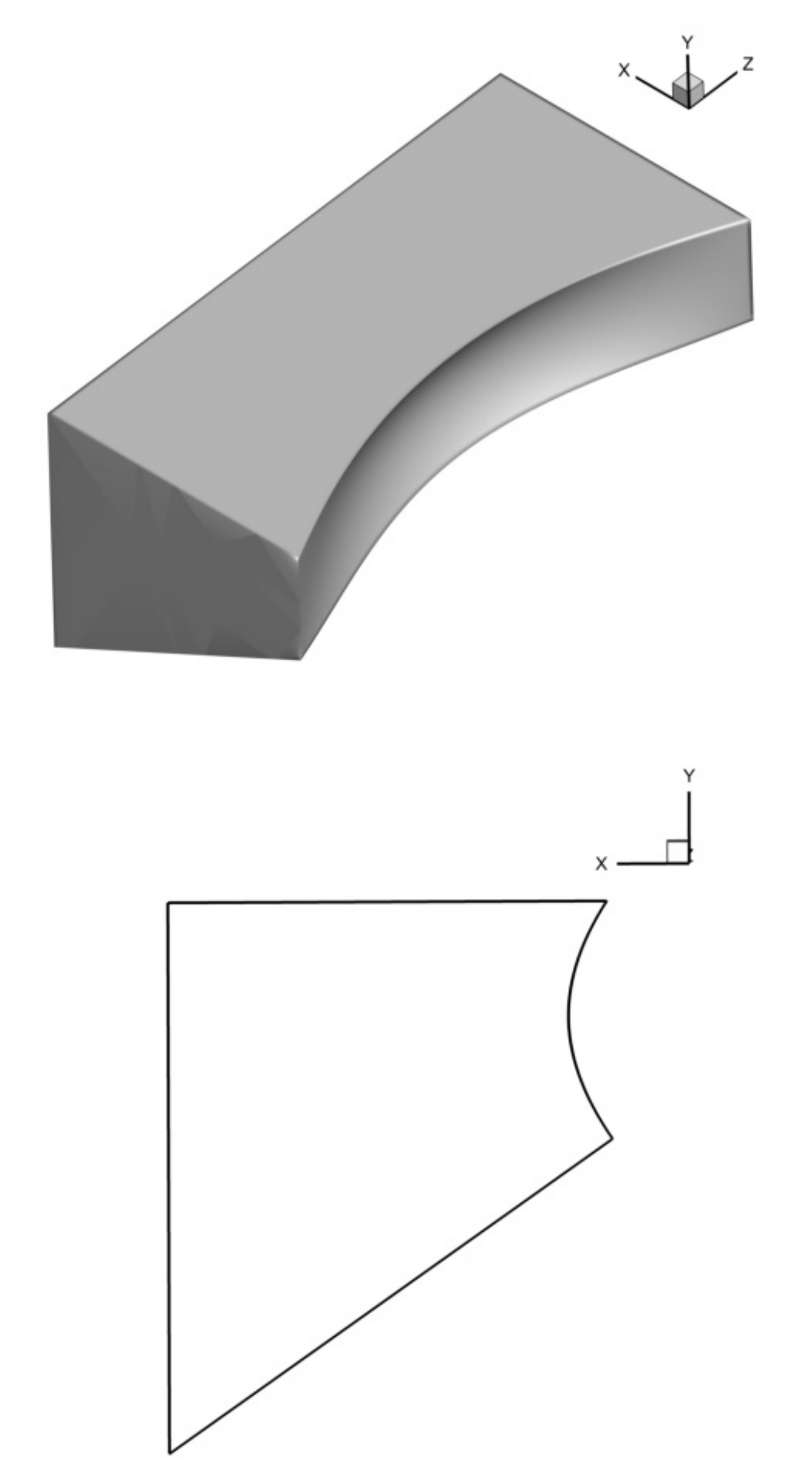}{.33\textwidth}{(c) $\frac{1}{4}$Mode $(1,0)$+$\frac{3}{4}$Mode $(1,1)$}
\fig{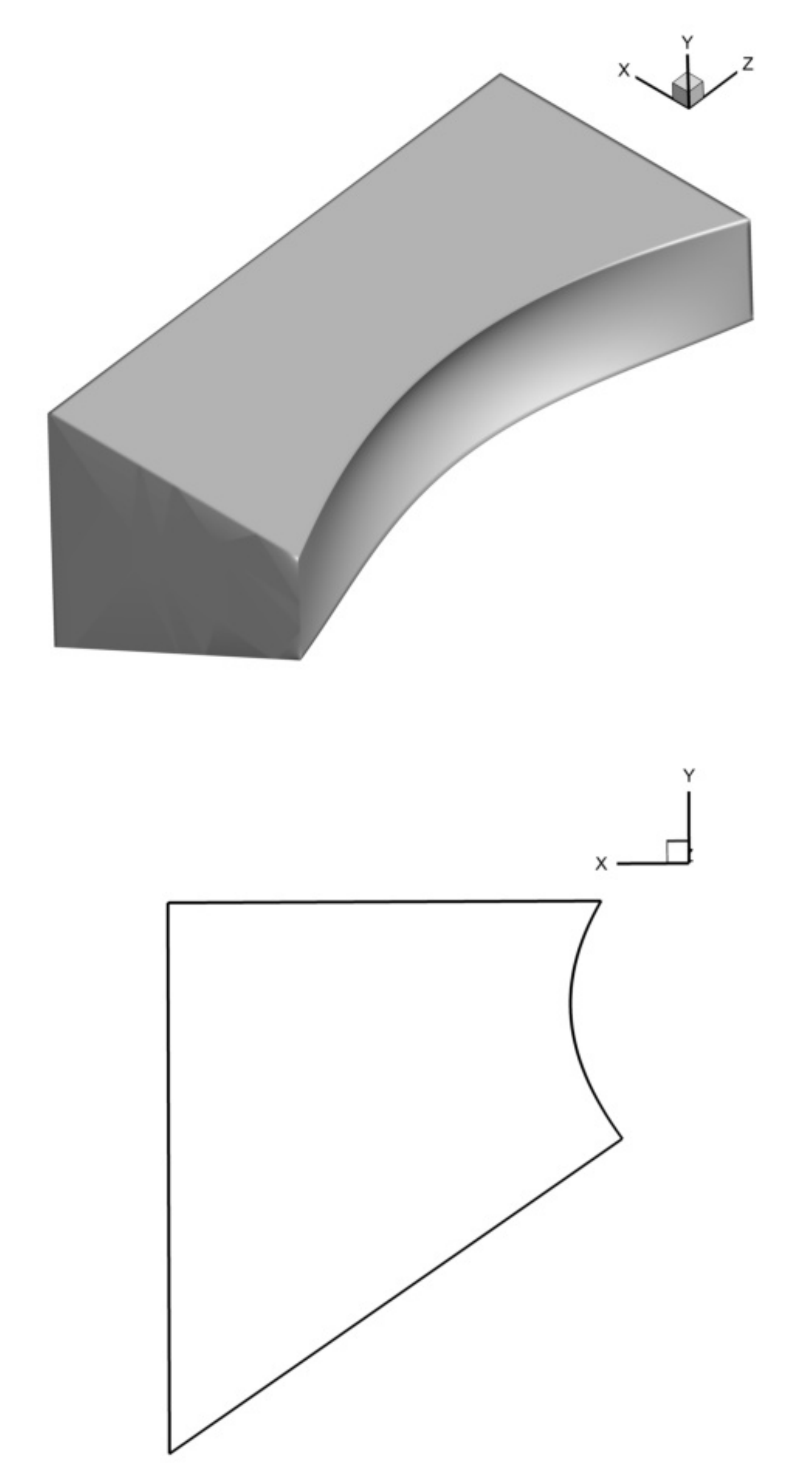}{.33\textwidth}{(d) $\frac{1}{2}$Mode $(1,0)$+$\frac{1}{2}$Mode $(1,1)$}
\fig{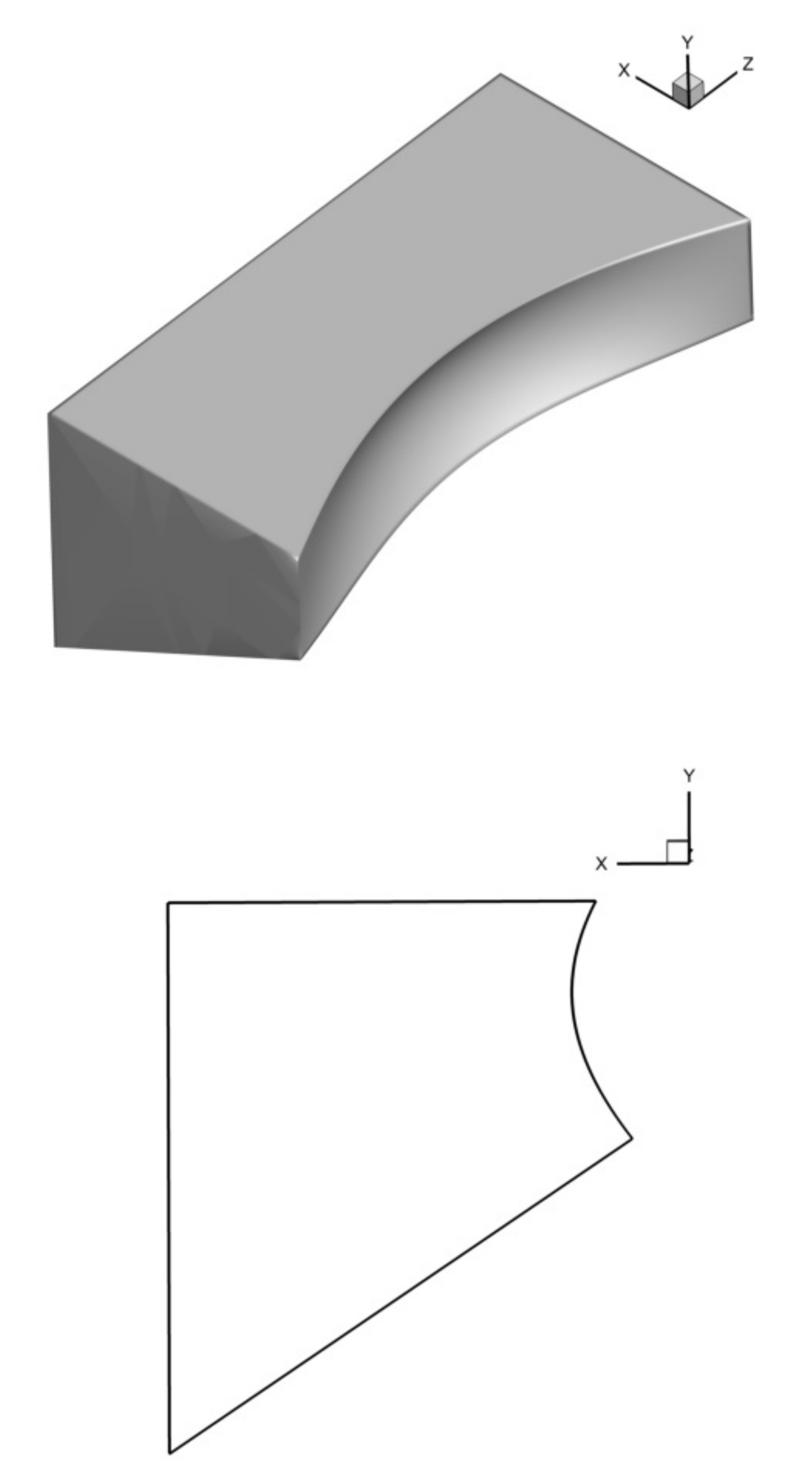}{.33\textwidth}{(e) $\frac{3}{4}$Mode $(1,0)$+$\frac{1}{4}$Mode $(1,1)$}}
\caption{Modal shape of the right-half vocal fold (Each 2D profile corresponds to mid-coronal plane).}
\label{fig:UKE_mode}
\end{figure*}

Past studies have shown that vocal fold vibrations are dominated by a few vibratory modes \cite{berry1994interpretation,berry2001mechanisms,dollinger2005medial}. Following the surface-wave approach in \cite{titze1988physics}, these modes can be described with a combination of (m,n) modes, where $m$ and $n$ correspond to the number of half-wavelengths in the anterior-posterior and inferior-superior directions, respectively. For normal phonation, the most dominant modes are the $(1,0)$ and $(1,1)$ modes, where $(1,0)$ represents the in-phase vibration and $(1,1)$ represents the out-of-phase vibration \cite{titze1988physics,smith2018vocal}. Taking the right-half vocal fold model in Figure \ref{fig:3layersGeo} as an example, the displacement of the medial surface over time can be represented by a linear combination of the modal displacement of these two modes,
\bea \label{eq:1}
\xi(y,z,t)=\alpha \xi(y, z, t)_{(1, 0)} + (1-\alpha) \xi(y, z, t)_{(1, 1)},
\eea
where the subscripts $(1,0)$ and $(1,1)$ respectively refer to modes $(1,0)$ and $(1,1)$, and $\alpha$ is the weight coefficient of mode $(1,0)$. An equivalent equation exists for the left-half vocal fold. An example of the modal shape of the right-half vocal fold is illustrated in Figure \ref{fig:UKE_mode}, where the subfigures (a) and (b) respectively represent the modal shape $(1,0)$ and $(1,1)$ of the vocal fold, and the subfigures (c), (d) and (e) represent the actual shape of the vocal fold obtained from Eq. (\ref{eq:1}) with $\alpha=\frac{1}{4}$, $\frac{1}{2}$ and $\frac{3}{4}$, respectively. Note that in our study, to simplify the model, only the lateral ($x$) vibration is allowed and the vertical ($y$) motion is fixed.

In \cite{smith2018vocal}, based on the surface-wave approach and small-angle approximation \cite{titze1988physics}, the modal displacement of the medial surface of the vocal fold at any instant in time were defined as, 
\bea \label{eq:2}
\xi(y, z, t)_{(m, n)}=\xi_{m}\sin(m\pi z/L)[\sin\omega t-n(\omega/c)(y-y_{m})\cos\omega t],
\eea
where $\xi_m$ is the modal displacement amplitude, $y_m$ is the inflection point for the vertical half wavelength, $\omega$ is angular frequency, and $c$ is the speed of the mucosal wave.

The displacement of the medial surface of the vocal fold over time in Eq. (\ref{eq:1}) can then be expressed as,
\bea \label{eq:3}
\xi(y,z,t)=\xi_{m}\sin(\pi z/L)[\sin\omega t-(1-\alpha)(\omega/c)(y-y_{m})\cos\omega t],
\eea
where $y_m$ is the inflection point for the vertical half wavelength \cite{smith2018vocal}. Note that our later FSI simulation results reflected that the location of the inflection point changes along the anteior-posterior direction, therefore, the inflection location is modeled as
\bea \label{eq:3.1}
& & y_m=T-\beta (\sin \frac{\pi z}{L}+1),
\eea
where $0 \leq \beta \leq T/2$.

By superimposing the time-dependent displacement in Eq. (\ref{eq:3}) on the prephonatory geometry, the three-dimensional shape of the glottis at any time instant can be obtained. Eq. (\ref{eq:3}) is also termed as the universal kinematics equation (UKE) in this paper.

\section{\label{sec:3} Generalized Glottal Shape Library}

The vocal fold shape during vibration can be described by Eqs. (\ref{eq:3}) and (\ref{eq:3.1}) with the following parameters: the vibration amplitude $\xi_m$, weight coefficient of mode $(1,0)$ $\alpha$, inflection point factor $\beta$, phase $\phi=12\omega t/\pi$, and ratio between the angular frequency and mucosal wave speed $\omega/c$, which is related to the vibration frequency $f$. The estimated physiological range of these parameters for normal phonation \cite{smith2018vocal} are listed in Table \ref{RangeGovernParams}.

\begin{table}[!ht]
\caption{Estimated physiological range of the parameters in the UKE.}
\begin{center}
\label{RangeGovernParams}
\begin{tabular}{c c}
\hline\hline
Parameters & Range \\
\hline
$\xi_m$ & $(0,0.1cm]$ \\
$\alpha$ & $[0,1]$ \\
$\beta$ & $[0,T/2]$ \\
$\phi$ & $[0,24]$ \\
$f$ & $[100Hz,250Hz]$ \\
\hline\hline
\end{tabular}
\end{center}
\end{table}

In this section, we aim to verify that the UKE can be used as a generalized equation to represent any glottal shape during normal phonation. To have a good estimation of the possible glottal shapes during FSI, FSI simulations of vocal fold vibration under various subglottal pressures and material properties are conducted. The simulations employ the finite-element vocal fold model coupled with the Bernoulli equations for fast solutions \cite{geng2016effect}. A large number of glottal shapes are extracted from the simulation results and used to fit the UKE by using the genetic algorithm (GA) \cite{goldberg2006genetic,mitchell1998introduction,forrest1996genetic}. The fitting error is used to quantify the representative capability of the UKE. Finally, the probability density function (PDF) of each input parameter in the UKE is obtained and used to build the generalized glottal shape library through appropriate resampling.

\subsection{Bernoulli-FEM FSI Simulation} \label{subsec3:1}

The vocal fold tissue is modeled as the viscoelastic, transversely isotropic material. The baseline material properties of each layer of the vocal fold \cite{alipour2000finite, xue2012computational} are listed in Table \ref{BenchmarkMat}.

\begin{table}[!ht]
\centering
\begin{threeparttable}
\caption{Baseline material properties of each layer of the vocal fold.}
\label{BenchmarkMat}
\begin{tabular}{c c c c c c c}
\hline\hline
 & $\rho (g/cm^3)$ & $E_{p} (kPa)$ & $\nu_{p}$ & $E_{pz}^{0} (kPa)$ & $\nu_{pz}$ & $G_{pz}^{0} (kPa)$ \\
\hline
Cover & 1.043 & 2.01 & 0.9 & 40 & 0.0 & 10\\
Ligament & 1.043 & 3.31 & 0.9 & 66 & 0.0 & 40\\
Body & 1.043 & 3.99 & 0.9 & 80 & 0.0 & 20\\
\hline\hline
\end{tabular}
\begin{tablenotes}
      \small
      \item $\rho$ is the tissue density; $E_{p}$ and $E_{pz}^{0}$ are the transversal and longitudinal Young’s Modulus, respectively; $\nu_{p}$ and $\nu_{pz}$ are the in-plane transversal and longitudinal Poisson ratio, respectively; $G_{pz}^{0}$ is the longitudinal shear modulus \cite{alipour2000finite, xue2012computational}. 
\end{tablenotes}
\end{threeparttable}
\end{table}

Based on the baseline material properties listed in Table \ref{BenchmarkMat}, the ranges of the material properties for each layer can be obtained by simultaneously multiplying the corresponding $E_{pz}^{0}$ and $G_{pz}^{0}$ with a factor $k$, where the physiological range of $k$ is $[0.5,5.0]$ with an increment size $\Delta k=0.5$. Note that the value of $k$ for the cover layer and ligament layer are always the same. The various material property factors of the cover-ligament layers and body layer under selected subglottal pressure conditions at $P_0=0.5kPa, 0.75kPa, 1.0kPa$ can be respectively expressed as 
\bea \label{eq:00}
k_{CL}=m\Delta k, \quad m=1,2,...,10,
\eea
\bea \label{eq:01}
k_{B}=n\Delta k, \quad n=1,2,...,10,
\eea
where the subscript $CL$ and $B$ indicate the cover-ligament layers and body layer, respectively.

\begin{figure}[!ht]
  \begin{center}
	\includegraphics[width=1.0\textwidth]{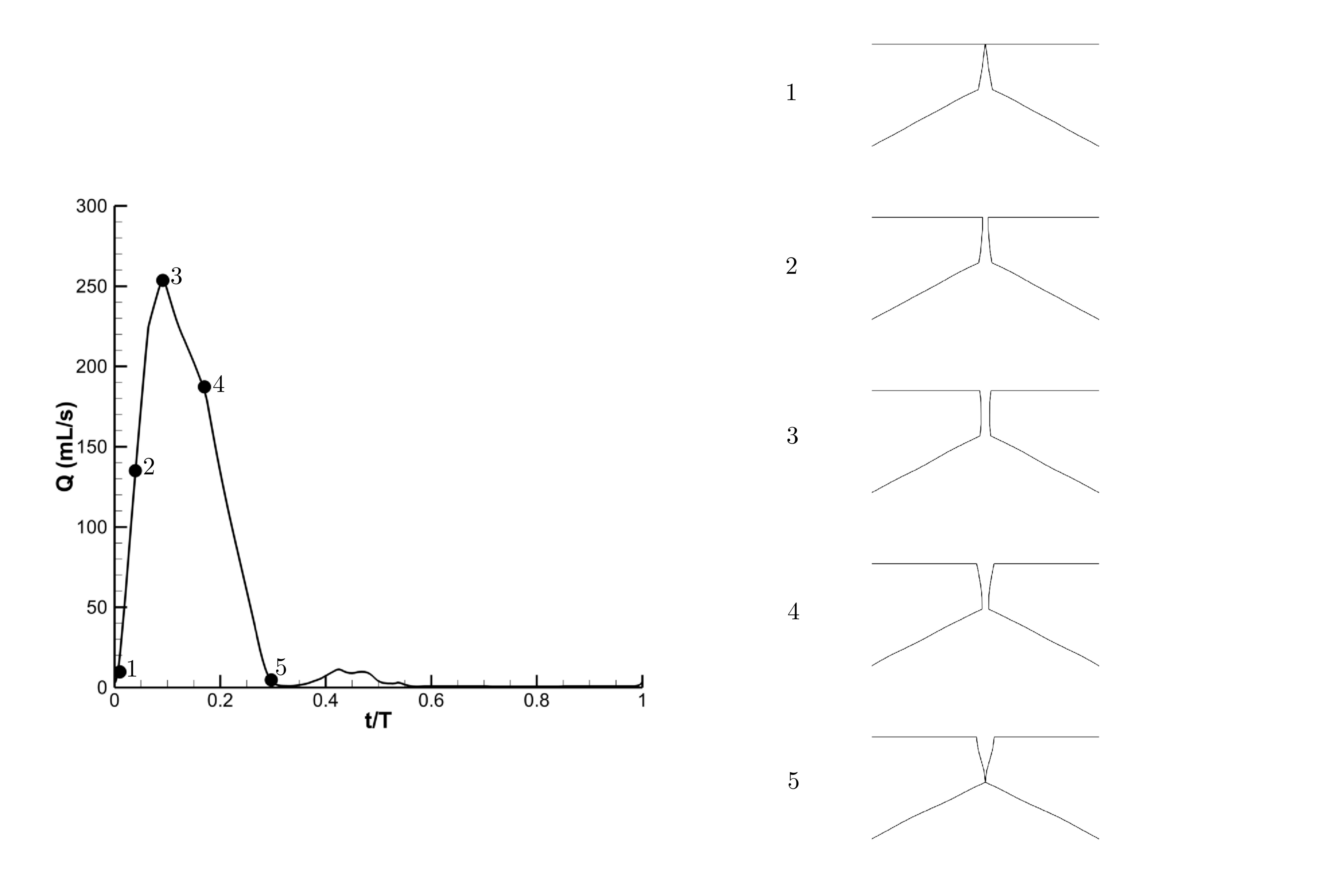}
  \end{center}
  \caption{Vibration pattern during one converged cycle at $P_0=1.0kPa$, $k_{CL}=1.0$, $k_{B}=4.0$.}
  \label{fig:fsiex2}
\end{figure}

By systematically varying $k_{CL}$, $k_{B}$ and $P_0$, a total of 300 cases are generated for the FSI simulations. For each case, the density and kinematic viscosity of the air are $1.145\times10^{-3} g/cm^3$ and $\nu=1.655\times10^{-1}cm^2/s$, respectively. The glottis are discretized with $N_S=69$ uniformly spaced cross sections along the inferior-superior direction such that the spacing is $0.01cm$. The contact surface is calculated as an average of the left and right surface coordinates. A uniform Rayleigh damping factor is used for each case. As an example, the vibration pattern of the vocal folds during one converged cycle at $P_0=1.0kPa$, $k_{CL}=1.0$, $k_{B}=4.0$ is illustrated in Figure \ref{fig:fsiex2}, where the left subfigure corresponds to the time history of the flow rate $Q$ during one converged cycle, and the right subfigure corresponds to the glottal shape at 5 representative phases probed from the left subfigure. The vibration shows a typical alternative convergent-divergent glottal shape change.  

\subsection{Glottal Shape Fitting with the GA}

In this subsection, we aim to verify that those glottal shapes extracted from FSI simulations in Subsection \ref{subsec3:1} can be represented by the UKE. The GA is employed to inversely determine the values of the fitting parameters from the range listed in Table \ref{RangeGovernParams} such that the difference between the optimized and target (FSI) values of the nodal displacement is minimal. In the optimization process, as the flow rate heavily relies on the minimum glottal area, an equal constraint between the optimized and target minimum cross-section area along the inferior-superior direction of the glottis is enforced. Therefore, the constrained minimization function for each glottal shape can be written as,
\bea \label{eq:4}
\begin{aligned}
& \xi_m,\alpha,\beta,\phi,f=\argmin
& & \frac{\sum_{i=1}^{n}[\xi_{optimized}^{i}(\xi_m,\alpha,\beta,\phi,f)-\xi_{target}^{i}]^2}{n} \\
& \text{subject to}
& & \argmin A_{j}^{optimized} = \argmin A_{j}^{target}, (A_{j}^{optimized})_{min}=(A_{j}^{target})_{min}
\end{aligned}
\eea
where the values of $\xi_m$, $\alpha$, $\beta$, $\phi$, $f$ are bounded by the corresponding ranges listed in Table \ref{RangeGovernParams}, $n$ is the number of nodal points of the glottis surface, and $A_{j}^{optimized}$ and $A_{j}^{target}$ are the optimized and target cross-section area function with $j$ the cross-section index, respectively. The constraints imply that the location and value of the optimized minimum cross-section area are equal to the target one.

\begin{figure}[!ht]
  \begin{center}
	\includegraphics[width=1.0\textwidth]{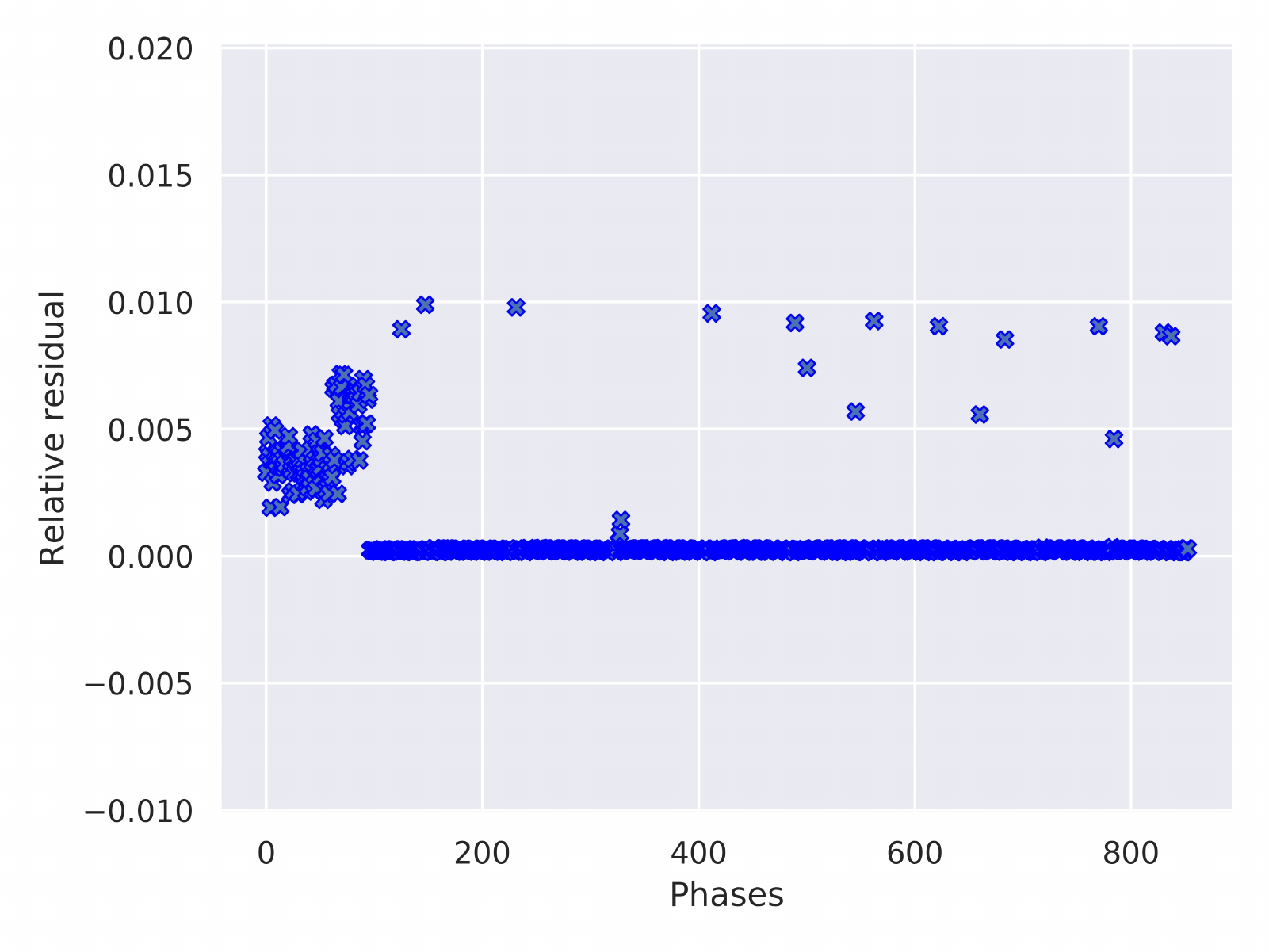}
  \end{center}
  \caption{Relative residual of the fitness function of GA.}
  \label{fig:GA_residual}
\end{figure}

\begin{figure*}
\baselineskip=12pt
\vskip1.3in
\figline{\leftfig{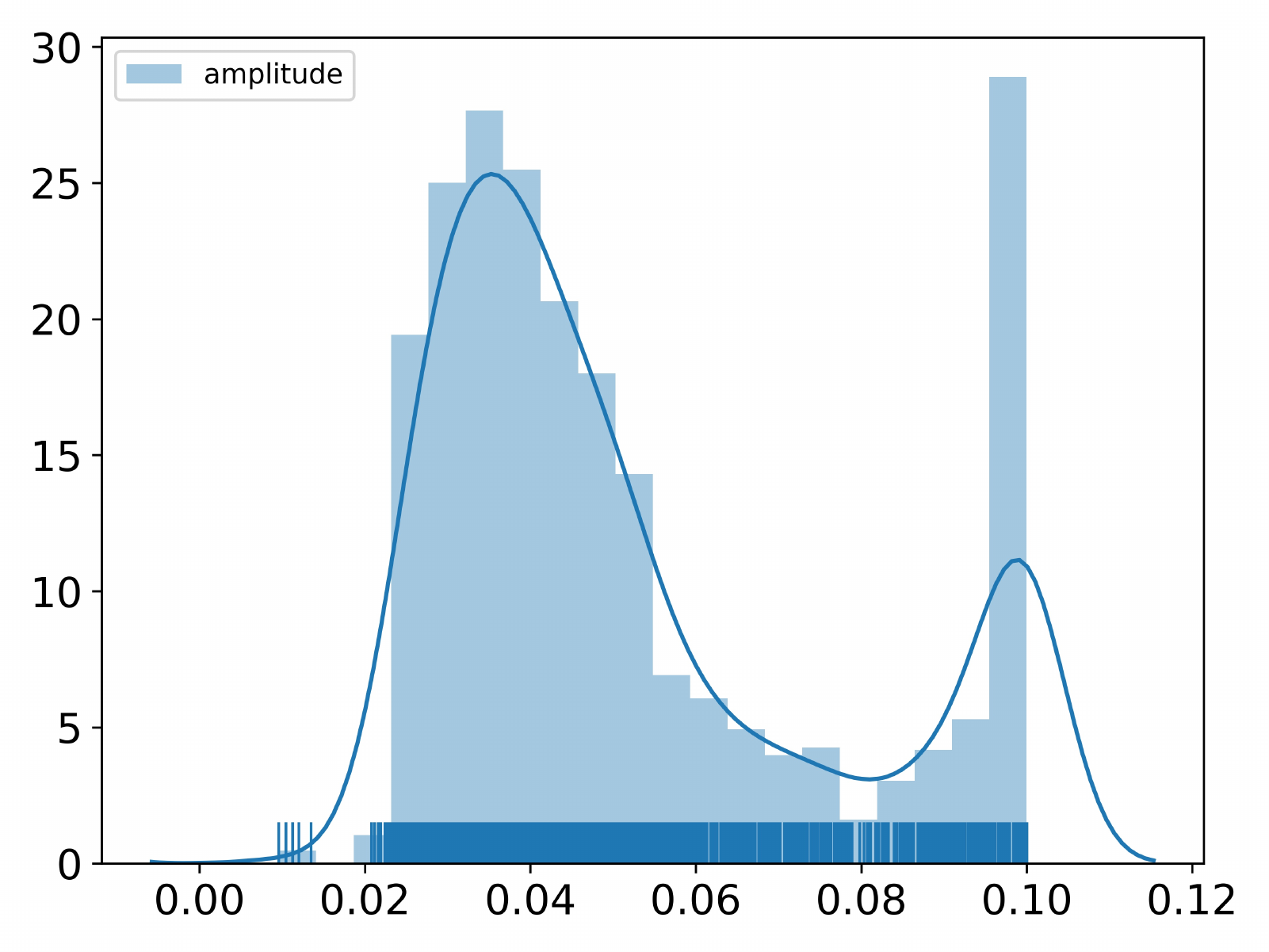}{.49\textwidth}{(a) $\xi_m$}
\leftfig{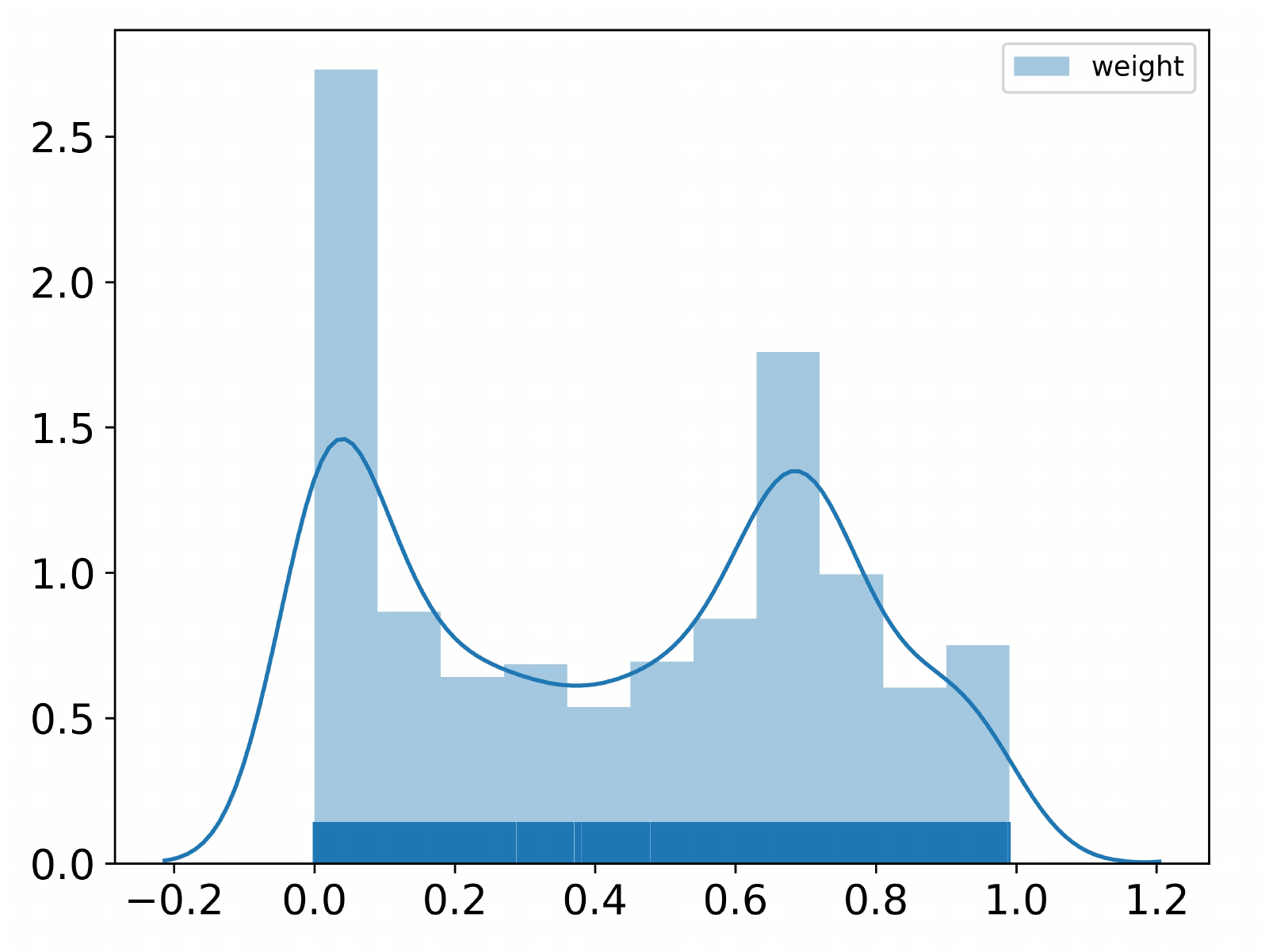}{.49\textwidth}{(b) $\alpha$}\hfill}
\figline{\leftfig{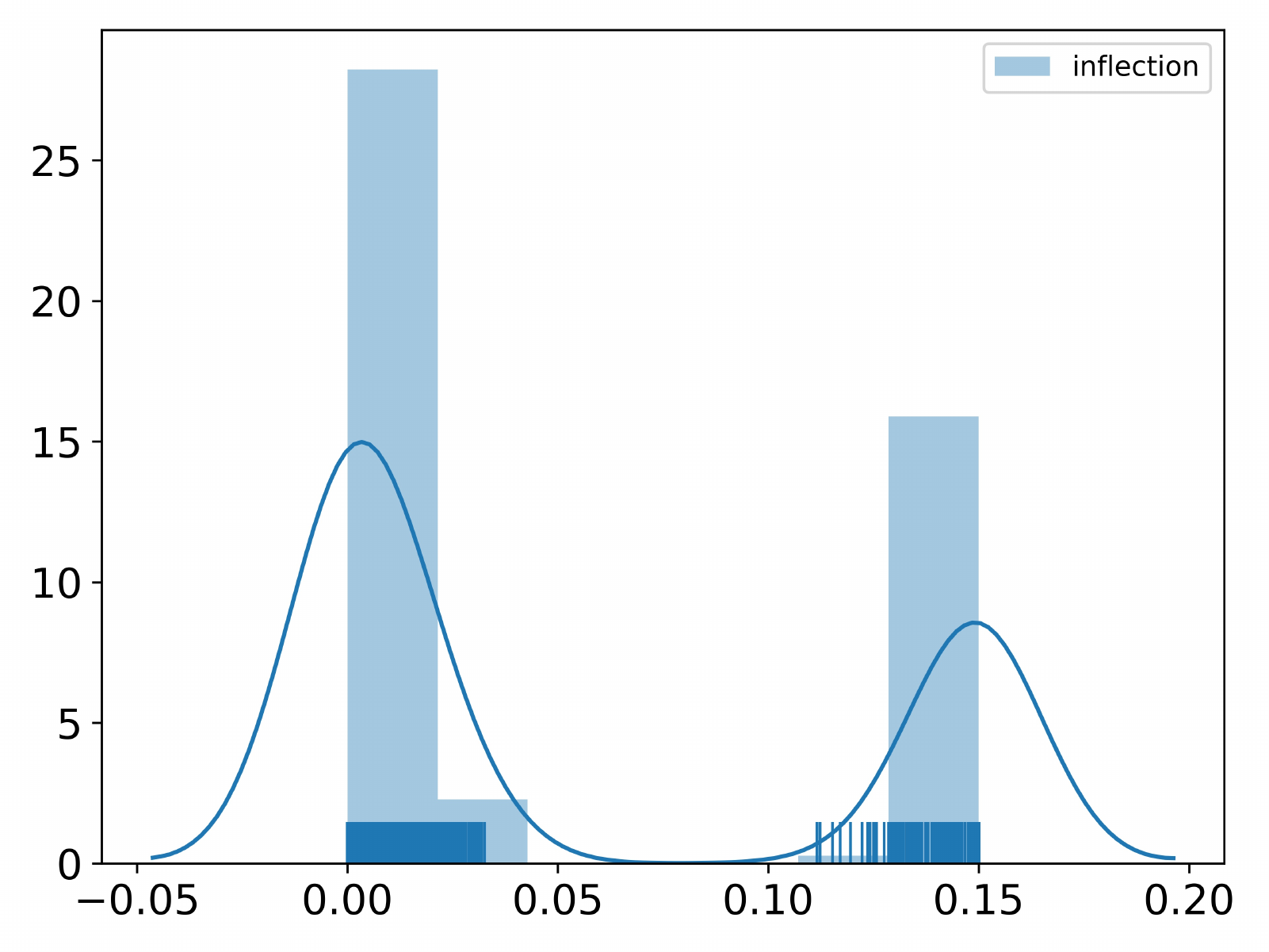}{.49\textwidth}{(c) $\beta$}
\leftfig{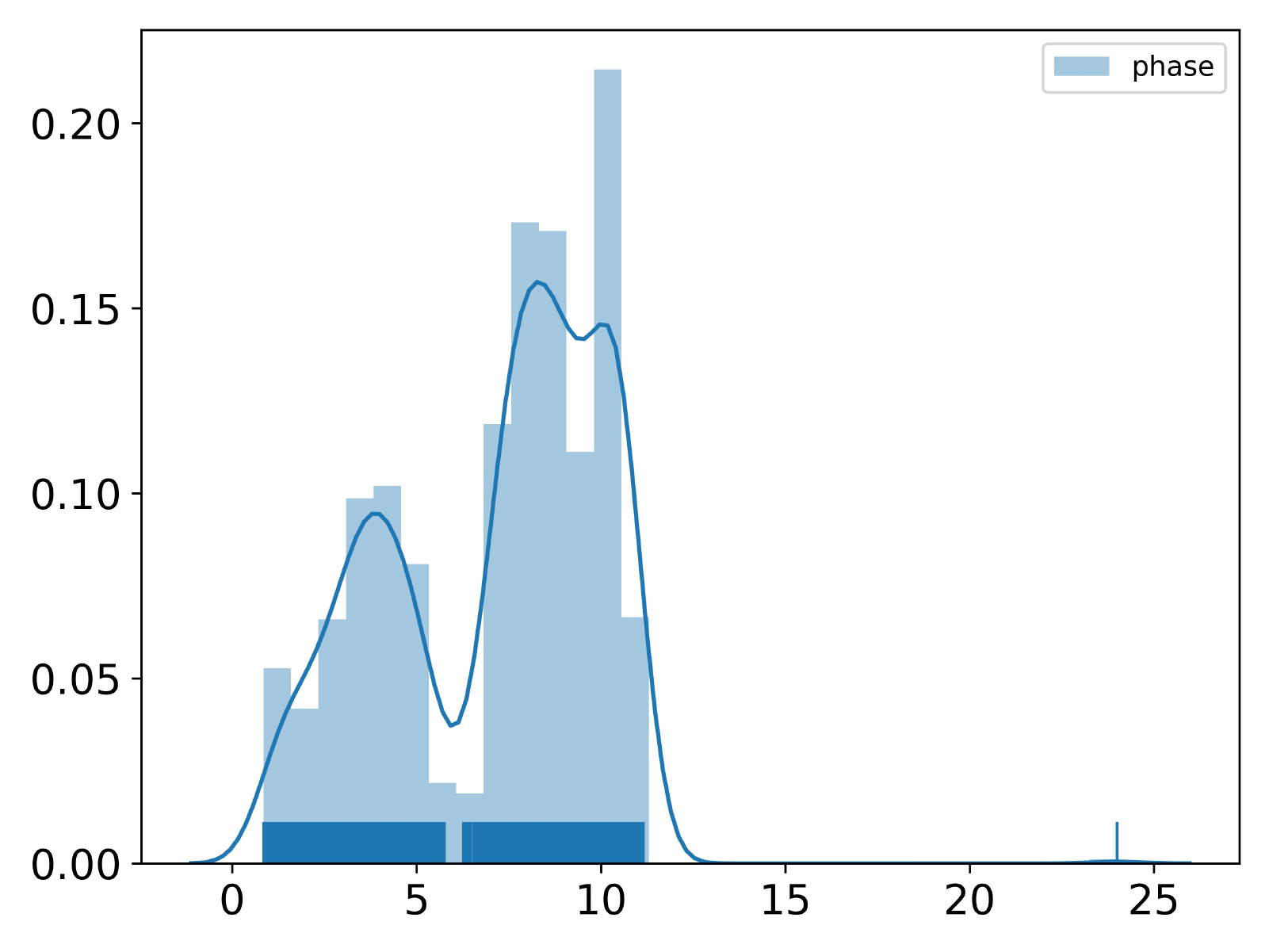}{.49\textwidth}{(d) $\phi$}\hfill}
\caption{PDF of optimized input parameters for $P_0=0.75kPa$.}
\label{fig:PDFP0.75}
\vskip36pt
\end{figure*}

The population size and the number of generation for the GA are chosen as $160$ and $100$, respectively. The overall residual of the fitness function extracted from the FSI cases in Subsection \ref{subsec3:1} is plotted in Figure \ref{fig:GA_residual}. The residual for each phase is normalized by the corresponding maximum nodal displacement. The relative residuals for most of the phases are close to 0 and the maximum relative residual among all the phases is around 0.01, indicating that GA converges well for each glottal shape and therefore the UKE can be used a generalized equation to represent the extracted glottal shapes. Furthermore, the kernel density estimation \cite{freedman2007statistics} is used as a non-parametric way to estimate the probability density function (PDF) of the fitting parameters, and the corresponding PDF for $P_0=0.75kPa$ is plotted in Figure \ref{fig:PDFP0.75}. The PDF for $P_0=0.5kPa$ and $P_0=1.0kPa$ are highly similar and thus not shown.  Note that the PDF of the optimized frequency is not plotted in those figures because the values for all cases are similar and the corresponding PDFs are concentrated at $f=210Hz$. Therefore, to reduce the number of redundant shapes, we fix the value of the optimized frequency to be $f=210Hz$. Based on the PDFs, the generalized glottal shape library can be built by appropriately resampling the parameters. Concretely, we first locate the parameter values with the  local maximum probabilities from each PDF, and then with this located value as the center value, conduct the uniform resampling from each PDF such that the majority of the representative glottal shapes can be included in this library. The re-sampled values of the input parameters under different subglottal pressure conditions are listed in Table \ref{SampleParams}. Note that for different subglottal pressure values, only the amplitude $\xi_m$ is different, and the other parameters are all the same. A total of $N_L=3960$ different shapes are generated by substituting the values in Table \ref{SampleParams} into the UKE, and these shapes constitute the generalized glottal shape library which are used as the raw data for training the DNN in the next section.

\begin{table}[h]
\caption{Resampled values of input parameters.}
\centering
\label{SampleParams}
\begin{tabular}{ccccc}
\hline\hline
$P_0 (kPa)$ & $\xi_{m}$ & $\alpha$ & $\beta$ & $\phi$ \\
\hline
\multicolumn{1}{c}{0.5} & \multicolumn{1}{c}{\begin{tabular}{@{}c@{}}0.02, 0.03, \\ 0.04, 0.1\end{tabular}} & \multirow{5}{*}{\begin{tabular}{@{}c@{}}0.0, 0.2, 0.4, \\ 0.6, 0.8, 1.0\end{tabular}} & \multirow{5}{*}{\begin{tabular}{@{}c@{}}0.0, 0.015, 0.03, \\ 0.135, 0.15\end{tabular}} & \multirow{5}{*}{\begin{tabular}{@{}c@{}c@{}}1, 2, 3, 4, \\ 5, 6, 7, 8, \\ 9, 10, 11\end{tabular}}\\
\multicolumn{1}{c}{0.75} & \multicolumn{1}{c}{\begin{tabular}{@{}c@{}}0.025, 0.04, \\ 0.055, 0.1\end{tabular}} &  &  & \\
\multicolumn{1}{c}{1.0} & \multicolumn{1}{c}{\begin{tabular}{@{}c@{}}0.035, 0.055, \\ 0.075, 0.1\end{tabular}} &  &  & \\
\hline\hline
\end{tabular}
\end{table}

\section{\label{sec:4} Implementation of the DNN Model}

For each shape in the generalized glottal shape library, the subglottal pressure $P_0$ and the parameters $\xi_m$, $\alpha$, $\beta$ and $\phi$ are the input features, and the corresponding output targets are the flow rate $Q$ and the pressure distribution $P_i$, where $i$ is the index of the discretized cross sections in the inferior-superior direction of the vocal folds. The ground truth values of the flow rate $Q$ and pressure distribution $P_i$ are obtained by solving the N-S equations. Then, the mapping relationship between the input features and the corresponding output targets can be established by a fully-connected DNN as follows:
\bea \label{eq:000}
Q, P_i=f(P_0,\xi_m,\alpha,\beta,\phi;\theta)
\eea
where $f$ is the function representing the overall DNN, and $\theta$ denotes all learnable parameters of the DNN. With this trained DNN, the flow rate and pressure distribution along any glottal shape generated by the UKE can be well predicted.

\subsection{N-S Solution of the Output Targets}
\begin{figure}[!ht]
  \begin{center}
	\includegraphics[width=0.8\textwidth]{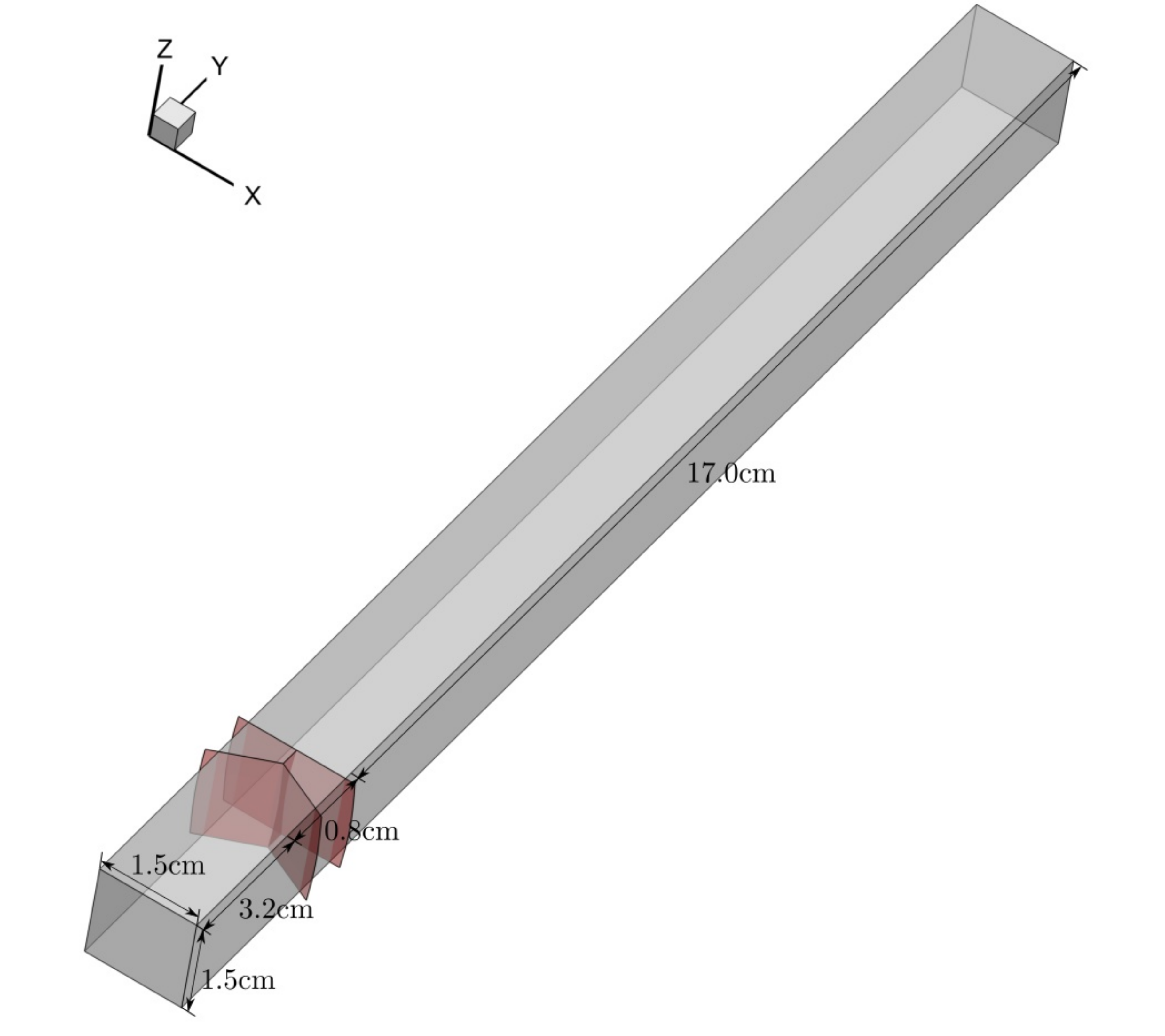}
  \end{center}
  \caption{Setup of the computational domain.}
  \label{fig:ComputDomain}
\end{figure}

\begin{figure*}
\baselineskip=12pt
\figline{\fig{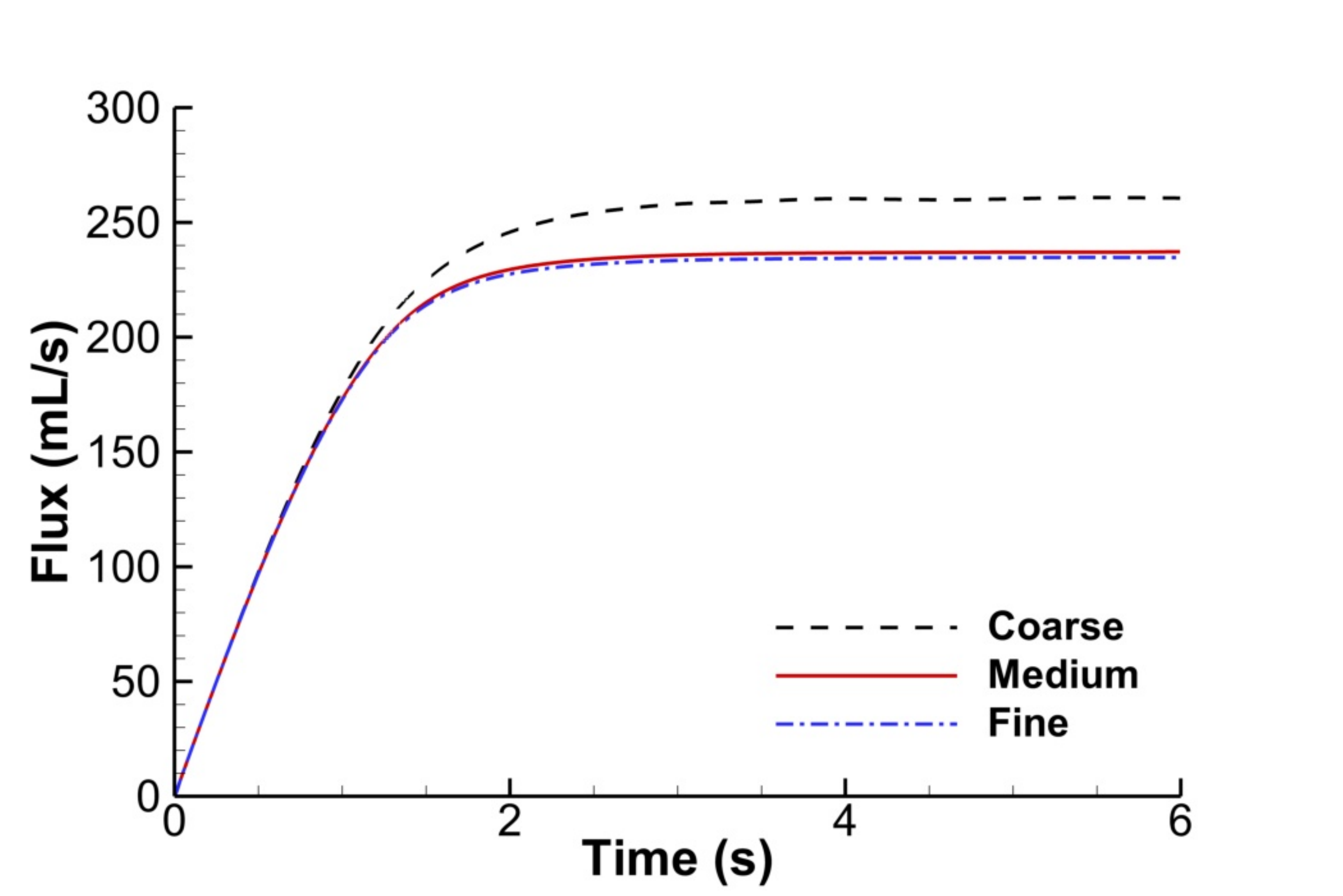}{.45\textwidth}{(a) flow rate}
\fig{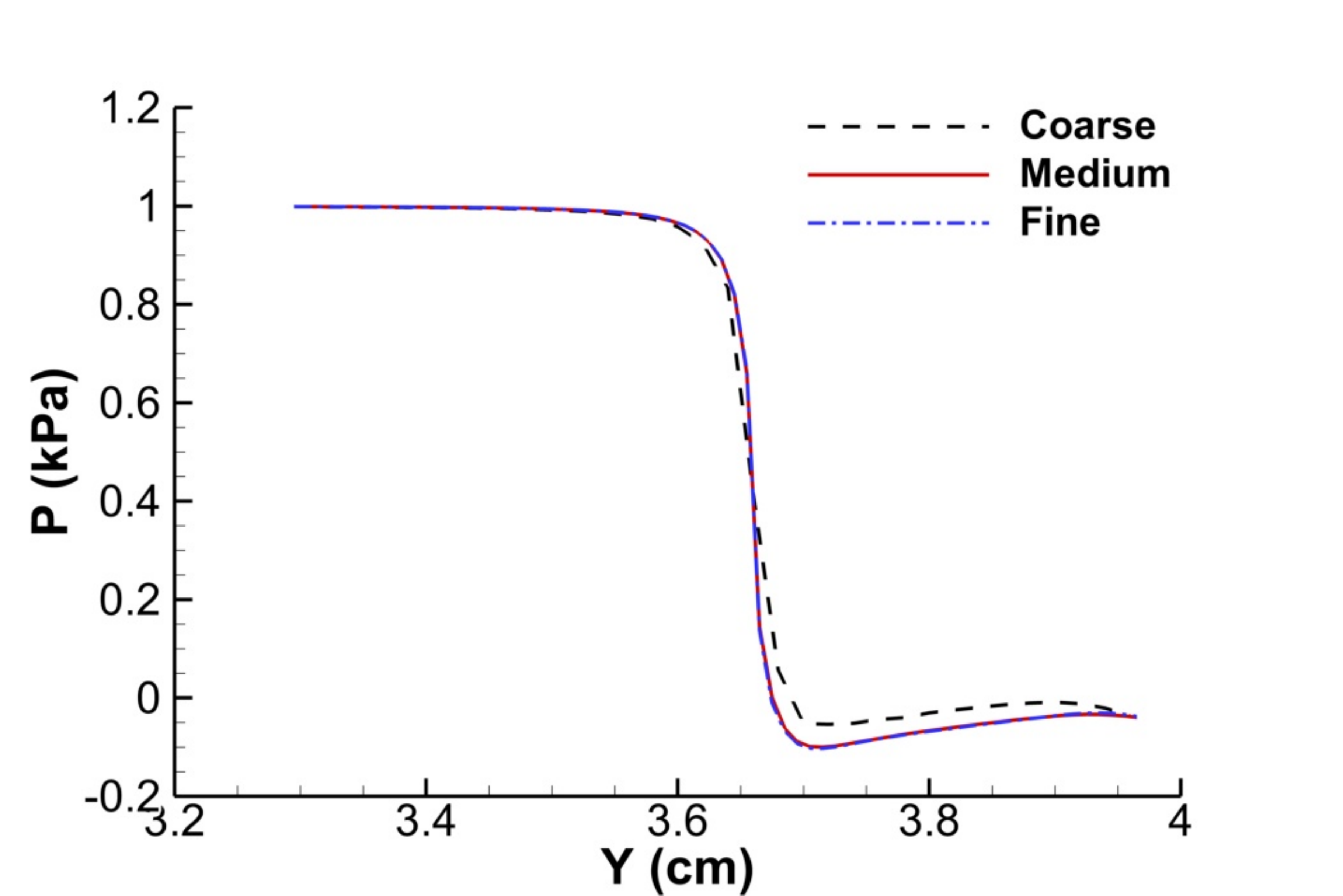}{.45\textwidth}{(b) average pressure distribution}}
\caption{Grid independence results.}
\label{fig:GridIndependent}
\end{figure*}

The fluid flow is governed by the incompressible N-S equations as follows,
\bea
& & \frac{\partial u_i}{\partial x_i} = 0  \\
& & \frac{\partial u_i}{ \partial t} + \frac{\partial u_i u_j}{ \partial x_j} = -\frac{1}{\rho_f}\frac{\partial p }{\partial x_i} + \nu_f\frac{\partial^{2}u_i }{\partial x_j \partial x_j},
\eea
where $u_i$, $\rho$, $p$, $\nu$ are the incompressible flow velocity, density, pressure, and kinematic viscosity, respectively. An in-house sharp-interface immersed-boundary N-S flow solver \cite{zheng2010coupled} is used to obtain the ground truth solution of the output targets. The setup of the computational domain is depicted in Figure \ref{fig:ComputDomain}. The size of the computational domain is $1.5cm\times21.0cm\times1.5cm$ in the $x$ (lateral), $y$ (inferior-superior) and $z$ (anterior-posterior) direction. The vocal folds are placed $3.2cm$ and $17.0cm$ away from the inlet and outlet of the computational domain, respectively. The grid independence study is performed by comparing the flow rate and average pressure distribution on coarse, medium and fine meshes with fixed $CFL$ number. The mesh number $N_x \times N_y \times N_z$ on the coarse, medium and fine meshes are $64 \times 64 \times 24$, $128 \times 128 \times 48$, and $256 \times 256 \times 96$ in the $x$, $y$ and $z$ direction, respectively, where $N_x$, $N_y$ and $N_z$ are the number of mesh nodes in the $x$, $y$ and $z$ direction, respectively. The mesh is stretched to the far field in the $x$ and $y$ direction, while uniformly distributed in the $z$ direction. The grid independence results of the flow rate and average pressure distribution are illustrated in Figure \ref{fig:GridIndependent}. From this figure, we can see that the medium mesh is adequate to obtain the ground truth solution of the output targets from the shape library. The relative error of the flow rate obtained on the coarse and medium mesh with respect to that obtained on the fine mesh are $12.1\%$ and $1.0\%$, respectively. The minimum interval of the medium mesh is $0.003cm$ and $0.01cm$ in the $x$ and $y$ direction, respectively. Moreover, the total CPU time required for convergence on the coarse, medium and fine meshes are respectively 0.2, 2.3 and 35 hours on a parallel computer with 32 CPUs. 

\subsection{Implementation Details of the DNN}

As mentioned above, the input features and corresponding output targets extracted from the shape library can be organized as a vector $\boldsymbol{x}$ and $\boldsymbol{y}$, respectively,

\bea
\boldsymbol{x}= \begin{bmatrix}
P_0 \\
\xi_{m} \\
\alpha \\
\beta \\
\phi \\
\end{bmatrix},\quad
\boldsymbol{y}= \begin{bmatrix}
Q \\
P_1 \\
P_2 \\
\vdots \\
\vdots \\
P_{N_P} \\
\end{bmatrix},
\eea
where $N_P=68$ is the dimension of the output pressure distribution.

The mapping relationship between the input features $\boldsymbol{x}$ and corresponding output targets $\boldsymbol{y}$ can be established by a fully-connected DNN \cite{goodfellow2016deep,lecun2015deep}. In the fully-connected DNN, the input and output layers are denoted as $\boldsymbol{z}_0$ and $\boldsymbol{z}_L$, respectively. The layers between the input and output layers are called the hidden layers $\boldsymbol{z}_l$, where $l=1,...,L-1$. Neurons in the hidden layer $\boldsymbol{z}_l$ have connections to all neurons of the previous layer $\boldsymbol{z}_{l-1}$,
\bea
& & \boldsymbol{z}_{l} = \sigma_{l}(\boldsymbol{W}_{l}^{T} \boldsymbol{z}_{l-1} + \boldsymbol{b}_{l})
\eea
where $\boldsymbol{W}_{l}$ is the learnable weights, $\boldsymbol{b}_{l}$ is the additive bias, and $\sigma_{l}$ is the nonlinear activation function.

The loss function $J$ of the DNN is
\bea
& & J=\frac{1}{N}\sum\left\|\boldsymbol{z}_{L}-\boldsymbol{y}\right\|^2_2+\lambda \left\|\boldsymbol{W}\right\|_2
\eea
where $\boldsymbol{z}_{L}$ is the predicted value and $\lambda$ is the regularization coefficient to prevent the overfitting of the DNN model.

Note that the range of values of $Q$ and $P_i$ are different, i.e., $Q\geq 0$ while $P_i/P_0\leq 1$, therefore for the ease of training the DNN, the input features $\boldsymbol{x}$ are respectively mapped to the subsets of the output targets $\boldsymbol{y}$ (i.e., $Q$ and $P_i$) with different architectures of the DNN.

The whole data set from the shape library is randomly split into the training and test sets. To avoid the overfitting of the model, we use $5$-fold cross validation \cite{goodfellow2016deep} to fine tune the architecture and hyperparameters of the DNN, such as the number of hidden layers, the number of neurons on each hidden layer, the initialization of the weights, the activation function, the optimization method, the mini-batch size, and the number of epochs \cite{goodfellow2016deep}. The final architecture and hyperparameters of the DNN are chosen from those that have the lowest errors on the validation set. The final DNN model is then trained on the full training set, and the prediction performance of the trained model is evaluated on the test set.

\begin{figure}[h]
\baselineskip=12pt
\figcolumn{
\fig{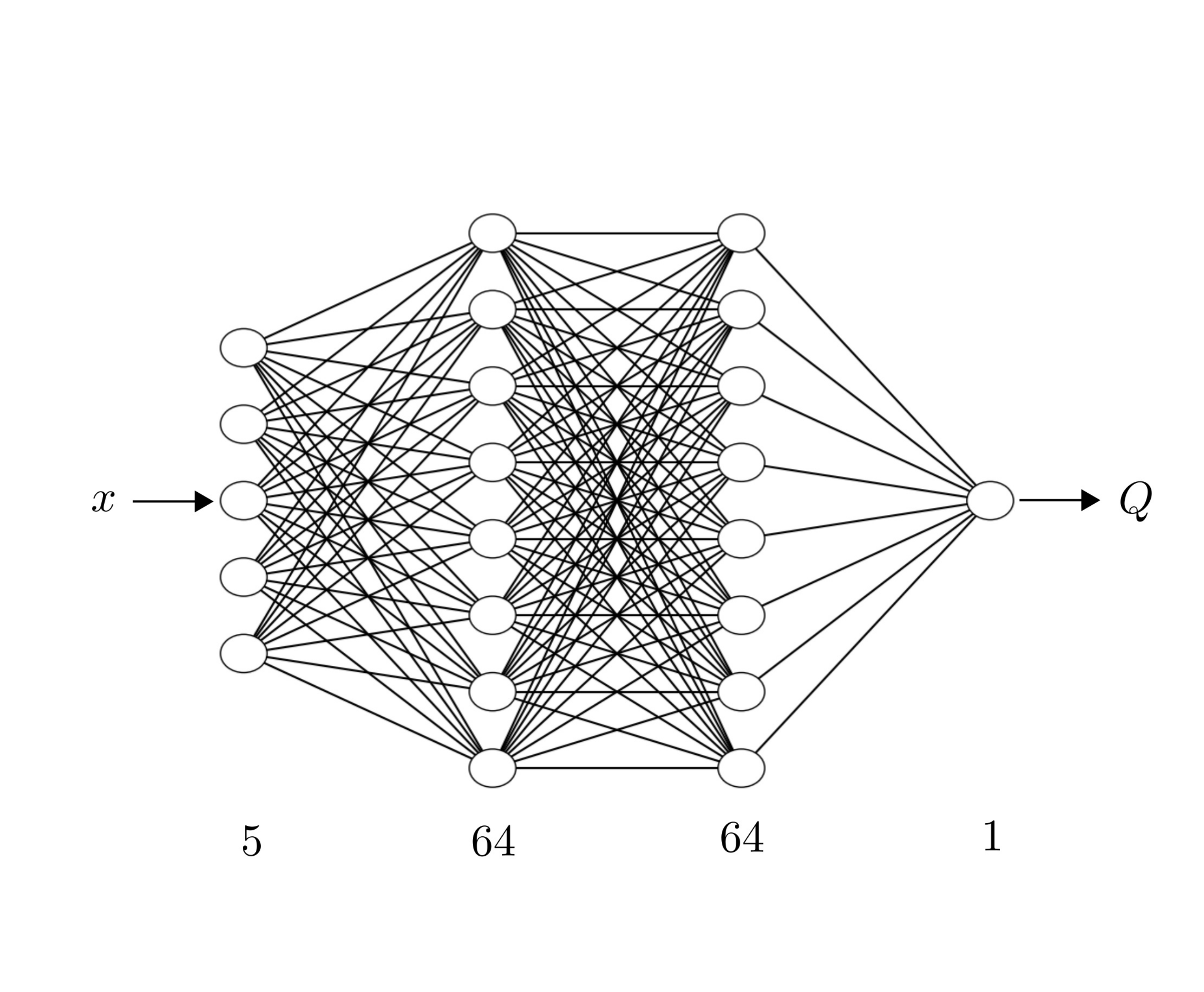}{.7\textwidth}{(a) DNN-Q}
\fig{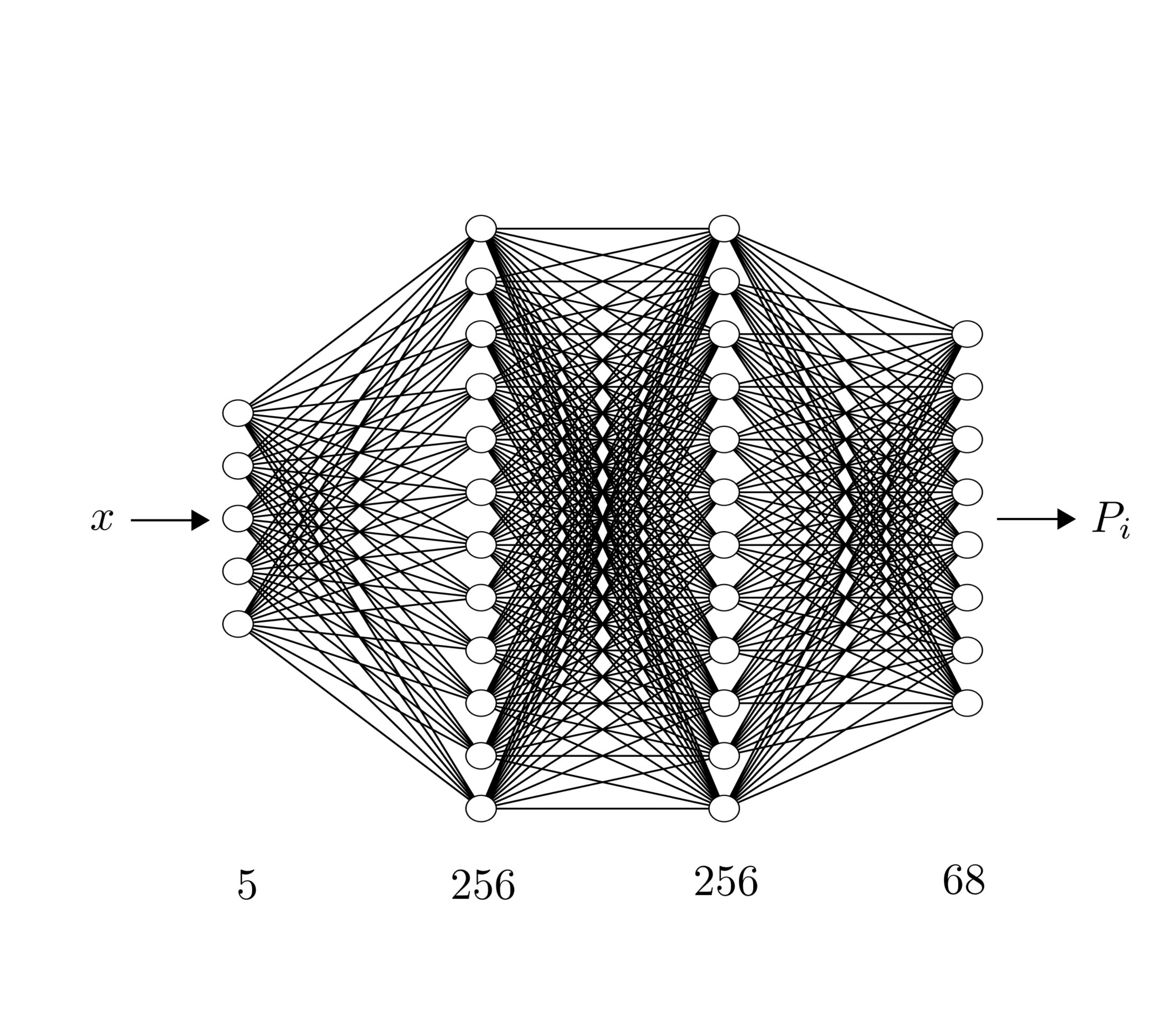}{.7\textwidth}{(b) DNN-P}
}
\caption{Architecture of the DNN.}
\label{fig:NeuralNetwork}
\end{figure}

The final architectures of the DNN for $Q$ and $P_i$ are illustrated in Figure \ref{fig:NeuralNetwork} and denoted as DNN-Q and DNN-P, respectively. The input layer for both DNNs has 5 neurons which correspond to the dimension of the input vector. The output layer of DNN-Q has a single neuron which corresponds to the ground truth value of the flow rate $Q$, while that of DNN-P has 68 neurons which correspond to the ground truth value of the pressure distribution on the discretized cross sections along the inferior-superior direction of the vocal folds. Since $Q$ and $P_i$ are bounded by different ranges ($Q\geq 0$ and $P_i/P_0\leq 1$), the softplus and tanh activation function \cite{goodfellow2016deep} are used on the output layer of DNN-Q and DNN-P, respectively. Besides the input layer and output layer, there are two hidden layers for both DNNs. The number of neurons on the hidden layers of DNN-Q are 64, and the softplus activation function is used on each hidden layer, whereas the number of neurons on the hidden layers of DNN-P are 256, and the relu activation function \cite{goodfellow2016deep} is used on each hidden layer. All of the weights on each layer are initialized with a random normal distribution. Both of the DNN models are optimized using a mean-squared loss function with an adaptive version of the stochastic gradient descent algorithm called Nadam (Nesterov Adam) \cite{ruder2016overview}. Both of the DNN models are trained with 10000 epochs, where one epoch consists of one full training cycle on the training set, and the mini-batch size is 128 for each epoch. The DNN models are implemented on the open-source machine learning platform Keras \cite{chollet2015keras} using TensorFlow \cite{tensorflow2015-whitepaper} as the backend. 

\subsection{Evaluation of the Trained DNN Models}
The relative percent difference (RPD) between the true and predicted outcomes is used to evaluate the trained DNN models. The expression of the RPD for $Q$ and $P_i$ for each glottal shape in the training data are as follows:
\bea
E_Q = \frac{\abs{Q-\hat{Q}}}{max(\abs{Q},\abs{\hat{Q}})} 
\eea
\bea
E_P = \frac{\sum_{i=1}^{N_P}\frac{\abs{P_i-\hat{P_i}}}{max(\abs{P_i},\abs{\hat{P_i}})}}{N_P}
\eea
where $Q, P_i$ and $\hat{Q}, \hat{P_i}$ are respectively the true and predicted outcomes.

\begin{figure}[h]
\baselineskip=12pt
\figcolumn{
\fig{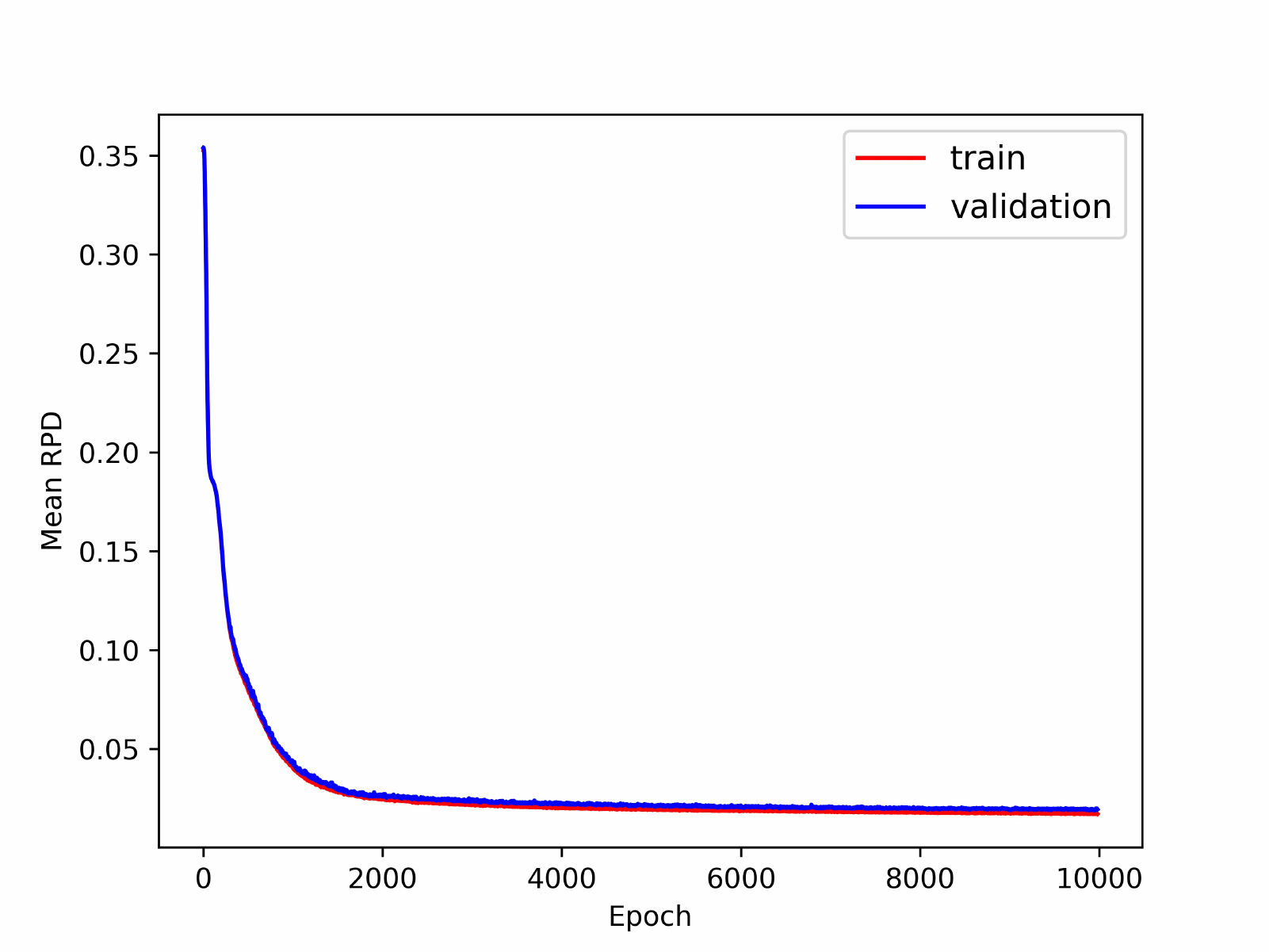}{.7\textwidth}{(a) DNN-Q}
\fig{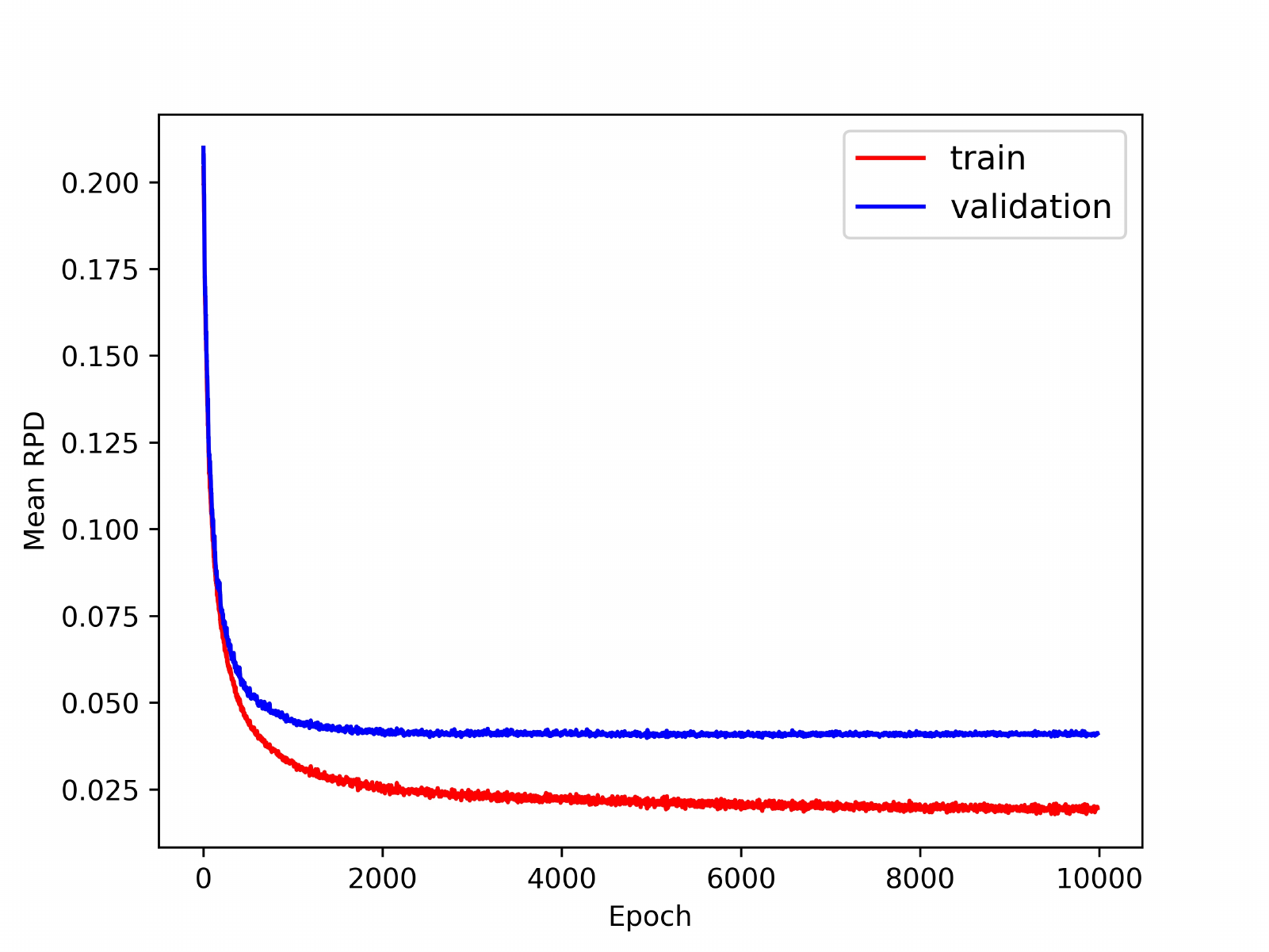}{.7\textwidth}{(b) DNN-P}
}
\caption{5-fold cross validation results.}
\label{fig:TitzeKfoldCV}
\end{figure}

\begin{figure}[h]
\baselineskip=12pt
\figcolumn{
\fig{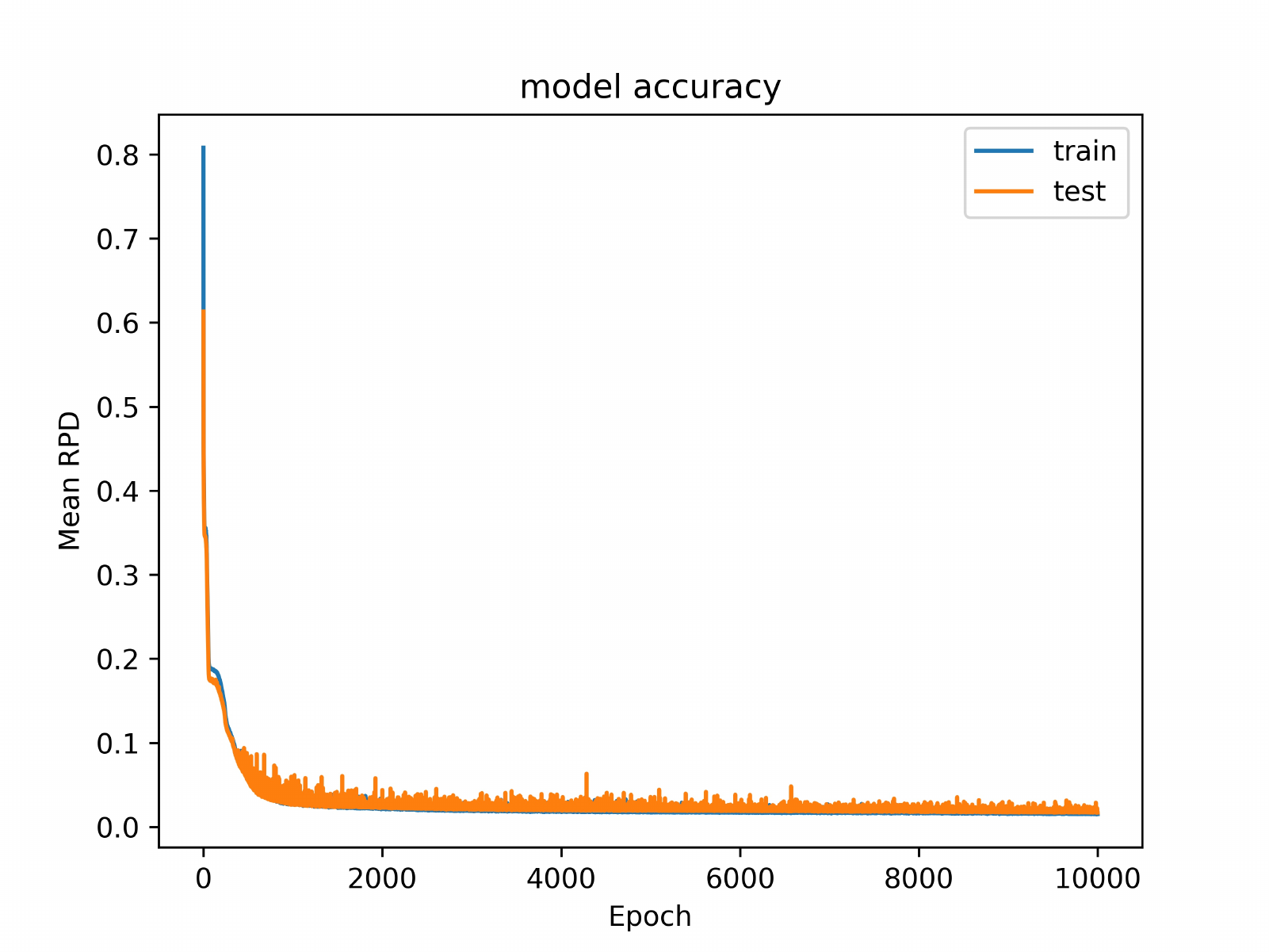}{.7\textwidth}{(a) History of the model accuracy}
\fig{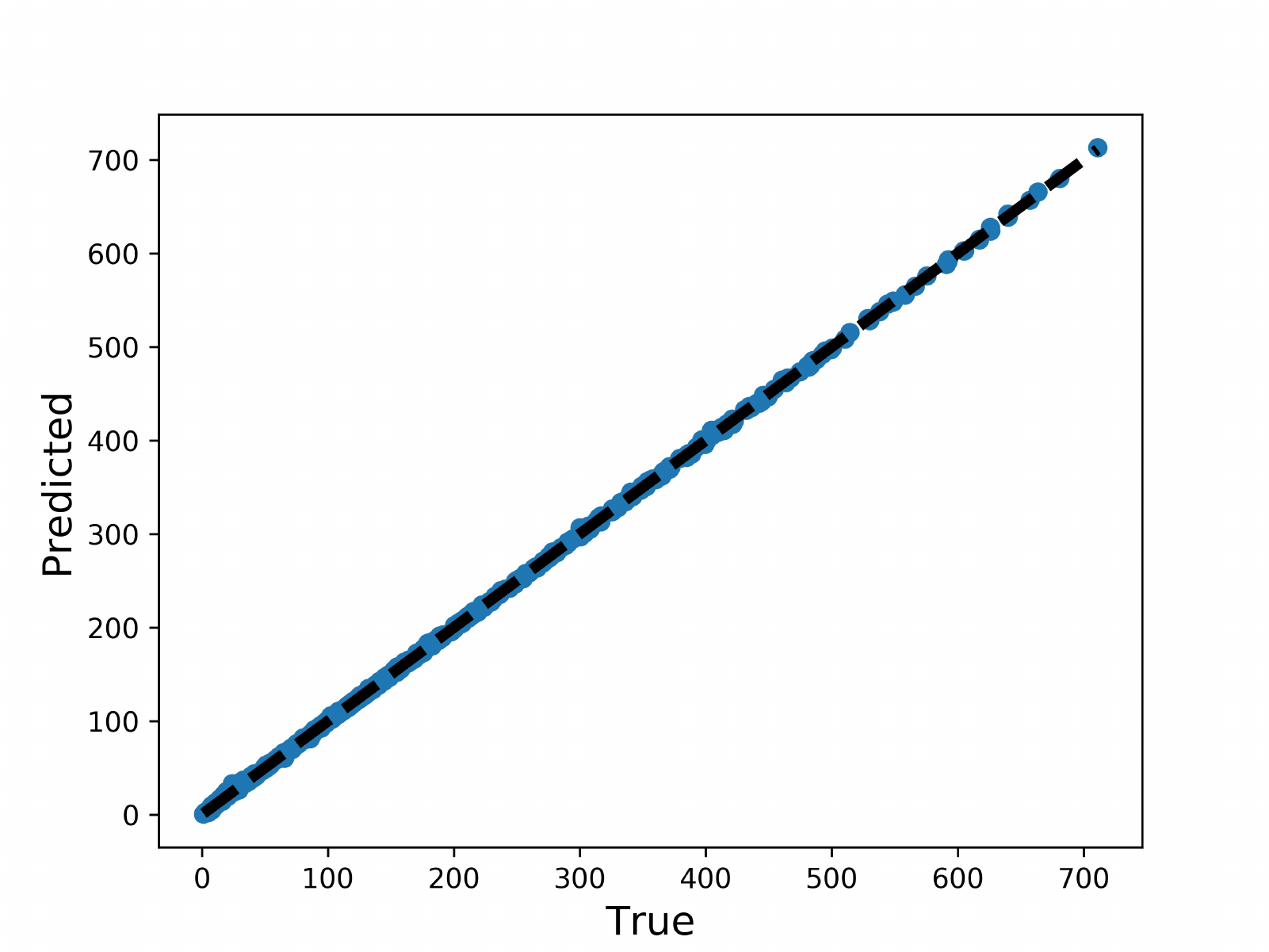}{.7\textwidth}{(b) Scatter plot}
}
\caption{Performance of the trained DNN-Q model on the test set.}
\label{fig:testQ}
\end{figure}

\begin{figure}[h]
\baselineskip=12pt
\figcolumn{
\fig{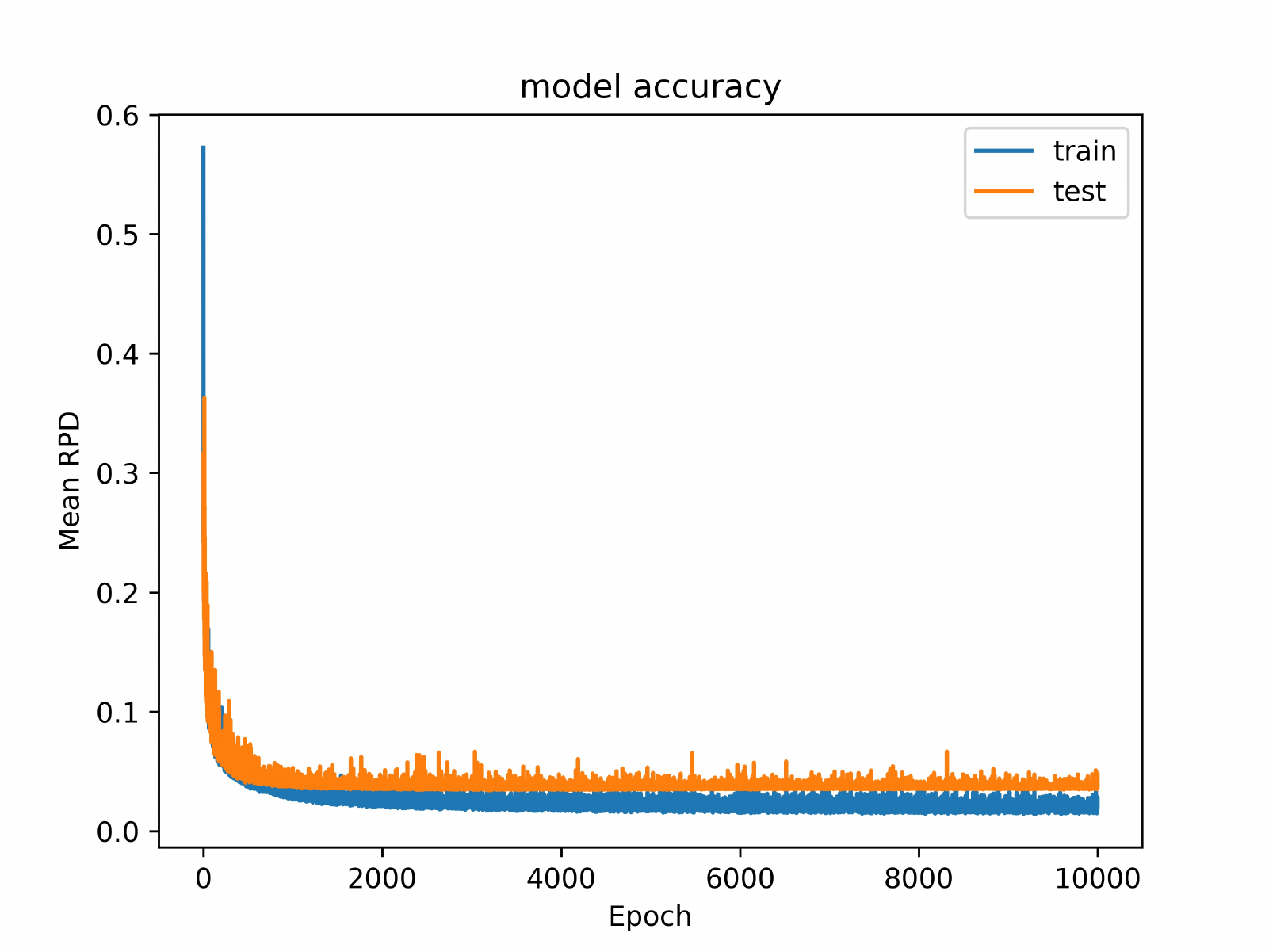}{.7\textwidth}{(a) History of the model accuracy}
\fig{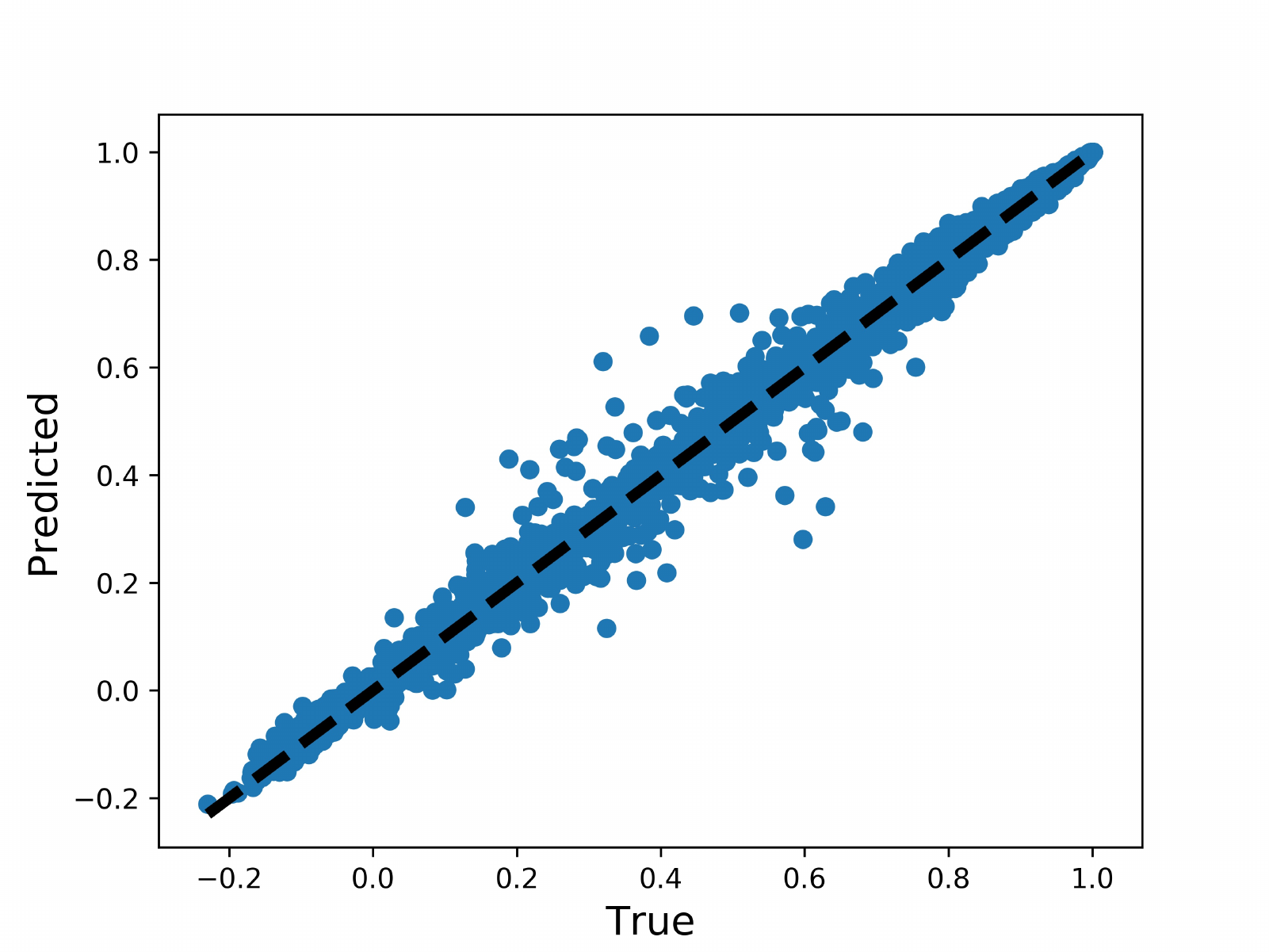}{.7\textwidth}{(b) Scatter plot}
}
\caption{Performance of the trained DNN-P model on the test set.}
\label{fig:testP}
\end{figure}

The history of the 5-fold cross validation results for DNN-Q and DNN-P is plotted in Figure \ref{fig:TitzeKfoldCV}. The horizontal axis corresponds to the number of epochs, and the vertical axis corresponds to the mean RPD between the true and predicted outcomes. The comparison is between the training and validation sets. It took 10000 epochs for the mean RPD on the training and validation sets to converge for DNN-Q and DNN-P. The converged mean RPD on the training and validation sets are $1.71\%$ and $1.89\%$ for DNN-Q, and $1.97\%$ and $4.12\%$ for DNN-P, respectively. The performance of the trained DNN-Q and DNN-P on the test set is plotted in Figure \ref{fig:testQ} and \ref{fig:testP}, respectively. The first subfigure for each figure shows the history of the model accuracy where the horizontal and vertical axes correspond to the number of epochs and mean RPD, respectively. The comparison is between the full training and test sets. After running 10000 epochs, the mean RPD on the test set converges at $1.74\%$ and $3.52\%$ for DNN-Q and DNN-P, respectively. The second subfigure illustrates the scatter plot of the true and predicted outcomes on the test set, and the good prediction performance on the test set for both DNN-Q and DNN-P can be observed. The final mean RPD on the training, validation and test sets for DNN-Q and DNN-P are summarized in Table \ref{meanRPDerror}.

\begin{table}[!ht]
\caption{Mean RPD on the training, validation and test sets.}
\begin{center}
\label{meanRPDerror}
\begin{tabular}{c c c c}
\hline\hline
 & train & validation & test\\
\hline
$Q$ & $1.71\%$ & $1.89\%$ & $1.74\%$ \\
$P_i$ & $1.97\%$ & $4.12\%$ & $3.52\%$ \\
\hline\hline
\end{tabular}
\end{center}
\end{table}

\begin{figure*}
\baselineskip=12pt
\figline{\leftfig{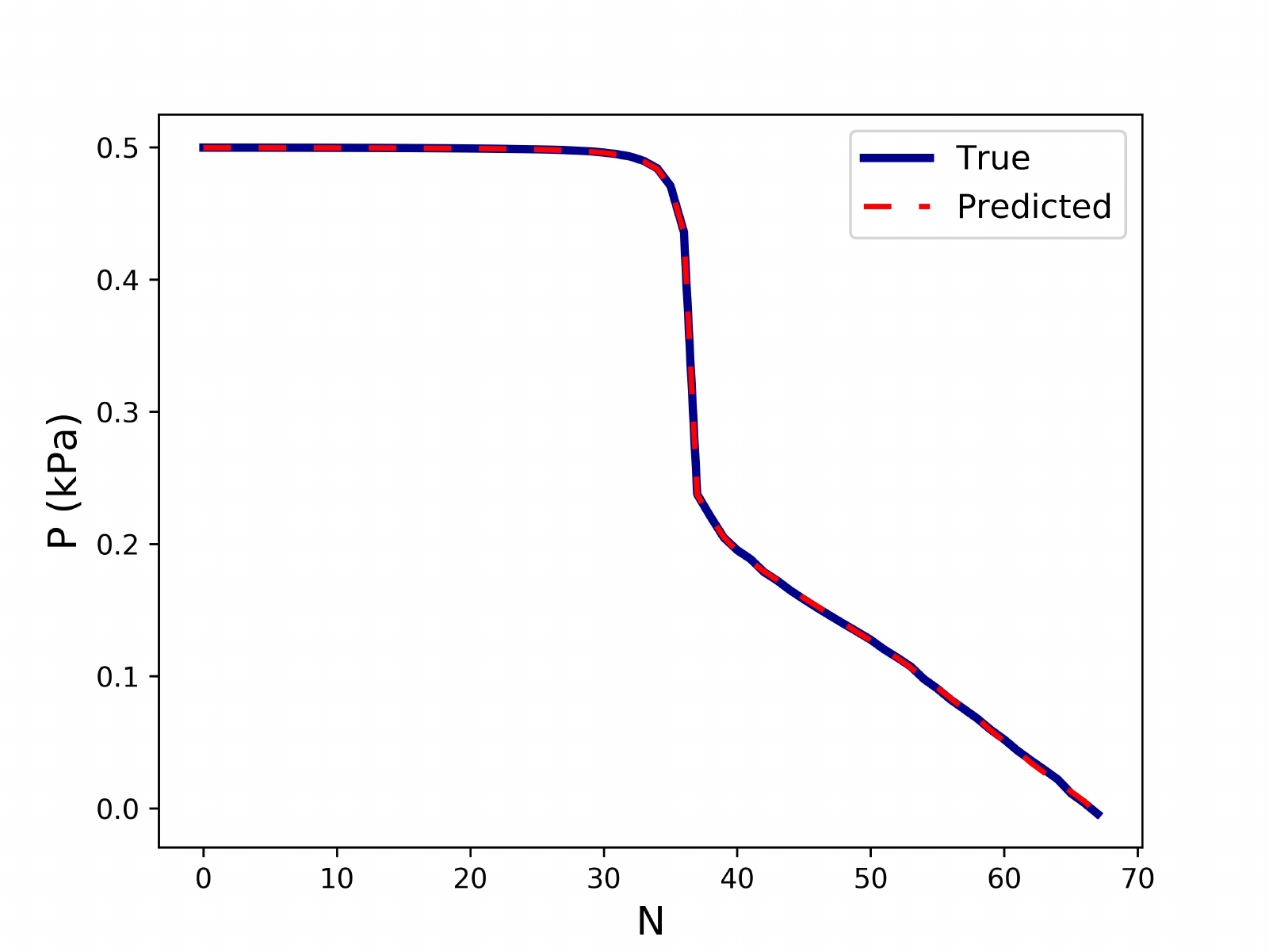}{.49\textwidth}{(a)}
\leftfig{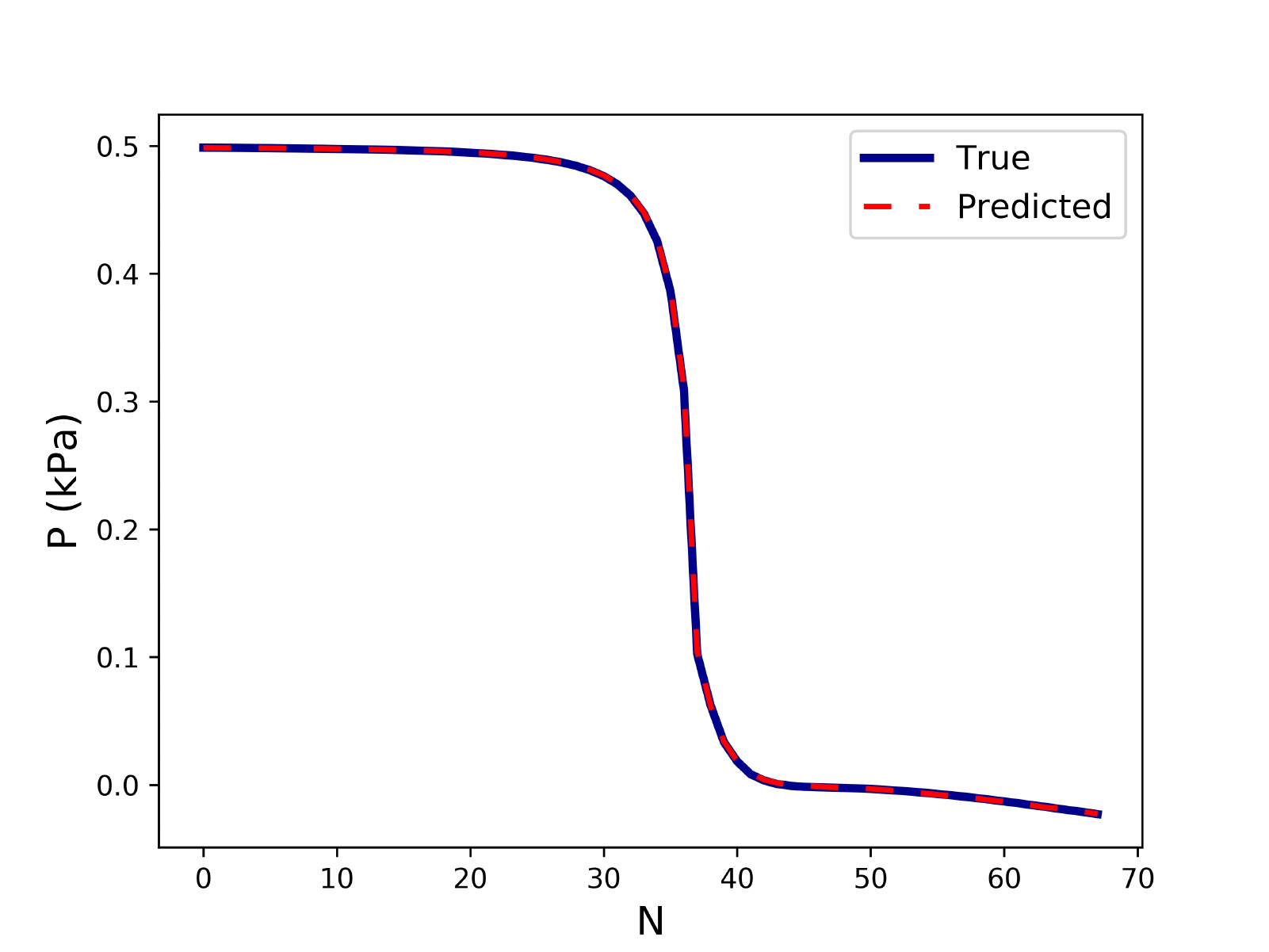}{.49\textwidth}{(b)}\hfill}
\figline{\leftfig{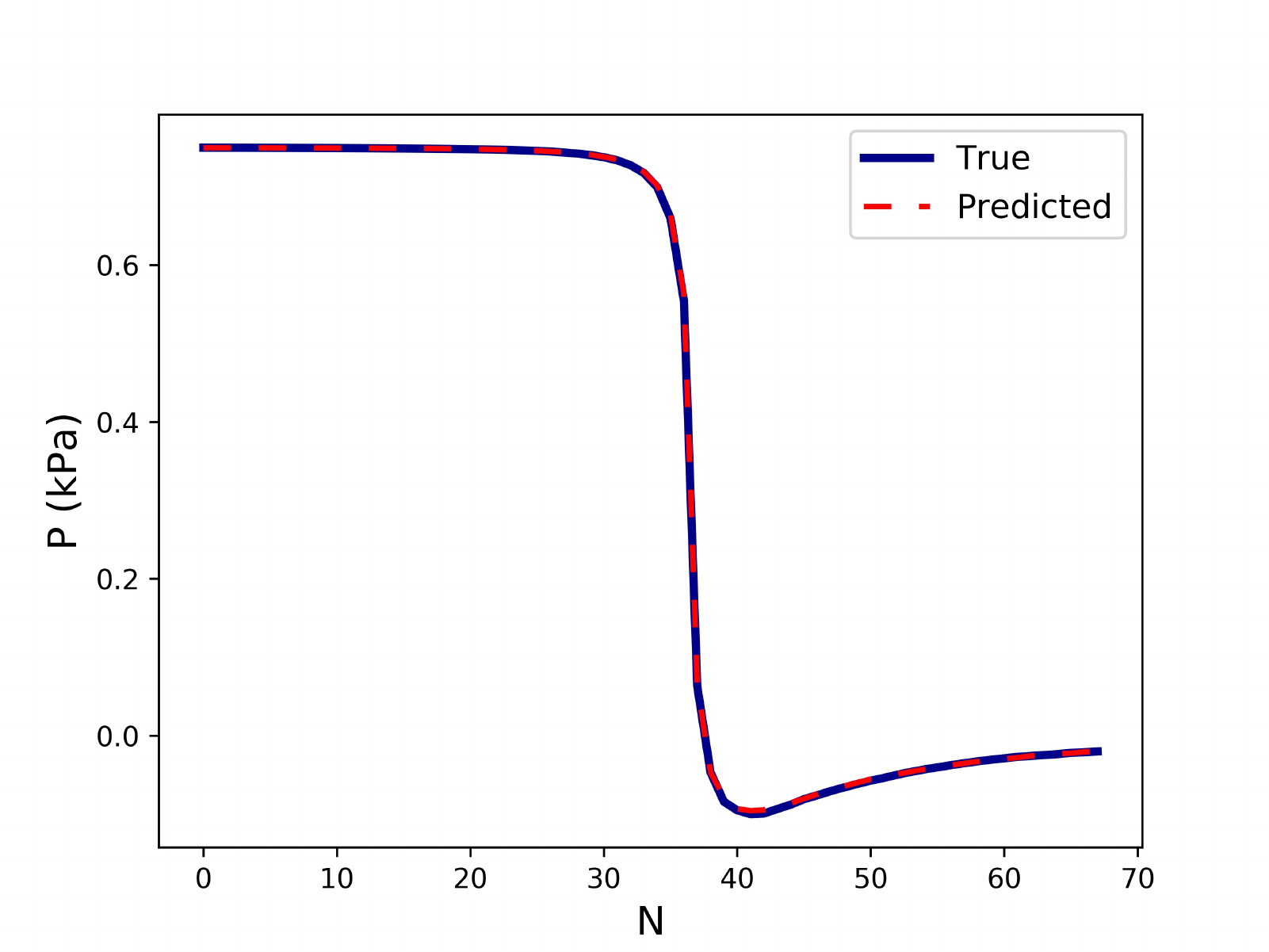}{.49\textwidth}{(c)}
\leftfig{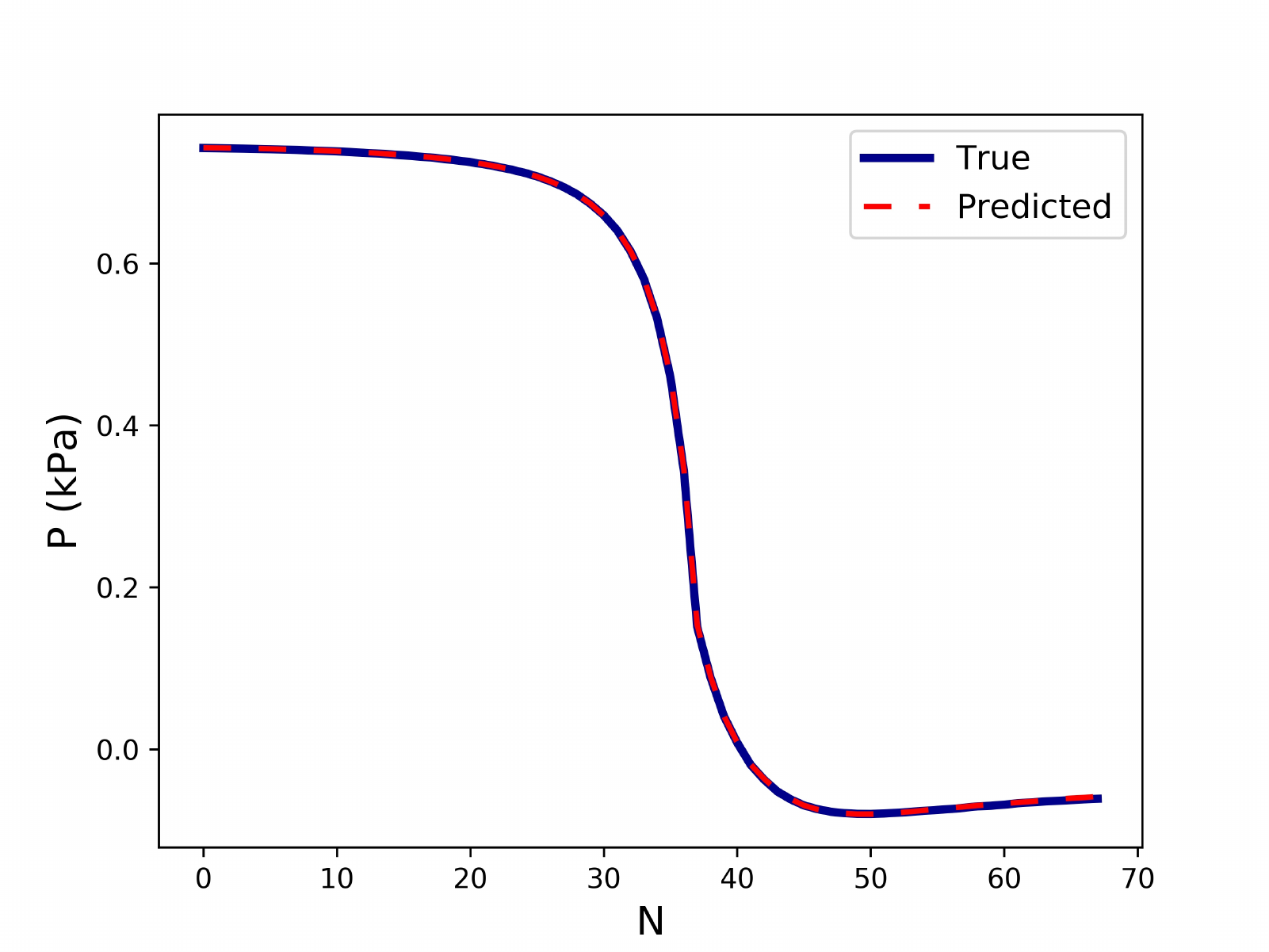}{.49\textwidth}{(d)}\hfill}
\figline{\leftfig{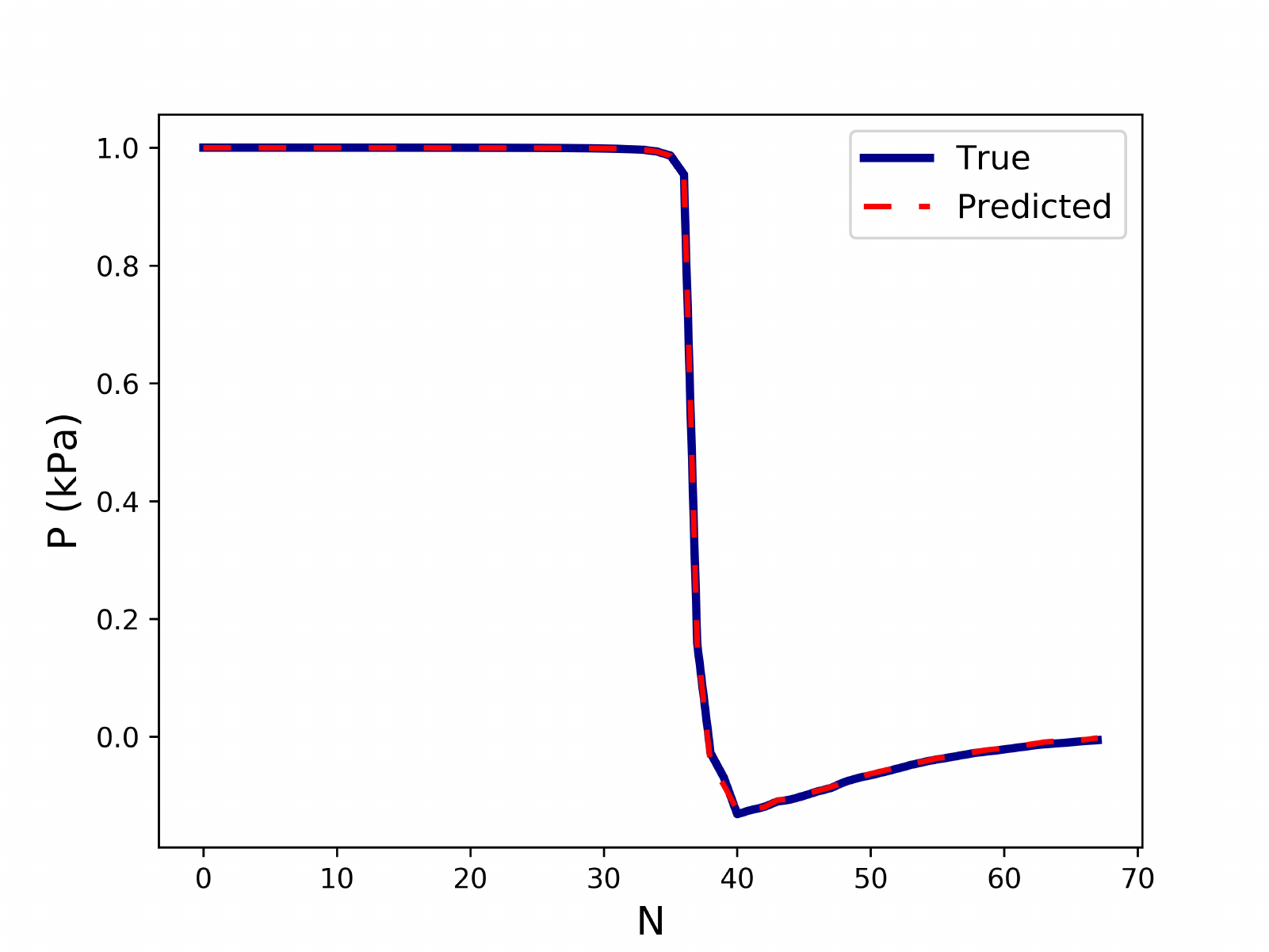}{.49\textwidth}{(e)}
\leftfig{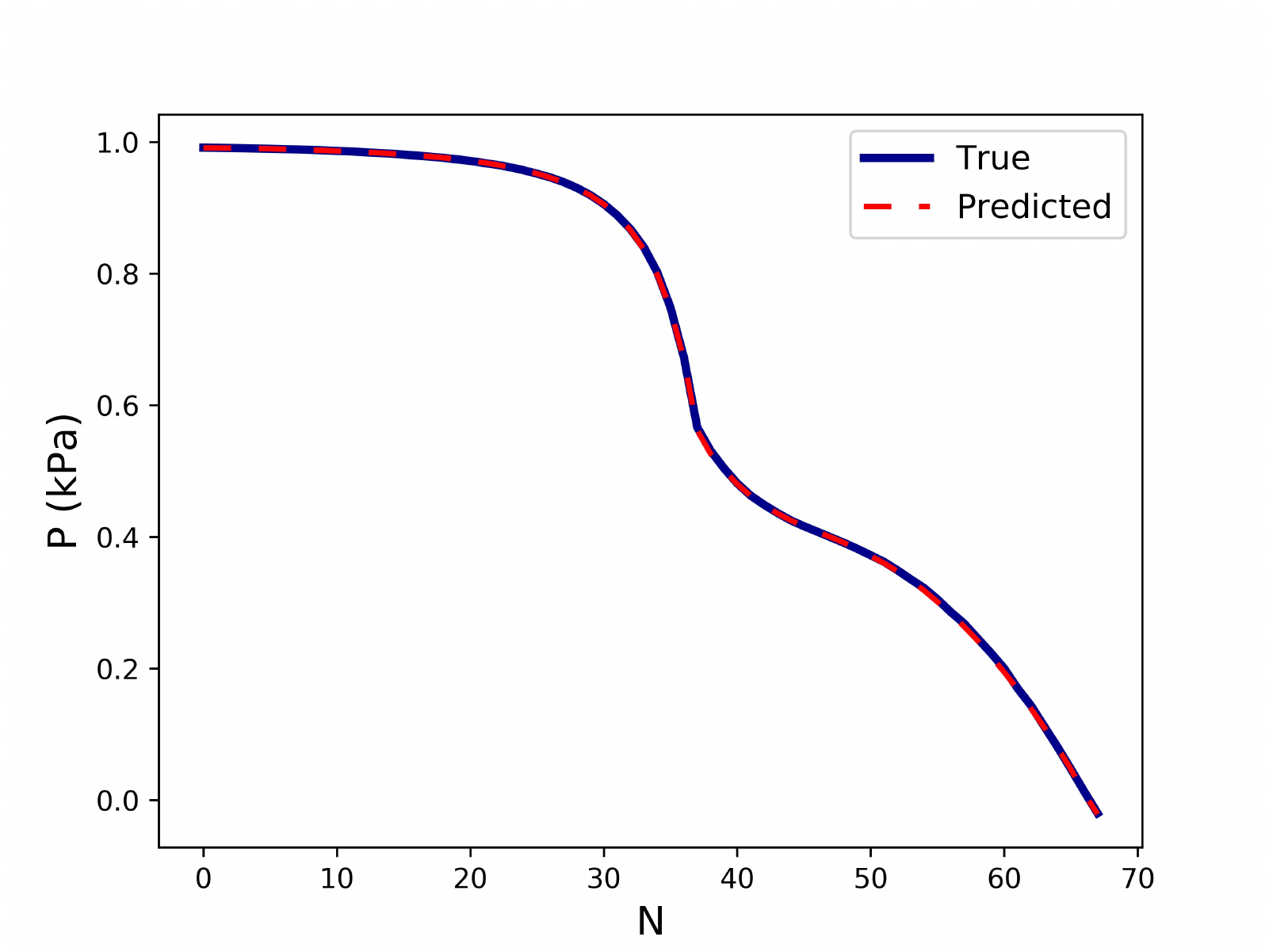}{.49\textwidth}{(f)}\hfill}
\caption{Comparison of the true and predicted pressure distribution.}
\label{fig:DNNPCompare}
\end{figure*}

Furthermore, 6 shapes under different subglottal pressures are randomly selected from the test set, and the comparison of the true and predicted pressure distribution of these shapes are shown in Figure \ref{fig:DNNPCompare}. From these figures, we can observe that the pressure distribution can be well predicted by the trained DNN-P model.

\begin{figure}[!ht]
  \begin{center}
	\includegraphics[width=1.0\textwidth]{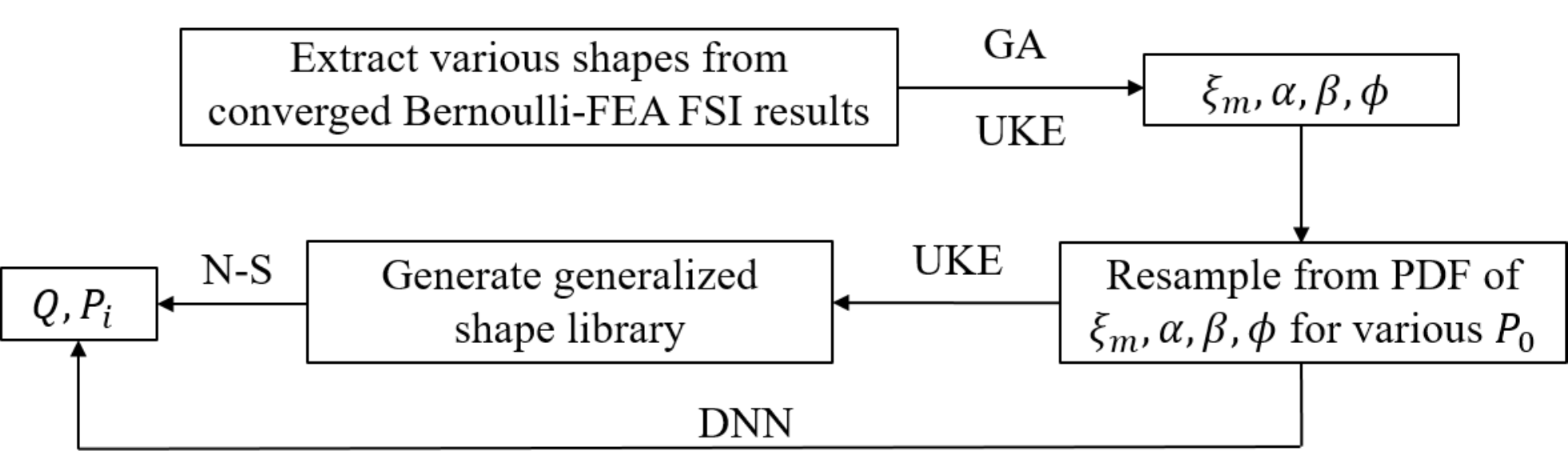}
  \end{center}
  \caption{Diagram of the implementation of the reduced-order flow solver.}
  \label{fig:workflow_rom_fluid}
\end{figure}

To summarize, the diagram of the implementation of the present reduced-order flow solver is illustrated in Figure \ref{fig:workflow_rom_fluid}. Concretely, it is divided into the following steps: firstly, various glottal shapes are extracted from 300 converged Bernoulli-FEM FSI results under different subglottal pressure and material properties. Secondly, these extracted shapes are fitted with the UKE using the GA and the PDF of the fitted input parameters of the UKE are determined. Thirdly, 3960 different glottal shapes are generated by appropriate resampling from the PDF of the input parameters with high probabilities and then substituting them into the UKE, which constitute the generalized shape library. Fourthly, for each shape in the library, the ground truth values of the flow rate $Q$ and pressure distribution $P_i$ are obtained by solving the N-S equation. Finally, the mapping relationship between the input parameters together with the subglottal pressure (input features) and the corresponding flow rate and pressure distribution along the inferior-superior direction of the glottal shape (output targets) are established by the fully-connected DNN. With this reduced-order flow solver, for any glottal shape, the input features can be extracted from the UKE with the GA and then the flow rate and pressure distribution can be predicted with the trained DNNs. The implementation procedure of the reduced-order flow solver can be summarized in Algorithm \ref{algo:1}.

\begin{figure}[!ht]
  \begin{center}
	\includegraphics[width=0.5\textwidth]{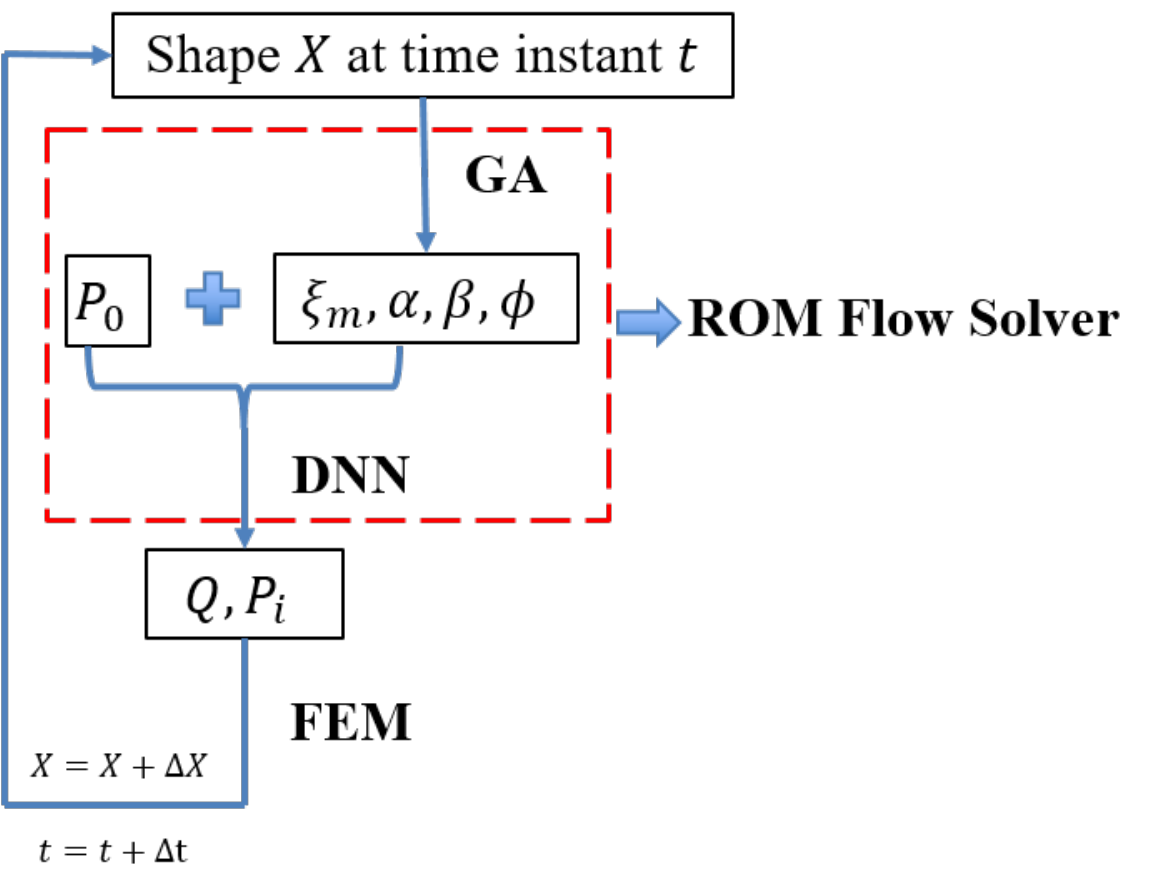}
  \end{center}
  \caption{Workflow of the reduced-order model for FSI simulation.}
  \label{fig:workflow_FSI}
\end{figure}

The developed reduced-order flow model is then coupled with the FEM based solid dynamics solver for FSI simulation. The abstract workflow of the ROM for FSI simulation is illustrated in Figure \ref{fig:workflow_FSI}. First, the flow rate $Q$ and pressure distribution $P_i$ of the glottal shape $X$ at a certain time instant $t$ can be obtained by the present reduced-order flow solver, then the pressure load is fed into the FEM solid solver to calculate the corresponding deformation of the glottis $\Delta X$, finally the updated glottal shape $X+\Delta X$ is used as the initial shape of the glottis at the next time instant $t+\Delta t$. The reduced-order flow solver and FEM based solid solver are coupled in a weak manner.

\begin{algorithm}[H]
Extract various shapes from converged Bernoulli-FEA FSI results;\\
Fit these extracted shapes with the UKE using the GA;\\
Obtain the PDF of the fitted parameters of the UKE: $\xi_m$, $\alpha$, $\beta$ and $\phi$;\\
Resample the PDF of  $\xi_m$, $\alpha$, $\beta$ and $\phi$ for various $P_0$;\\
Substitute the resampled values into the UKE to generate the generalized shape library;\\
Obtain the ground-truth values of $Q$ and $P_i$ for each shape in the library;\\
Establish the mapping relationship Eq.(\ref{eq:000}) with a fully-connected DNN 
\caption{Implementation of the reduced-order flow solver}
\label{algo:1}
\end{algorithm}

\section{\label{sec:5} Evaluation of the Performance of the Generalized ROM for FSI Simulation}

To evaluate the prediction performance of the present generalized ROM for FSI simulation, the ROM-FSI results are first compared with the FOM quasi-static (QS) results and the correlation and agreement between these results are analyzed, and then compared with the FOM-FSI results in terms of the voice quality-related parameters and CPU time. Detailed discussions are given as below.

\subsection{Comparison with FOM-QS Results}
\begin{figure}[!ht]
  \begin{center}
	\includegraphics[width=1.0\textwidth]{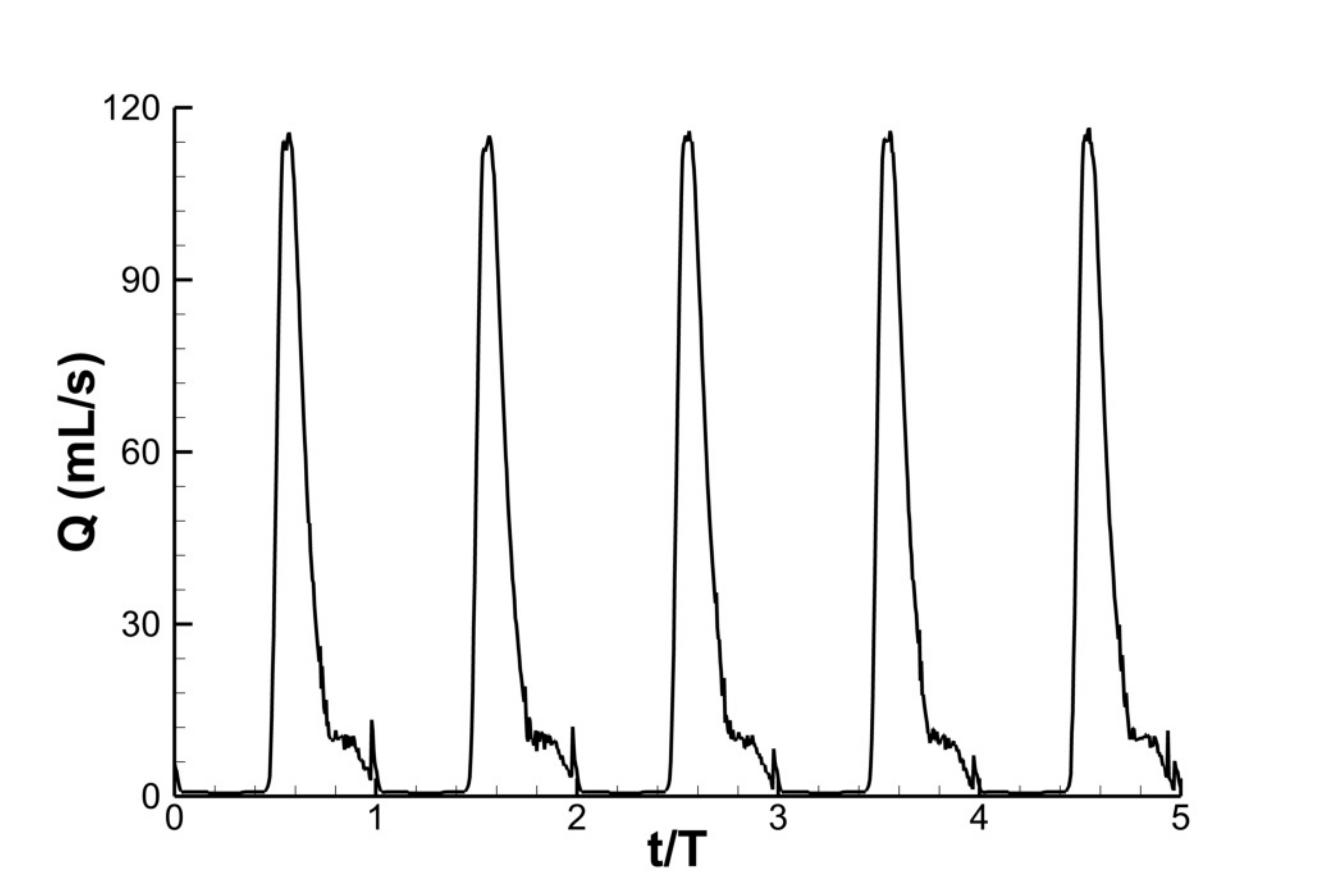}
  \end{center}
  \caption{Example of the converged time history of the predicted flow rate $Q$ at $P_0=0.8kPa$, $k_{CL}=4.75$, $k_B=3.75$.}
  \label{fig:Qhist_eval}
\end{figure}

A series of new subglottal pressure and material properties are simulated using the ROM-FSI model to generate the glottal shapes that are not in the shape library and evaluate the corresponding prediction performance. The values of the selected subglottal pressure and material properties are listed in Table \ref{Selected4Validation}. The simulation setup is the same as described in Subsection \ref{subsec3:1}. An example of the converged time history of the flow rate $Q$ at $P_0=0.8kPa$, $k_{CL}=4.75$, $k_B=3.75$ predicted by the ROM is illustrated in Figure \ref{fig:Qhist_eval}. Note that some fluctuations at the end of the closing phase can be observed, and this is likely due to the unsatisfactory representation of these shapes by the UKE due to the contact issue.

\begin{table}[!ht]
\caption{Selected subglottal pressure and material properties for evaluation.}
\begin{center}
\label{Selected4Validation}
\begin{tabular}{c c c}
\hline\hline
$P_0$(kPa) & $k_{CL}$ & $k_{B}$ \\
\hline
0.625 & \multirow{4}{*}{1.75, 2.75, 3.75, 4.75} & \multirow{4}{*}{1.75, 3.75}\\
0.7 &  & \\
0.8 &  & \\
0.875 &  & \\
\hline\hline
\end{tabular}
\end{center}
\end{table}

Various glottal shapes are extracted from the converged FSI results of the cases listed in Table \ref{Selected4Validation}. By excluding the fully-closed and nearly-closed shapes which may not be well represented by the UKE due to the contact issue, the total number of the extracted shapes for evaluation is 1582.

For each FSI case $n$ in Table \ref{Selected4Validation}, at each time step of the steady-cycle ROM-FSI result, the flow rate $Q_{ROM}^{n,k}$ and pressure distribution $P_{i,ROM}^{n,k}$ are respectively extracted, and the corresponding reference values of $Q_{FOM}^{n,k}$ and $P_{i,FOM}^{n,k}$ can be computed by the FOM, where $k$ is the index of the time step for each case. The time-averaged error of $Q$ and $P_i$ for each FSI case, designated as $E_Q^n$ and $E_P^n$, can be calculated as follows:

\bea
E_Q^n = \frac{1}{n_t\bar{Q}_{FOM}^{n}}\sum_{k=1}^{n_t}\abs{Q_{FOM}^{n,k}-Q_{ROM}^{n,k}}\\
E_P^n = \sum_{k=1}^{n_t}\sum_{i=1}^{N_P}\frac{\abs{P_{i,FOM}^{n,k}-P_{i,ROM}^{n,k}}}{P_0}
\eea
where $n_t$ and $\bar{Q}_{FOM}^{n}$ are the number of extracted time instants and the time-averaged reference values of the flow rate for each case, respectively.

The overall average error of $Q$ and $P_i$, designated as $E_Q$ and $E_P$, can be calculated as:
\bea
E_Q = \frac{1}{n_c}\sum_{n=1}^{n_c}E_Q^n\\
E_P = \frac{1}{n_c}\sum_{n=1}^{n_c}E_P^n
\eea
where $n_c$ is the number of cases listed in Table \ref{Selected4Validation}. The overall average error of $Q$ and $P_i$ are $7.87\%$ and $1.68\%$, respectively.

\begin{figure}[h]
\baselineskip=12pt
\figcolumn{
\fig{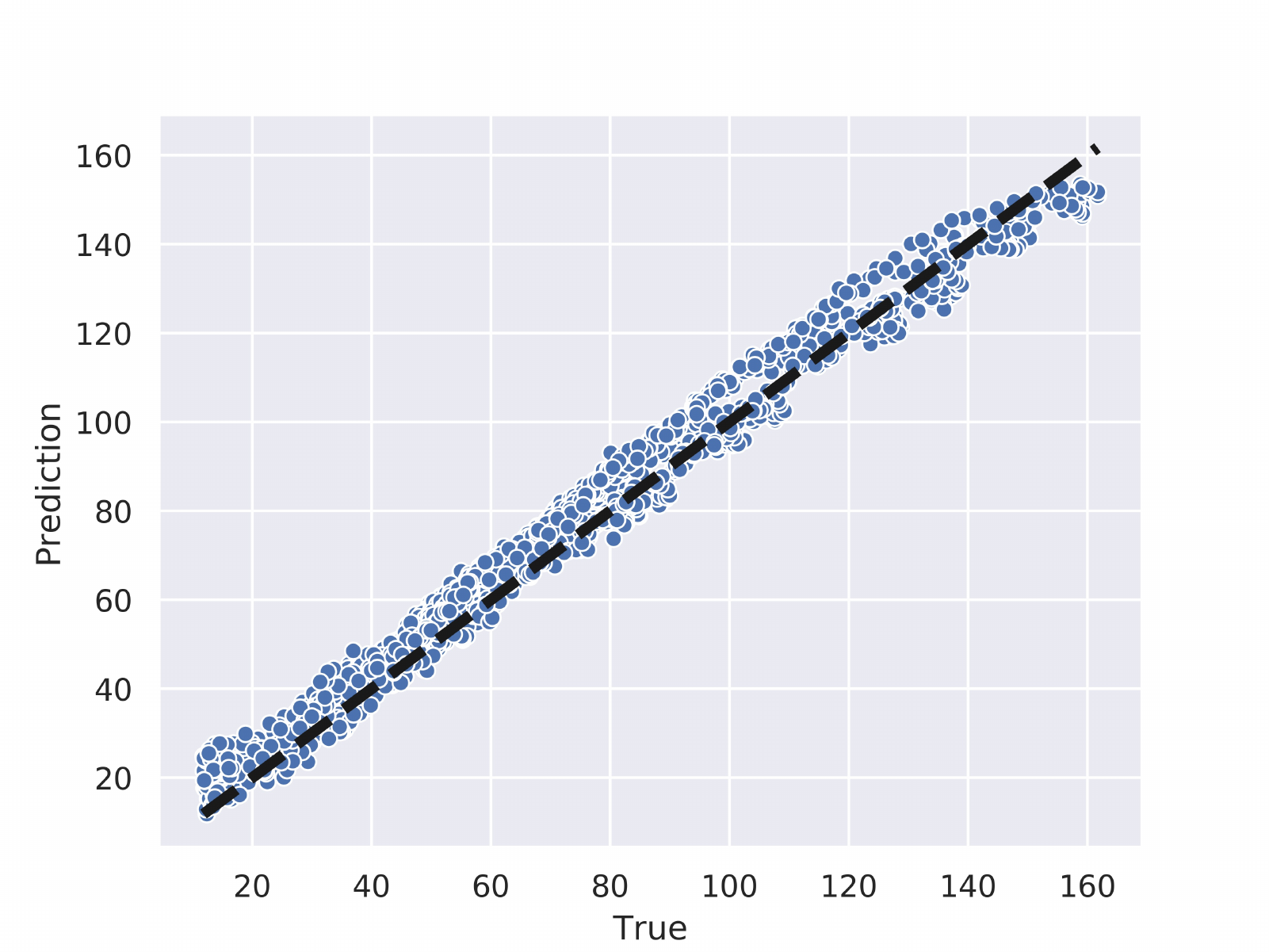}{.7\textwidth}{(a) Scatter plot}
\fig{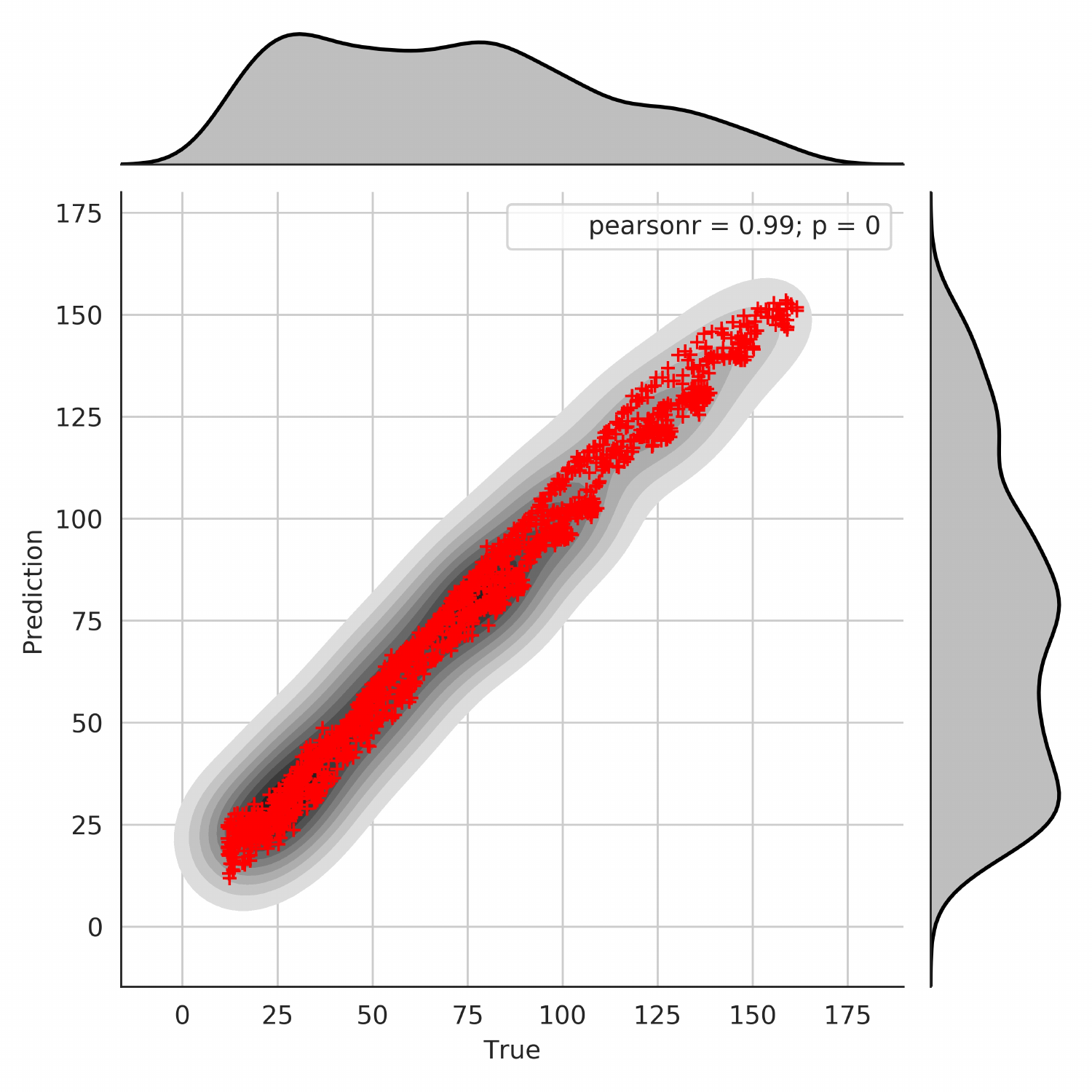}{.7\textwidth}{(b) Correlation plot}
}
\caption{Scatter and correlation plot of $Q$.}
\label{fig:CorreQ}
\end{figure}

\begin{figure}[!ht]
  \begin{center}
	\includegraphics[width=1.0\textwidth]{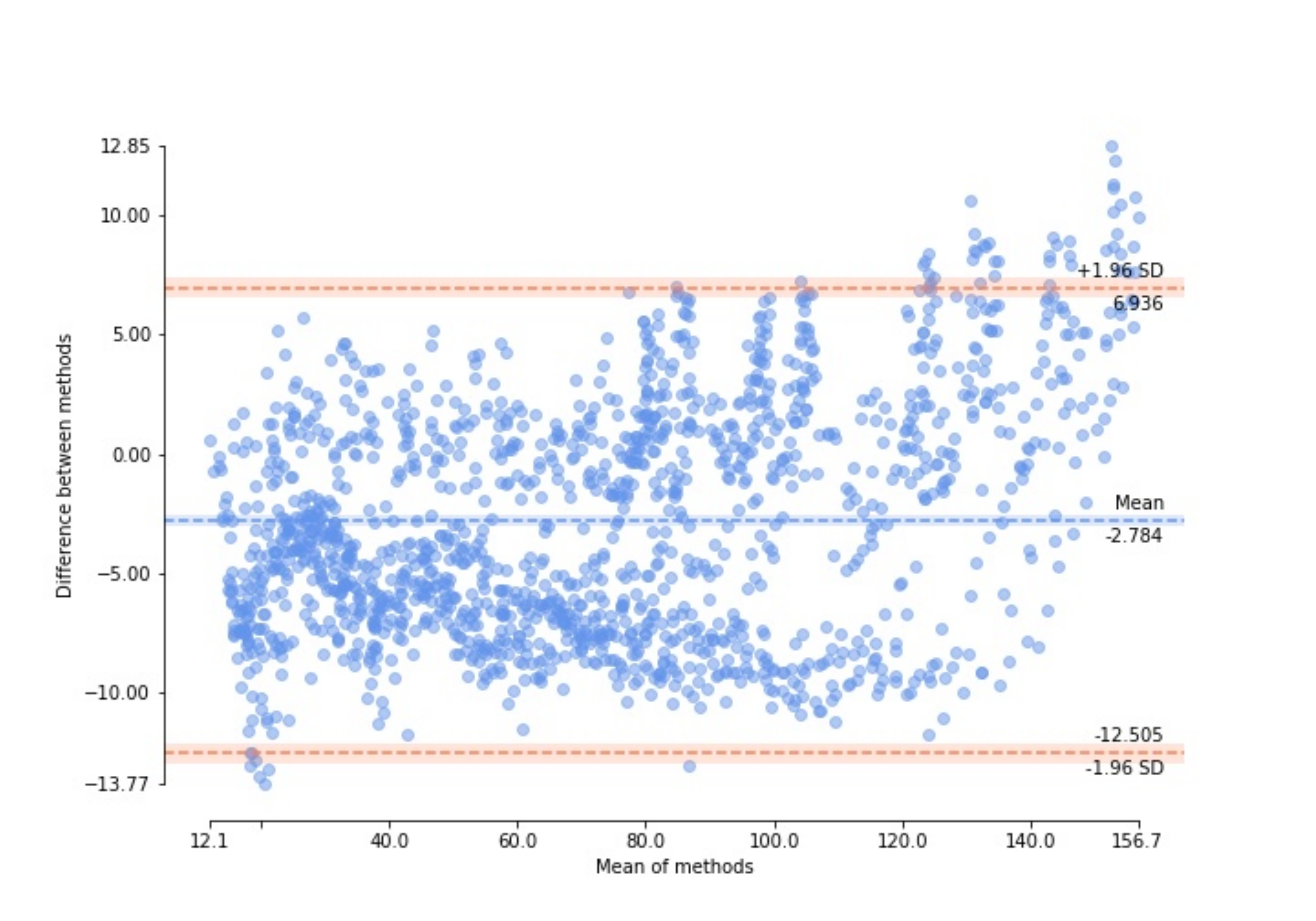}
  \end{center}
  \caption{Bland-Altman analysis plot of $Q$.}
  \label{fig:Q_BlandAltman}
\end{figure}

Additionally, the correlation and agreement between the true and predicted $Q$ and $P_i$ for the extracted 1582 glottal shapes are quantified. In terms of $Q$, the Pearson correlation coefficient \cite{freedman2007statistics} between $Q_{FOM}$ and $Q_{ROM}$ is excellent (0.993, $P<0.0005$). The scatter and correlation plots are also depicted in Figure \ref{fig:CorreQ}, where the horizontal and vertical axes correspond to the true ($Q_{FOM}$) and predicted ($Q_{ROM}$) values, respectively. The Bland-Altman plot \cite{altman1983measurement} is used to analyze the agreement between $Q_{FOM}$ and $Q_{ROM}$. The result is plotted in Figure \ref{fig:Q_BlandAltman}. As can be seen from this figure, the mean difference between $Q_{FOM}$ and $Q_{ROM}$ is $-2.784mL/s$, and the $95\%$ limits of agreement (LoA) between them is from $-12.505mL/s$ to $6.936mL/s$. The $95\%$ confidence interval (CI) of the mean difference, upper LoA and lower LoA between $Q_{FOM}$ and $Q_{ROM}$ is $[-3.0288mL/s, -2.5401mL/s]$, $[6.5177mL/s, 7.3539mL/s]$ and $[-12.9288mL/s, -12.0866mL/s]$, respectively. The number of the outliers is 38, and the percentage of the outliers is $2.40\%$. 

\begin{figure}[h]
\baselineskip=12pt
\figcolumn{
\fig{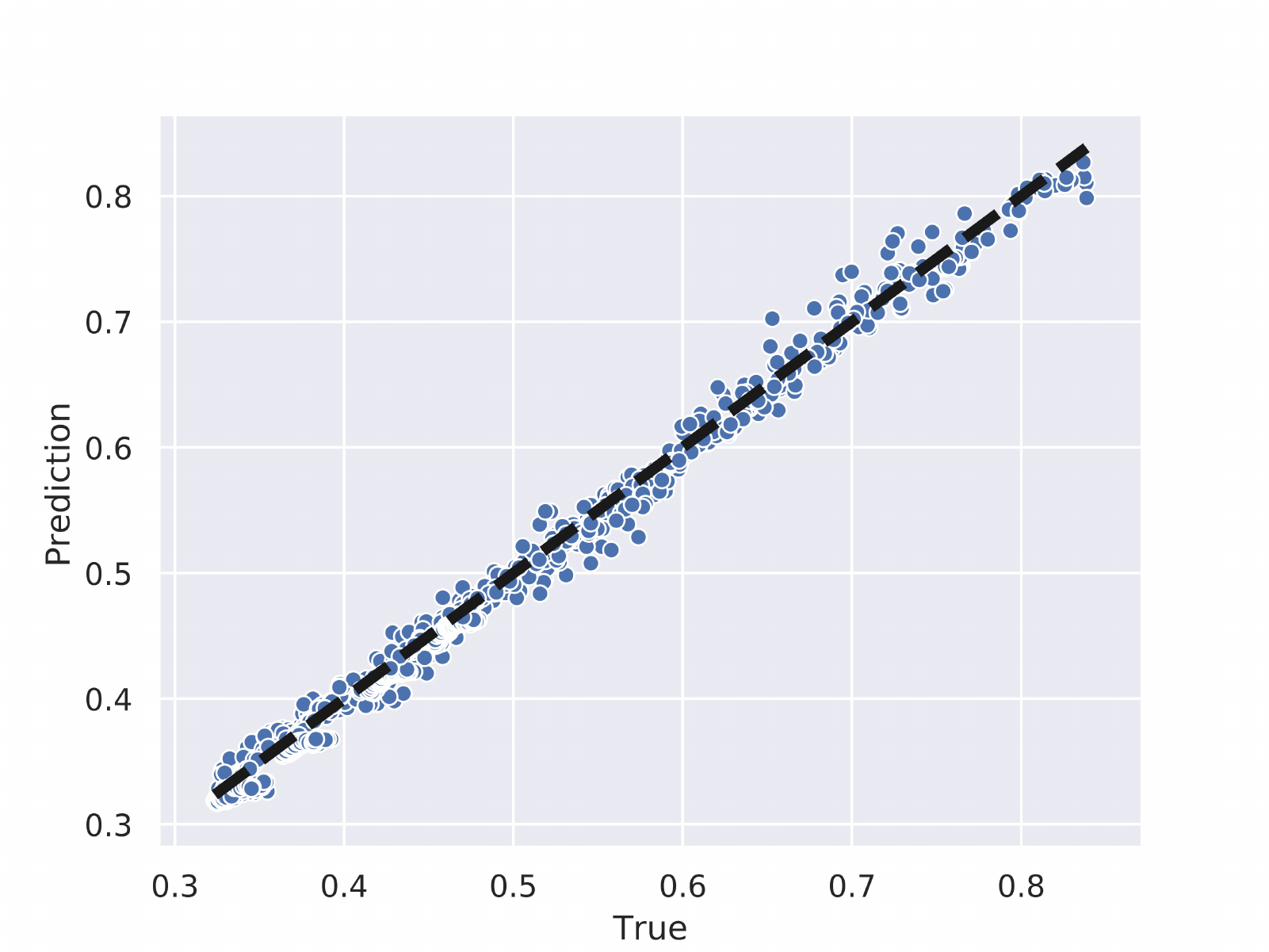}{.7\textwidth}{(a) Scatter plot}
\fig{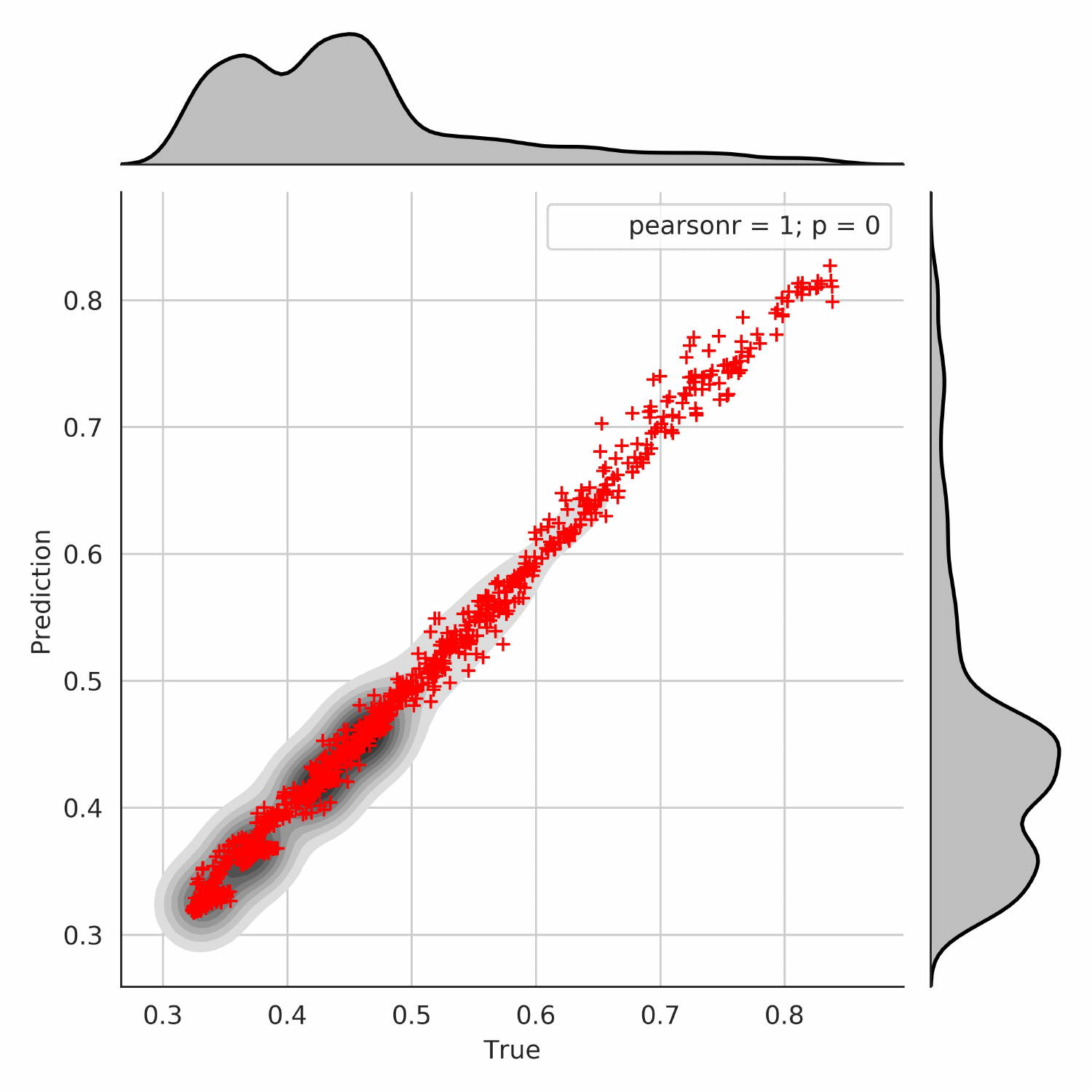}{.7\textwidth}{(b) Correlation plot}
}
\caption{Scatter and correlation plot of $P_i$.}
\label{fig:CorreP}
\end{figure}

\begin{figure}[!ht]
  \begin{center}
	\includegraphics[width=1.0\textwidth]{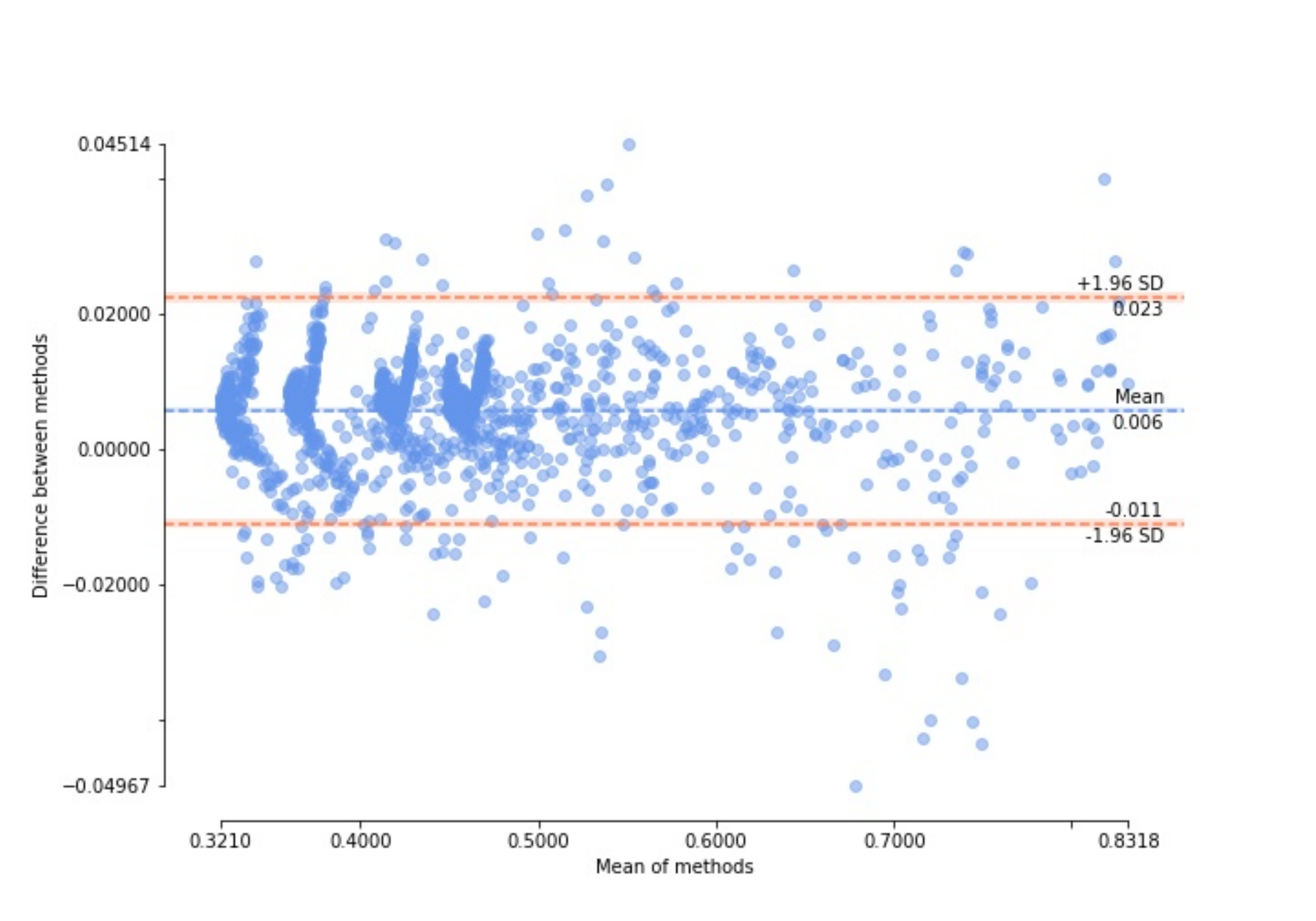}
  \end{center}
  \caption{Bland-Altman analysis plot of $P_i$.}
  \label{fig:P_BlandAltman}
\end{figure}

Similarly, in terms of $P_i$, the Pearson correlation coefficient between $P_{i,FOM}$ and $P_{i,ROM}$ is excellent (0.997, $P<0.0005$). The scatter and correlation plots are also depicted in Figure \ref{fig:CorreP}, where the horizontal and vertical axes correspond to the true ($P_{i,FOM}$) and predicted ($P_{i,ROM}$) values, respectively. The Bland-Altman analysis between $P_{i,FOM}$ and $P_{i,ROM}$ is plotted in Figure \ref{fig:P_BlandAltman}. From this figure, we can observe that the mean difference between $P_{i,FOM}$ and $P_{i,ROM}$ is $0.006kPa$, and the $95\%$ LoA between them is from $-0.011kPa$ to $0.023kPa$. The $95\%$ CI of the mean difference, upper LoA and lower LoA between $P_{i,FOM}$ and $P_{i,ROM}$ is $[0.0053kPa, 0.0062kPa]$, $[0.0218kPa, 0.0232kPa]$ and $[-0.0117kPa, -0.0103kPa]$, respectively. The number of the outliers is 87, and the percentage of the outliers is $5.50\%$.

The above correlation and agreement analysis results between the true and predicted $Q$ and $P_i$ for various glottal shapes indicate that the present ROM-FSI results agree very well with the corresponding FOM-QS results.

\subsection{Comparison with FOM-FSI Results}

\begin{figure}[h]
\baselineskip=12pt
\figcolumn{
\fig{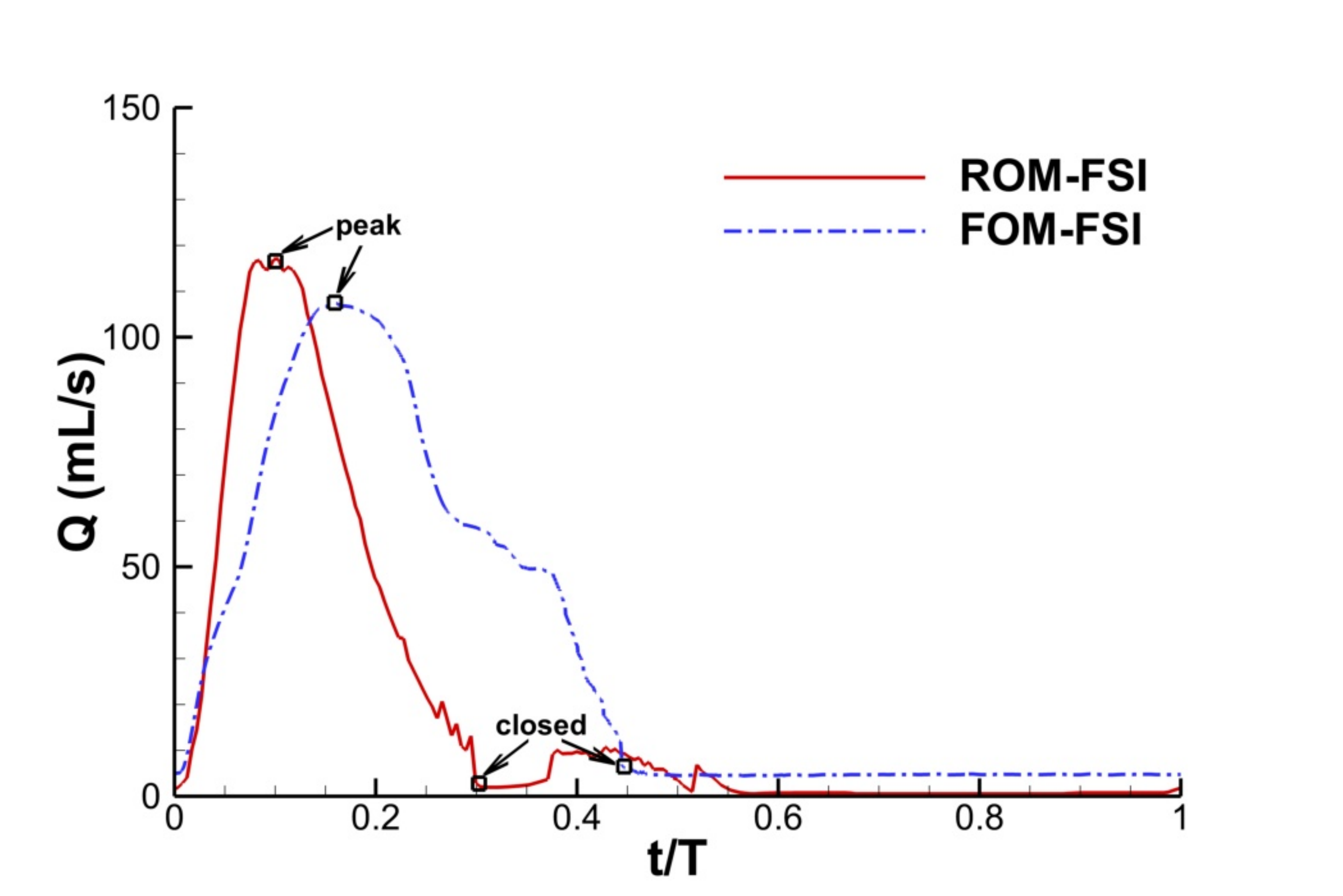}{.8\textwidth}{(a) Case 1}
\fig{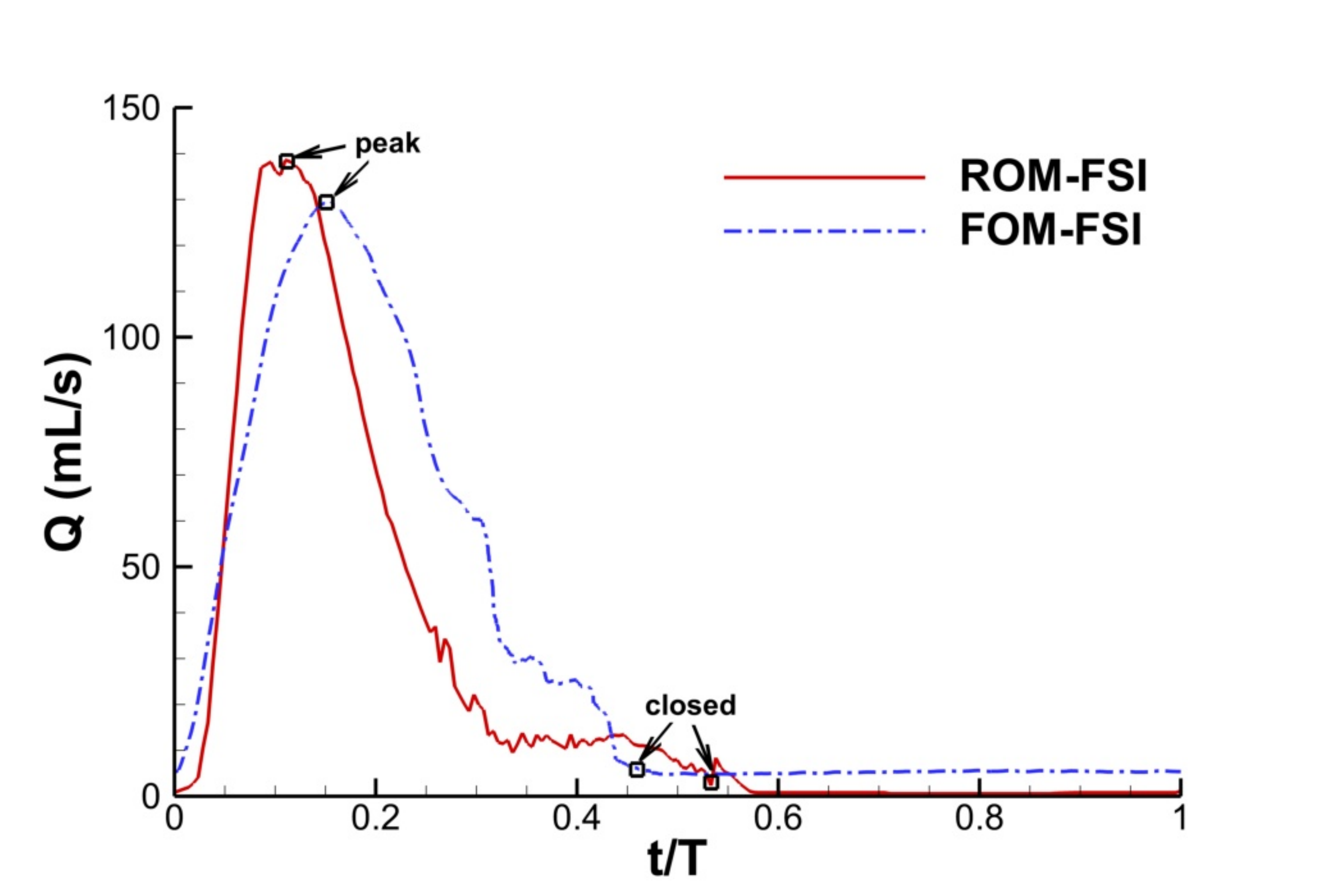}{.8\textwidth}{(b) Case 2}
}
\caption{Comparison of the phase-averaged time history of the flow rate.}
\label{fig:Qhist_compare_FSI}
\end{figure}

FSI simulations at $P_0=0.8kPa$, $k_{CL}=1.75$, $k_B=3.75$ (Case 1) and $P_0=0.875kPa$, $k_{CL}=3.75$, $k_B=3.75$ (Case 2) from Table \ref{Selected4Validation} are conducted by using both the ROM-FSI mdoel and FOM-FSI model. The comparison of the phase-averaged time history of the flow rate $Q$ for both cases are illustrated in Figure \ref{fig:Qhist_compare_FSI}. From this figure, we can observe that the peak flow rate, mean flow rate and the fundamental frequency are close to each other while the skewing of the waveform is different. Several important voice quality-related parameters \cite{xue2014subject} are computed from Figure \ref{fig:Qhist_compare_FSI} for both of the cases and the corresponding phase-averaged values are listed in Table \ref{VoiceQualityParams}. It can be seen from this table that the overall agreement between the values obtained by the ROM-FSI and FOM-FSI is satisfactory. The relative errors $\delta$ of $F_0$, $Q_{max}$, $Q_{mean}$ and $\xi_{m}$ between the ROM-FSI and FOM-FSI for both cases are within $10\%$, while the relative errors $\delta$ of $\tau_{0}$ and $\tau_{s}$ between the ROM-FSI and FOM-FSI for both of the cases are relatively larger. The difference could come from two sources: (a) in the GA optimization process, although the desired location and value of the optimized minimum cross-section area are preset to be equal to the target one (Eq. (\ref{eq:4})), the actual optimized location of the minimum cross-section area may be shifted and the corresponding value may be changed especially for the divergent shape, which may affect the profile of the flow rate at the flow decreasing phase. To remedy this, further improvements on the UKE model may be considered, and (b) the ROM-FSI model is a quasi-steady model while the FOM-FSI is a fully unsteady model. The quasi-steady assumption might also contribute to the differences between the two models.

\begin{table}[!ht]
\centering
\begin{threeparttable}
\caption{Comparison of voice quality-related parameters.}
\label{VoiceQualityParams}
\begin{tabular}{c c c c c c c}
\hline\hline
 & \begin{tabular}{@{}c@{}}ROM-FSI \\ Case 1\end{tabular} & \begin{tabular}{@{}c@{}}FOM-FSI \\ Case 1\end{tabular} & $\delta_1$ & \begin{tabular}{@{}c@{}}ROM-FSI \\ Case 2\end{tabular} & \begin{tabular}{@{}c@{}}FOM-FSI \\ Case 2\end{tabular} & $\delta_2$ \\
\hline
$F_0$ (Hz) & 210.8 & 216.3 & $2.5\%$ & 212.0 & 222.5 & $4.7\%$ \\
$Q_{max}$ (mL/s) & 117.0 & 107.3 & $9.0\%$ & 138.5 & 129.5 & $6.9\%$\\
$Q_{mean}$ (mL/s) & 54.8 & 56.6 & $3.2\%$ & 63.6 & 59.5 & $6.9\%$\\
$\tau_{0}$ & 0.30 & 0.45 & $33.3\%$ & 0.53 & 0.46 & $15.2\%$\\
$\tau_{s}$ & 0.49 & 0.55 & $10.9\%$ & 0.26 & 0.49 & $46.9\%$\\
$\xi_{m}$ (cm) & 0.060 & 0.055 & $9.1\%$ & 0.069 & 0.063 & $9.5\%$\\
\hline\hline
\end{tabular}
\begin{tablenotes}
      \small
      \item $F_0$ is the fundamental frequency; $Q_{max}$ and $Q_{mean}$ are the peak and mean glottal flow rate of the open quotient, respectively; $\tau_{0}$ is the open quotient, defined as $\tau_{0}=T_{0}/T$, where $T_0$ is the duration of the glottal open phase and $T$ is the cycle period; $\tau_{s}$ is the skewing quotient, defined as $\tau_{s}=T_{p}/T_{n}$ where $T_p$ is the duration of the flow increasing phase and $T_n$ is the duration of the flow decreasing phase \cite{xue2014subject}; $\xi_m$ is the vibration amplitude; $\delta_1$ and $\delta_2$ are the absolute value of the relative error between the ROM-FSI and FOM-FSI results for Case 1 and Case 2, respectively.
\end{tablenotes}
\end{threeparttable}
\end{table}

\begin{figure*}
\baselineskip=12pt
\figline{\fig{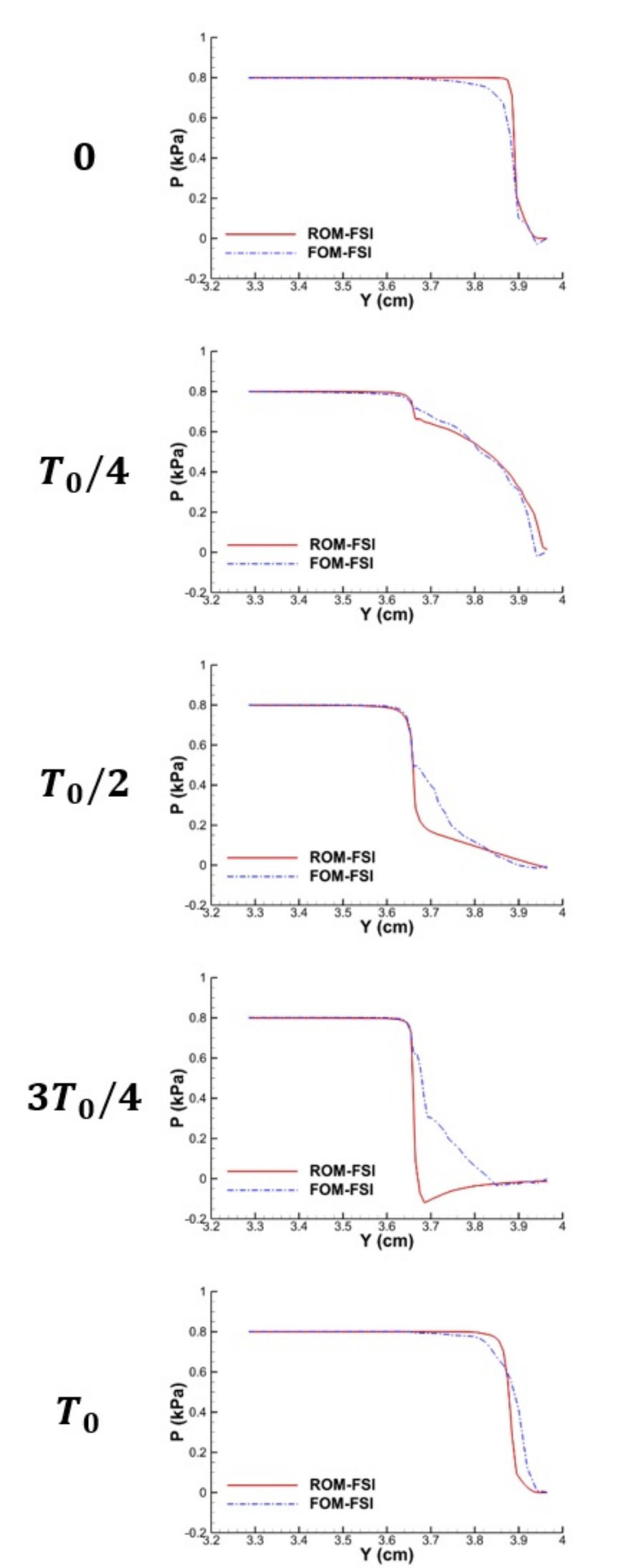}{.49\textwidth}{(a) Case 1}
\fig{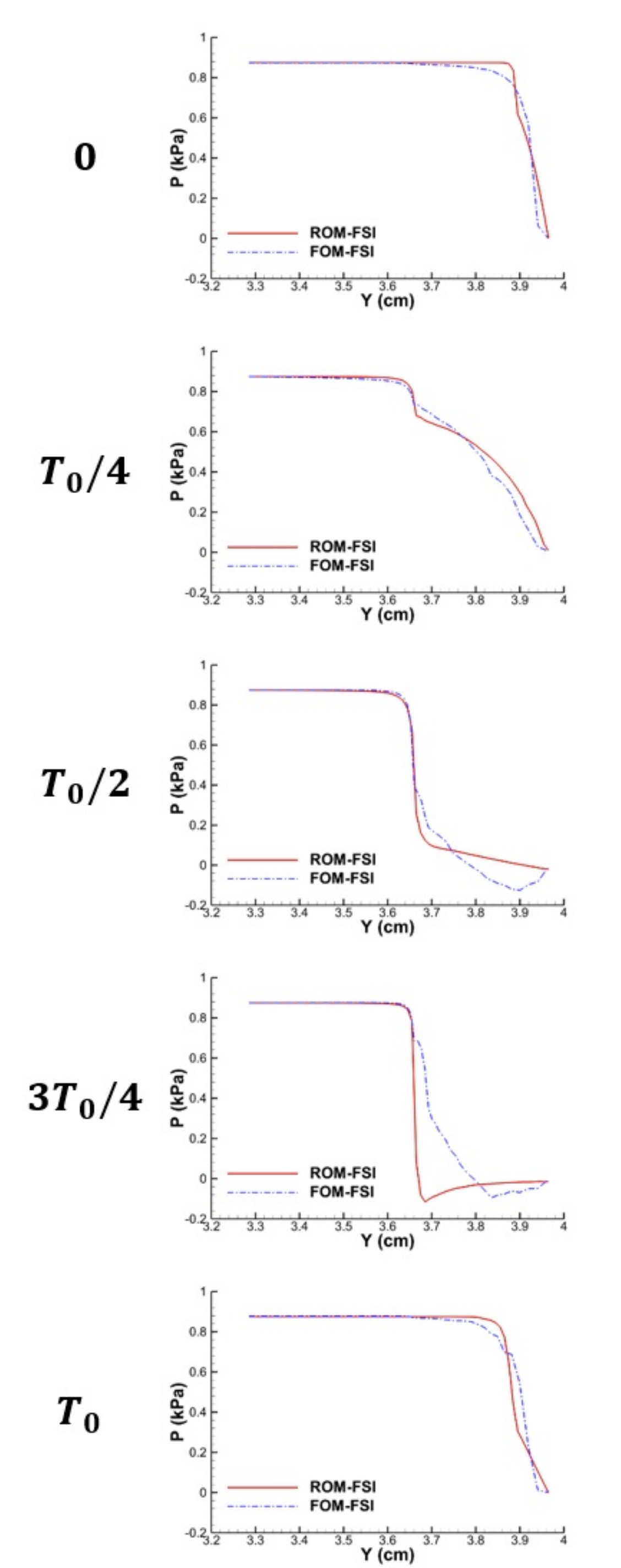}{.49\textwidth}{(b) Case 2}}
\caption{Comparison of the phase-averaged pressure distribution.}
\label{fig:pres_compare_cases}
\end{figure*}

\begin{figure*}
\baselineskip=12pt
\figline{\fig{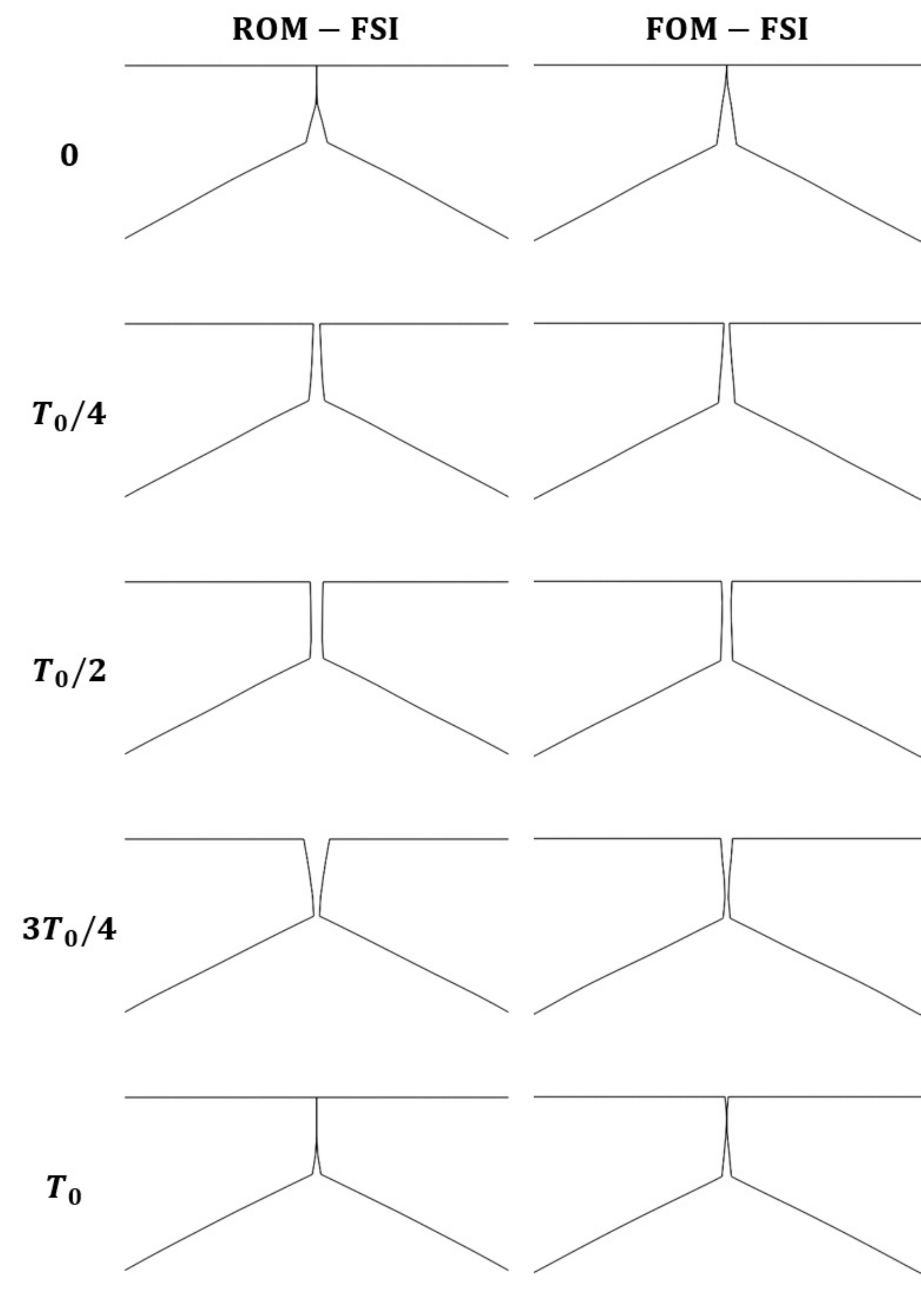}{.49\textwidth}{(a) Case 1}
\fig{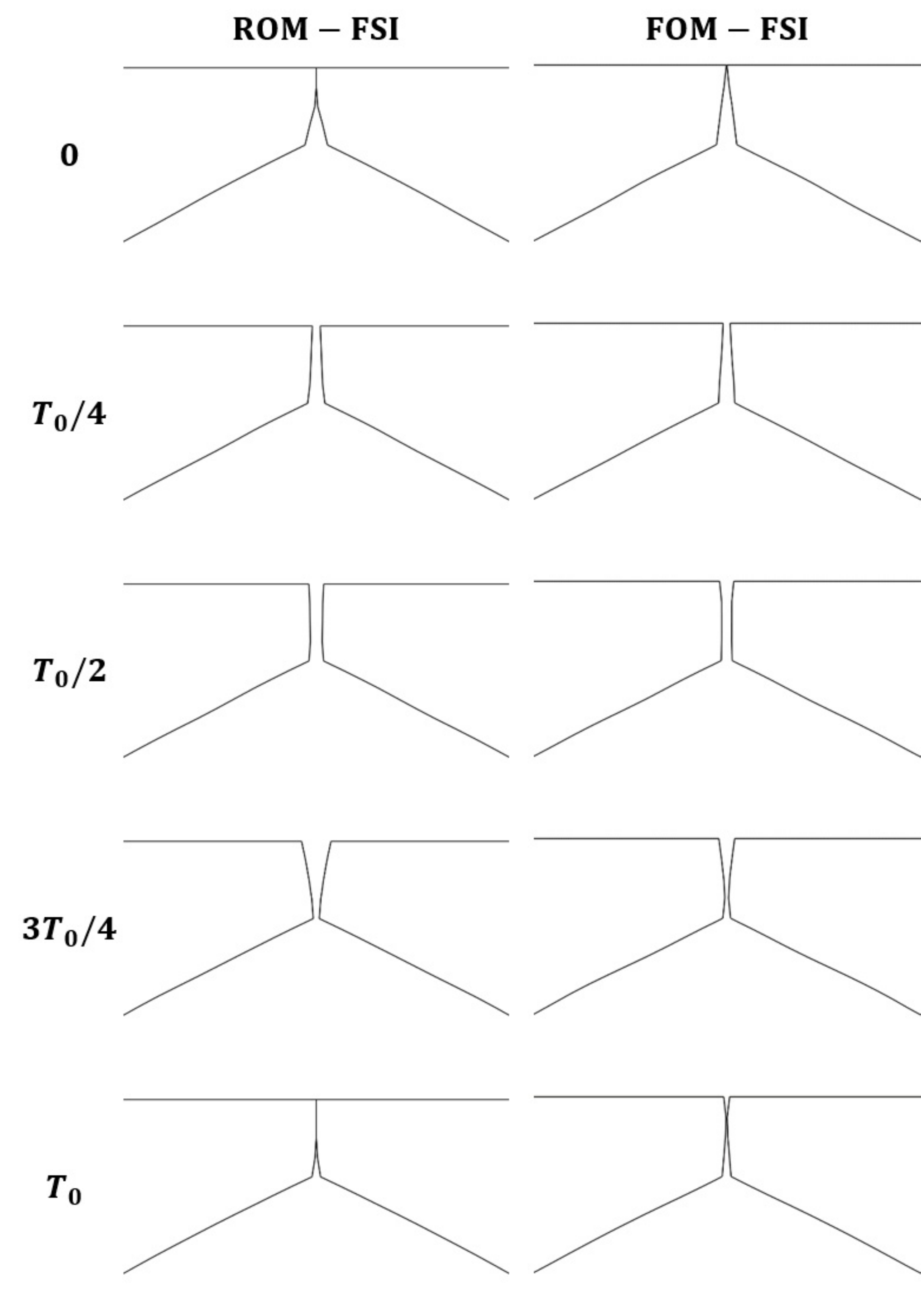}{.49\textwidth}{(b) Case 2}}
\caption{Comparison of the vibration pattern at different phases in the coronal view.}
\label{fig:vib_compare_cases}
\end{figure*}

The comparison of the phase-averaged pressure distribution $P_i$ for both cases are illustrated in Figure \ref{fig:pres_compare_cases}. Note that $T_0$ is the duration of the glottal open phase probed from Figure \ref{fig:Qhist_compare_FSI} for each case. The overall agreement is good except at the flow decreasing phase. The glottal vibration patterns at the correspondent phases for both cases obtained by the ROM-FSI and FOM-FSI are compared in Figure \ref{fig:vib_compare_cases}. From the figure, we can see that the glottal vibration patterns obtained by the ROM-FSI agree well with those obtained by the FOM-FSI except at the flow decreasing phase. The discrepancies of the pressure distribution as well as the vibration patterns are consistent with those of the phase-averaged flow rate shown in Figure \ref{fig:Qhist_compare_FSI}.

\begin{figure*}
\baselineskip=12pt
\figline{\fig{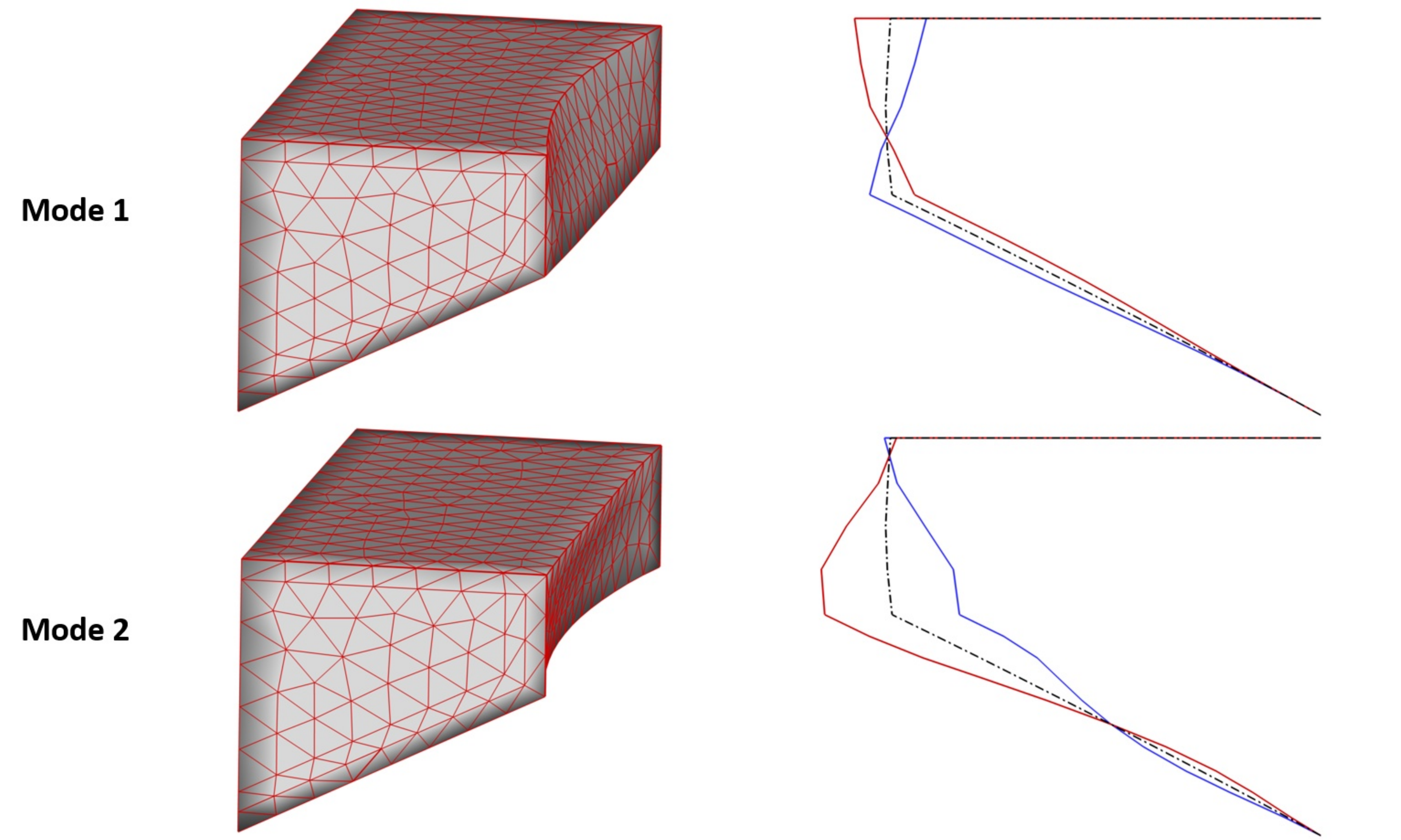}{.49\textwidth}{(a) Case 1}
\fig{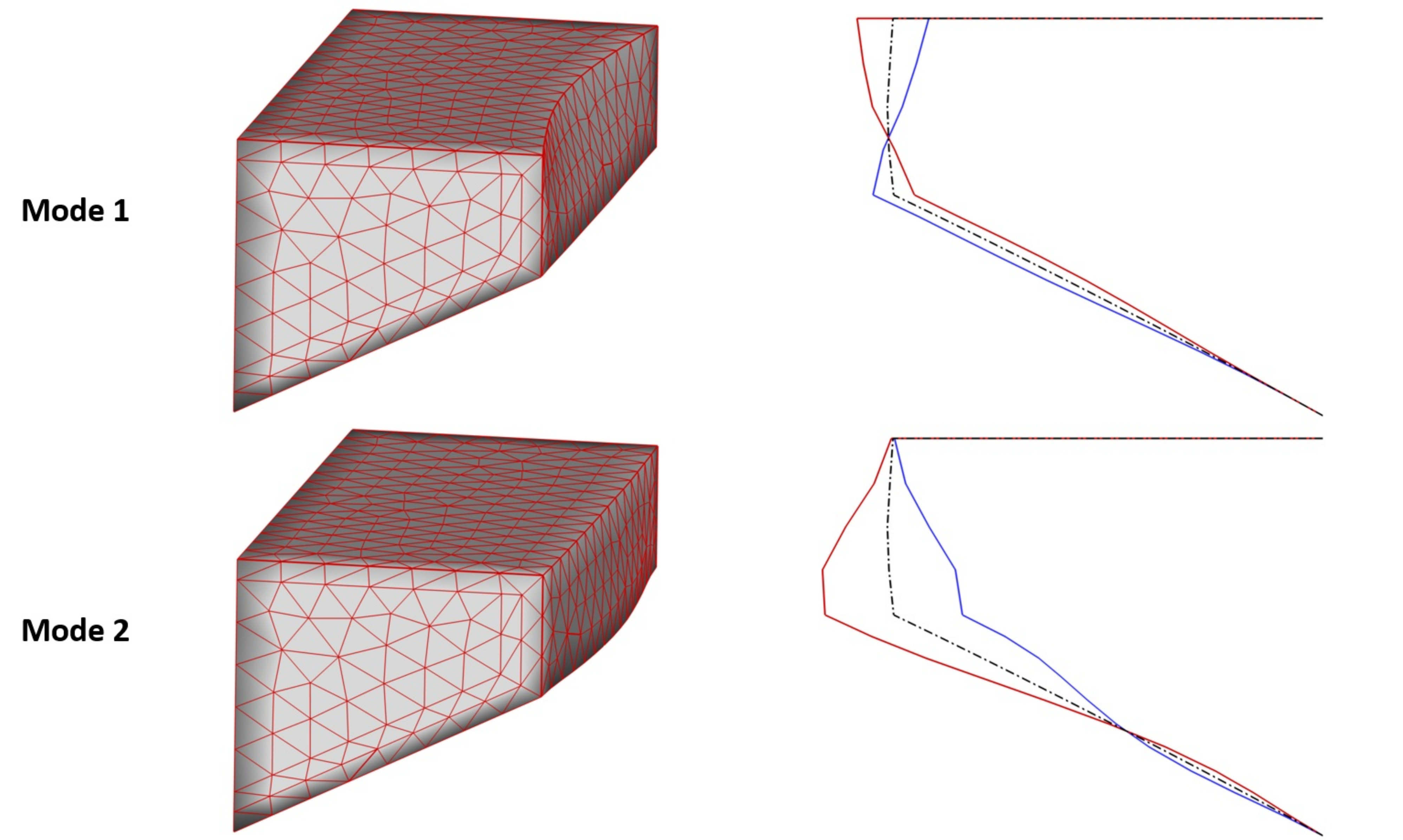}{.49\textwidth}{(b) Case 2}}
\caption{Three-dimensional and mid-coronal profile of the most two energetic empirical eigen functions at two extreme phases (dash dot line: equilibrium position).}
\label{fig:ROM_POD_cases}
\end{figure*}

\begin{figure}[!ht]
  \begin{center}
	\includegraphics[width=0.8\textwidth]{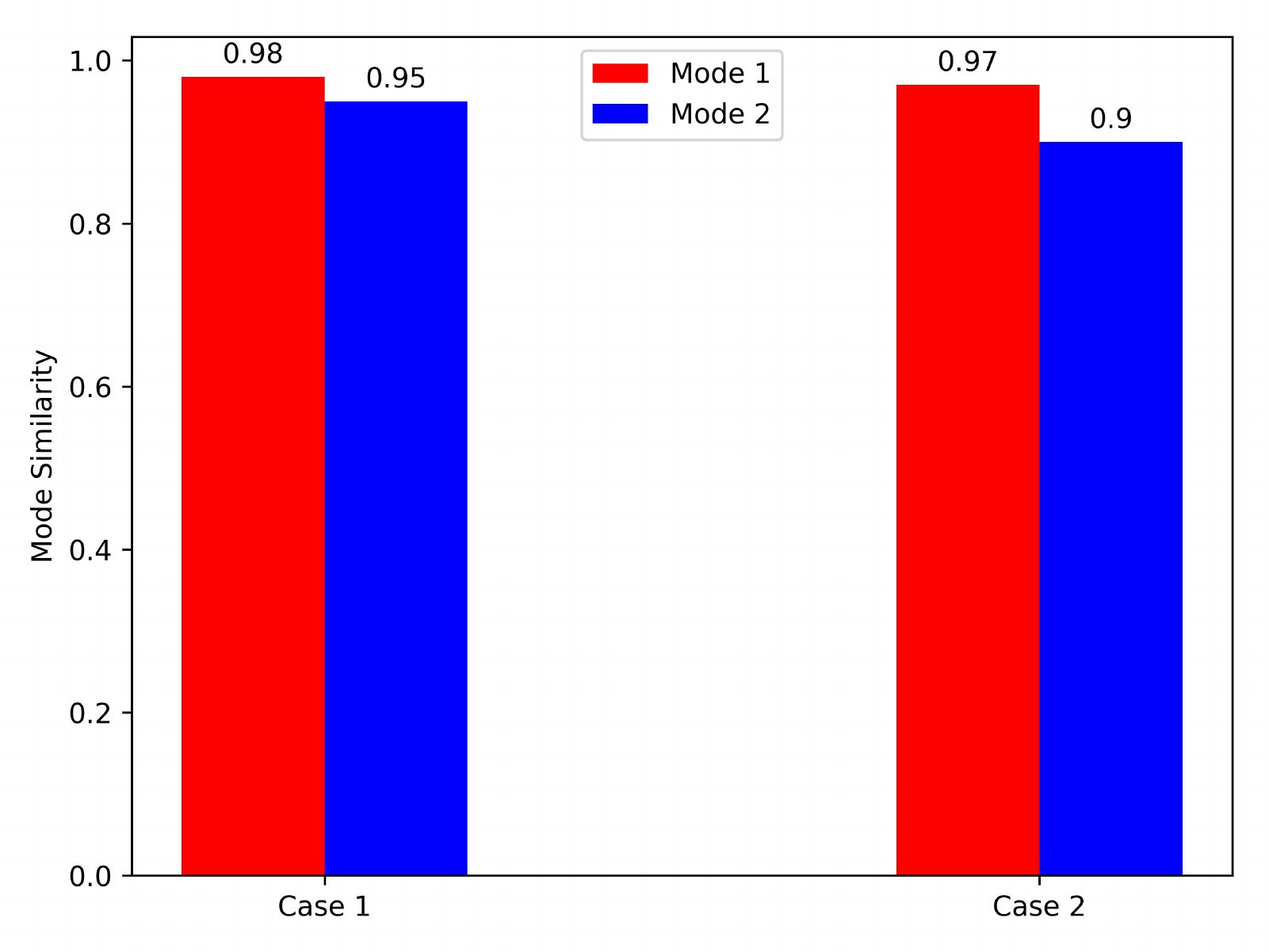}
  \end{center}
  \caption{Mode similarity.}
  \label{fig:mode_similarity}
\end{figure}

The proper orthogonal decomposition (POD) analysis \cite{liang2002proper} is employed to extract the most energetic empirical eigen-modes from the snapshots of ROM-FSI and FOM-FSI results for both cases. The three-dimensional and mid-coronal profile of the two most energetic empirical eigen functions at two extreme phases for both cases are illustrated in Figure \ref{fig:ROM_POD_cases}. For both cases, these two modes contain around $98\%$ of the total energy. To precisely quantify the similarity between the two modes obtained by the ROM-FSI and FOM-FSI, the dot-product between the corresponding normalized eigenmode \cite{xue2011sensitivity} for both cases is computed and plotted in Figure \ref{fig:mode_similarity}. The dot-product of any two normalized modes is indicative of the similarity between the two modes with the value of one corresponding to an exact match, and zero indicating orthogonality. From the figure, we can observe that modes 1 and 2 obtained by the ROM-FSI are highly similar to the corresponding modes obtained by the FOM-FSI for both cases which indicates a good prediction performance of the present ROM for FSI simulation of the vocal fold vibration.

Furthermore, the average CPU time required for one vibration cycle of the ROM-FSI and FOM-FSI simulation is compared. In order to obtain one vibration cycle, the average CPU time required for the ROM-FSI simulation is 1.5 hours per CPU on a single CPU, while that required for the FOM-FSI simulation is 20 hours per CPU on a parallel computer with 64 CPUs, which indicating the high efficiency of the present ROM for FSI simulation of the glottal flow.

\section{\label{sec:6}Conclusion}

A deep-learning based generalized reduced-order model (ROM) that can provide fast and accurate prediction of the dynamics of the glottal flow during normal phonations is proposed in this paper. 

The approach is based on the assumption that the vocal fold kinematics can be approximated by a few vibration modes as described by the surface-wave approach. Therefore, the vibration of the vocal folds during normal phonations can be represented by a universal kinematics equation (UKE) which is a linear combination of the dominant two modes. To verify that the UKE can be used as a generalized equation to represent any glottal shape during normal phonation, A large number of glottal shapes are generated from Bernoulli-FEM FSI simulation under various subglottal pressure and material properties and are fitted with a UKE using the genetic algorithm (GA). Furthermore, the probability density function (PDF) for each fitting parameter is obtained and used to build the generalized glottal shape library by appropriately resampling the PDF of the parameters and substituting into the UKE. For each shape in the library, the ground truth value of the flow rate and pressure distribution are obtained from high-fidelity N-S solutions. A fully-connected deep neural network (DNN) is used to build the empirical mapping between input parameters (parameters in the UKE and subglottal pressure) and output parameters (flow rate and pressure distribution). K-fold cross validation is performed to fine tune the architecture and hyperparameters and evaluate the prediction performance of the DNN. The developed reduced order glottal flow model is therefore composed of two parts: (a) glottal shape parameterization using the UKE and GA, and (b) glottal flow rate and intraglottal pressure prediction using the trained DNN. The present reduced-order flow solver is directly coupled with a finite-element method (FEM) based solid dynamics solver for FSI simulation. The ROM-FSI results are compared with the full-order model (FOM) quasi-static (QS) and FSI results. For the comparison with the FOM-QS model, the ROM model shows an excellent agreement in terms of predicting the flow rate and pressure distribution. The average error of the prediction for the flow rate and pressure distribution are $7.87\%$ and $1.68\%$, respectively. For the comparison with the FOM-FSI model, the ROM model shows a good agreement on the frequency, peak and mean flow rate and vocal fold vibration pattern with the relative errors less than $10\%$. The ROM model shows a relatively larger error in predicting the opening quotient and skewness quotient. The comparison of the details of the intraglottal pressure distribution between the two models reflects that one of the reasons might be the inaccurate prediction of the location of the minimum area when the glottis has a divergent shape. It should be noted that the ROM-FSI model is a quasi-steady model while the FOM-FSI is a fully unsteady model. The quasi-steady assumption might also contribute to the differences between the two models. The overall good prediction performance of the present ROM in accuracy and efficiency indicates a great promise for future clinical use. The developed ROM can be further extended to predict the dynamics of the glottal flow during abnormal phonations with relative ease.

\begin{acknowledgments}
The project was supported by Grant Number 5R21DC016428 from the National Institute on Deafness and Other Communication Disorders (NIDCD). Numerical simulations were performed using resources of the Extreme Science and Engineering Discovery Environment (XSEDE) (allocation Award Nos. TG-BIO150055 and TG-CTS180004).
\end{acknowledgments}


\begin{thebibliography}{41}
\def\enquote#1{``#1,''}
\def\plainquote#1{``#1''}
\expandafter\ifx\csname natexlab\endcsname\relax\def\natexlab#1{#1}\fi
\providecommand{\dourl}[1]{\href{http://#1}{\nolinkurl{#1}}}
\providecommand{\bibinfo}[2]{#2}
\providecommand{\noopsort}[1]{}
\providecommand{\switchargs}[2]{#2#1}
  \def\eatspace #1{#1}

\bibitem[{Abadi \emph{et~al.}(2015)Abadi, Agarwal, Barham, Brevdo, Chen, Citro,
  Corrado, Davis, Dean, Devin, Ghemawat, Goodfellow, Harp, Irving, Isard, Jia,
  Jozefowicz, Kaiser, Kudlur, Levenberg, Man\'{e}, Monga, Moore, Murray, Olah,
  Schuster, Shlens, Steiner, Sutskever, Talwar, Tucker, Vanhoucke, Vasudevan,
  Vi\'{e}gas, Vinyals, Warden, Wattenberg, Wicke, Yu, and
  Zheng}]{tensorflow2015-whitepaper}
\bibinfo{author}{Abadi, M.}, \bibinfo{author}{Agarwal, A.},
  \bibinfo{author}{Barham, P.}, \bibinfo{author}{Brevdo, E.},
  \bibinfo{author}{Chen, Z.}, \bibinfo{author}{Citro, C.},
  \bibinfo{author}{Corrado, G.~S.}, \bibinfo{author}{Davis, A.},
  \bibinfo{author}{Dean, J.}, \bibinfo{author}{Devin, M.},
  \bibinfo{author}{Ghemawat, S.}, \bibinfo{author}{Goodfellow, I.},
  \bibinfo{author}{Harp, A.}, \bibinfo{author}{Irving, G.},
  \bibinfo{author}{Isard, M.}, \bibinfo{author}{Jia, Y.},
  \bibinfo{author}{Jozefowicz, R.}, \bibinfo{author}{Kaiser, L.},
  \bibinfo{author}{Kudlur, M.}, \bibinfo{author}{Levenberg, J.},
  \bibinfo{author}{Man\'{e}, D.}, \bibinfo{author}{Monga, R.},
  \bibinfo{author}{Moore, S.}, \bibinfo{author}{Murray, D.},
  \bibinfo{author}{Olah, C.}, \bibinfo{author}{Schuster, M.},
  \bibinfo{author}{Shlens, J.}, \bibinfo{author}{Steiner, B.},
  \bibinfo{author}{Sutskever, I.}, \bibinfo{author}{Talwar, K.},
  \bibinfo{author}{Tucker, P.}, \bibinfo{author}{Vanhoucke, V.},
  \bibinfo{author}{Vasudevan, V.}, \bibinfo{author}{Vi\'{e}gas, F.},
  \bibinfo{author}{Vinyals, O.}, \bibinfo{author}{Warden, P.},
  \bibinfo{author}{Wattenberg, M.}, \bibinfo{author}{Wicke, M.},
  \bibinfo{author}{Yu, Y.},  and \bibinfo{author}{Zheng, X.}
  (\textbf{\bibinfo{year}{2015}}). \plainquote{\bibinfo{title}{{TensorFlow}:
  Large-scale machine learning on heterogeneous systems}}
  \dourl{https://www.tensorflow.org/}, \bibinfo{note}{software available from
  tensorflow.org}.

\bibitem[{Alipour \emph{et~al.}(2000)Alipour, Berry, and
  Titze}]{alipour2000finite}
\bibinfo{author}{Alipour, F.}, \bibinfo{author}{Berry, D.~A.},  and
  \bibinfo{author}{Titze, I.~R.} (\textbf{\bibinfo{year}{2000}}).
  \enquote{\bibinfo{title}{A finite-element model of vocal-fold vibration}}
  \bibinfo{journal}{The Journal of the Acoustical Society of America}
  \textbf{108}(6), \bibinfo{pages}{3003--3012}.

\bibitem[{Altman and Bland(1983)}]{altman1983measurement}
\bibinfo{author}{Altman, D.~G.},  and \bibinfo{author}{Bland, J.~M.}
  (\textbf{\bibinfo{year}{1983}}). \enquote{\bibinfo{title}{Measurement in
  medicine: the analysis of method comparison studies}}
  \bibinfo{journal}{Journal of the Royal Statistical Society: Series D (The
  Statistician)} \textbf{32}(3), \bibinfo{pages}{307--317}.

\bibitem[{Berry(2001)}]{berry2001mechanisms}
\bibinfo{author}{Berry, D.~A.} (\textbf{\bibinfo{year}{2001}}).
  \enquote{\bibinfo{title}{Mechanisms of modal and nonmodal phonation}}
  \bibinfo{journal}{Journal of Phonetics} \textbf{29}(4),
  \bibinfo{pages}{431--450}.

\bibitem[{Berry \emph{et~al.}(1994)Berry, Herzel, Titze, and
  Krischer}]{berry1994interpretation}
\bibinfo{author}{Berry, D.~A.}, \bibinfo{author}{Herzel, H.},
  \bibinfo{author}{Titze, I.~R.},  and \bibinfo{author}{Krischer, K.}
  (\textbf{\bibinfo{year}{1994}}). \enquote{\bibinfo{title}{Interpretation of
  biomechanical simulations of normal and chaotic vocal fold oscillations with
  empirical eigenfunctions}} \bibinfo{journal}{The Journal of the Acoustical
  Society of America} \textbf{95}(6), \bibinfo{pages}{3595--3604}.

\bibitem[{Chollet \emph{et~al.}(2015)}]{chollet2015keras}
\bibinfo{author}{Chollet, F.} \emph{et~al.} (\textbf{\bibinfo{year}{2015}}).
  \plainquote{\bibinfo{title}{Keras}} .

\bibitem[{Deverge \emph{et~al.}(2003)Deverge, Pelorson, Vilain, Lagr{\'e}e,
  Chentouf, Willems, and Hirschberg}]{deverge2003influence}
\bibinfo{author}{Deverge, M.}, \bibinfo{author}{Pelorson, X.},
  \bibinfo{author}{Vilain, C.}, \bibinfo{author}{Lagr{\'e}e, P.-Y.},
  \bibinfo{author}{Chentouf, F.}, \bibinfo{author}{Willems, J.},  and
  \bibinfo{author}{Hirschberg, A.} (\textbf{\bibinfo{year}{2003}}).
  \enquote{\bibinfo{title}{Influence of collision on the flow through in-vitro
  rigid models of the vocal folds}} \bibinfo{journal}{The Journal of the
  Acoustical Society of America} \textbf{114}(6), \bibinfo{pages}{3354--3362}.

\bibitem[{D{\"o}llinger \emph{et~al.}(2005)D{\"o}llinger, Berry, and
  Berke}]{dollinger2005medial}
\bibinfo{author}{D{\"o}llinger, M.}, \bibinfo{author}{Berry, D.~A.},  and
  \bibinfo{author}{Berke, G.~S.} (\textbf{\bibinfo{year}{2005}}).
  \enquote{\bibinfo{title}{Medial surface dynamics of an in vivo canine vocal
  fold during phonation}} \bibinfo{journal}{The Journal of the Acoustical
  Society of America} \textbf{117}(5), \bibinfo{pages}{3174--3183}.

\bibitem[{Erath \emph{et~al.}(2011)Erath, Za{\~n}artu, Peterson, and
  Plesniak}]{erath2011nonlinear}
\bibinfo{author}{Erath, B.~D.}, \bibinfo{author}{Za{\~n}artu, M.},
  \bibinfo{author}{Peterson, S.~D.},  and \bibinfo{author}{Plesniak, M.~W.}
  (\textbf{\bibinfo{year}{2011}}). \enquote{\bibinfo{title}{Nonlinear vocal
  fold dynamics resulting from asymmetric fluid loading on a two-mass model of
  speech}} \bibinfo{journal}{Chaos: An Interdisciplinary Journal of Nonlinear
  Science} \textbf{21}(3), \bibinfo{pages}{033113}.

\bibitem[{Forrest(1996)}]{forrest1996genetic}
\bibinfo{author}{Forrest, S.} (\textbf{\bibinfo{year}{1996}}).
  \enquote{\bibinfo{title}{Genetic algorithms}} \bibinfo{journal}{ACM Computing
  Surveys (CSUR)} \textbf{28}(1), \bibinfo{pages}{77--80}.

\bibitem[{Freedman \emph{et~al.}(2007)Freedman, Pisani, and
  Purves}]{freedman2007statistics}
\bibinfo{author}{Freedman, D.}, \bibinfo{author}{Pisani, R.},  and
  \bibinfo{author}{Purves, R.} (\textbf{\bibinfo{year}{2007}}). International
  student edition \emph{\bibinfo{title}{Statistics: Fourth International
  Student Edition}} (\bibinfo{publisher}{W.W. Norton \& Company}),
  \dourl{https://books.google.com/books?id=mviJQgAACAAJ}.

\bibitem[{Geng \emph{et~al.}(2016)Geng, Xue, and Zheng}]{geng2016effect}
\bibinfo{author}{Geng, B.}, \bibinfo{author}{Xue, Q.},  and
  \bibinfo{author}{Zheng, X.} (\textbf{\bibinfo{year}{2016}}).
  \enquote{\bibinfo{title}{The effect of vocal fold vertical stiffness
  variation on voice production}} \bibinfo{journal}{The Journal of the
  Acoustical Society of America} \textbf{140}(4), \bibinfo{pages}{2856--2866}.

\bibitem[{Goldberg(2006)}]{goldberg2006genetic}
\bibinfo{author}{Goldberg, D.} (\textbf{\bibinfo{year}{2006}}).
  \emph{\bibinfo{title}{Genetic Algorithms}} (\bibinfo{publisher}{Pearson
  Education}), \dourl{https://books.google.com/books?id=6gzS07Sv9hoC}.

\bibitem[{Goodfellow \emph{et~al.}(2016)Goodfellow, Bengio, and
  Courville}]{goodfellow2016deep}
\bibinfo{author}{Goodfellow, I.}, \bibinfo{author}{Bengio, Y.},  and
  \bibinfo{author}{Courville, A.} (\textbf{\bibinfo{year}{2016}}).
  \emph{\bibinfo{title}{Deep Learning}} (\bibinfo{publisher}{MIT press}).

\bibitem[{Ishizaka and Flanagan(1972)}]{ishizaka1972synthesis}
\bibinfo{author}{Ishizaka, K.},  and \bibinfo{author}{Flanagan, J.~L.}
  (\textbf{\bibinfo{year}{1972}}). \enquote{\bibinfo{title}{Synthesis of voiced
  sounds from a two-mass model of the vocal cords}} \bibinfo{journal}{Bell
  System Technical Journal} \textbf{51}(6), \bibinfo{pages}{1233--1268}.

\bibitem[{Jiang and Zhang(2002)}]{jiang2002chaotic}
\bibinfo{author}{Jiang, J.~J.},  and \bibinfo{author}{Zhang, Y.}
  (\textbf{\bibinfo{year}{2002}}). \enquote{\bibinfo{title}{Chaotic vibration
  induced by turbulent noise in a two-mass model of vocal folds}}
  \bibinfo{journal}{The Journal of the Acoustical Society of America}
  \textbf{112}(5), \bibinfo{pages}{2127--2133}.

\bibitem[{LeCun \emph{et~al.}(2015)LeCun, Bengio, and Hinton}]{lecun2015deep}
\bibinfo{author}{LeCun, Y.}, \bibinfo{author}{Bengio, Y.},  and
  \bibinfo{author}{Hinton, G.} (\textbf{\bibinfo{year}{2015}}).
  \enquote{\bibinfo{title}{Deep learning}} \bibinfo{journal}{Nature}
  \textbf{521}(7553), \bibinfo{pages}{436}.

\bibitem[{Liang \emph{et~al.}(2002)Liang, Lee, Lim, Lin, Lee, and
  Wu}]{liang2002proper}
\bibinfo{author}{Liang, Y.}, \bibinfo{author}{Lee, H.}, \bibinfo{author}{Lim,
  S.}, \bibinfo{author}{Lin, W.}, \bibinfo{author}{Lee, K.},  and
  \bibinfo{author}{Wu, C.} (\textbf{\bibinfo{year}{2002}}).
  \enquote{\bibinfo{title}{Proper orthogonal decomposition and its
  applications?{P}art {I}: Theory}} \bibinfo{journal}{Journal of Sound and
  Vibration} \textbf{252}(3), \bibinfo{pages}{527--544}.

\bibitem[{Luo \emph{et~al.}(2008)Luo, Mittal, Zheng, Bielamowicz, Walsh, and
  Hahn}]{luo2008immersed}
\bibinfo{author}{Luo, H.}, \bibinfo{author}{Mittal, R.},
  \bibinfo{author}{Zheng, X.}, \bibinfo{author}{Bielamowicz, S.~A.},
  \bibinfo{author}{Walsh, R.~J.},  and \bibinfo{author}{Hahn, J.~K.}
  (\textbf{\bibinfo{year}{2008}}). \enquote{\bibinfo{title}{An
  immersed-boundary method for flow--structure interaction in biological
  systems with application to phonation}} \bibinfo{journal}{Journal of
  Computational Physics} \textbf{227}(22), \bibinfo{pages}{9303--9332}.

\bibitem[{Mitchell(1998)}]{mitchell1998introduction}
\bibinfo{author}{Mitchell, M.} (\textbf{\bibinfo{year}{1998}}).
  \emph{\bibinfo{title}{An Introduction to Genetic Algorithms}}
  (\bibinfo{publisher}{MIT press}).

\bibitem[{Mittal \emph{et~al.}(2011)Mittal, Zheng, Bhardwaj, Seo, Xue, and
  Bielamowicz}]{mittal2011toward}
\bibinfo{author}{Mittal, R.}, \bibinfo{author}{Zheng, X.},
  \bibinfo{author}{Bhardwaj, R.}, \bibinfo{author}{Seo, J.~H.},
  \bibinfo{author}{Xue, Q.},  and \bibinfo{author}{Bielamowicz, S.}
  (\textbf{\bibinfo{year}{2011}}). \enquote{\bibinfo{title}{Toward a
  simulation-based tool for the treatment of vocal fold paralysis}}
  \bibinfo{journal}{Frontiers in Physiology} \textbf{2}, \bibinfo{pages}{19}.

\bibitem[{Pelorson \emph{et~al.}(1994)Pelorson, Hirschberg, Van~Hassel,
  Wijnands, and Auregan}]{pelorson1994theoretical}
\bibinfo{author}{Pelorson, X.}, \bibinfo{author}{Hirschberg, A.},
  \bibinfo{author}{Van~Hassel, R.}, \bibinfo{author}{Wijnands, A.},  and
  \bibinfo{author}{Auregan, Y.} (\textbf{\bibinfo{year}{1994}}).
  \enquote{\bibinfo{title}{Theoretical and experimental study of
  quasisteady-flow separation within the glottis during phonation. application
  to a modified two-mass model}} \bibinfo{journal}{The Journal of the
  Acoustical Society of America} \textbf{96}(6), \bibinfo{pages}{3416--3431}.

\bibitem[{Ruder(2016)}]{ruder2016overview}
\bibinfo{author}{Ruder, S.} (\textbf{\bibinfo{year}{2016}}).
  \enquote{\bibinfo{title}{An overview of gradient descent optimization
  algorithms}} \bibinfo{journal}{arXiv preprint arXiv:1609.04747} .

\bibitem[{Ruty \emph{et~al.}(2007)Ruty, Pelorson, Van~Hirtum, Lopez-Arteaga,
  and Hirschberg}]{ruty2007vitro}
\bibinfo{author}{Ruty, N.}, \bibinfo{author}{Pelorson, X.},
  \bibinfo{author}{Van~Hirtum, A.}, \bibinfo{author}{Lopez-Arteaga, I.},  and
  \bibinfo{author}{Hirschberg, A.} (\textbf{\bibinfo{year}{2007}}).
  \enquote{\bibinfo{title}{An in vitro setup to test the relevance and the
  accuracy of low-order vocal folds models}} \bibinfo{journal}{The Journal of
  the Acoustical Society of America} \textbf{121}(1),
  \bibinfo{pages}{479--490}.

\bibitem[{Scherer \emph{et~al.}(1983)Scherer, Titze, and
  Curtis}]{scherer1983pressure}
\bibinfo{author}{Scherer, R.~C.}, \bibinfo{author}{Titze, I.~R.},  and
  \bibinfo{author}{Curtis, J.~F.} (\textbf{\bibinfo{year}{1983}}).
  \enquote{\bibinfo{title}{Pressure-flow relationships in two models of the
  larynx having rectangular glottal shapes}} \bibinfo{journal}{The Journal of
  the Acoustical Society of America} \textbf{73}(2), \bibinfo{pages}{668--676}.

\bibitem[{Smith and Titze(2018)}]{smith2018vocal}
\bibinfo{author}{Smith, S.~L.},  and \bibinfo{author}{Titze, I.~R.}
  (\textbf{\bibinfo{year}{2018}}). \enquote{\bibinfo{title}{Vocal fold contact
  patterns based on normal modes of vibration}} \bibinfo{journal}{Journal of
  Biomechanics} \textbf{73}, \bibinfo{pages}{177--184}.

\bibitem[{Steinecke and Herzel(1995)}]{steinecke1995bifurcations}
\bibinfo{author}{Steinecke, I.},  and \bibinfo{author}{Herzel, H.}
  (\textbf{\bibinfo{year}{1995}}). \enquote{\bibinfo{title}{Bifurcations in an
  asymmetric vocal-fold model}} \bibinfo{journal}{The Journal of the Acoustical
  Society of America} \textbf{97}(3), \bibinfo{pages}{1874--1884}.

\bibitem[{Story and Titze(1995)}]{story1995voice}
\bibinfo{author}{Story, B.~H.},  and \bibinfo{author}{Titze, I.~R.}
  (\textbf{\bibinfo{year}{1995}}). \enquote{\bibinfo{title}{Voice simulation
  with a body-cover model of the vocal folds}} \bibinfo{journal}{The Journal of
  the Acoustical Society of America} \textbf{97}(2),
  \bibinfo{pages}{1249--1260}.

\bibitem[{Tao and Jiang(2008)}]{tao2008chaotic}
\bibinfo{author}{Tao, C.},  and \bibinfo{author}{Jiang, J.~J.}
  (\textbf{\bibinfo{year}{2008}}). \enquote{\bibinfo{title}{Chaotic component
  obscured by strong periodicity in voice production system}}
  \bibinfo{journal}{Physical Review E} \textbf{77}(6), \bibinfo{pages}{061922}.

\bibitem[{Titze(1988)}]{titze1988physics}
\bibinfo{author}{Titze, I.~R.} (\textbf{\bibinfo{year}{1988}}).
  \enquote{\bibinfo{title}{The physics of small-amplitude oscillation of the
  vocal folds}} \bibinfo{journal}{The Journal of the Acoustical Society of
  America} \textbf{83}(4), \bibinfo{pages}{1536--1552}.

\bibitem[{Titze(1994)}]{ingo1994principles}
\bibinfo{author}{Titze, I.~R.} (\textbf{\bibinfo{year}{1994}}).
  \emph{\bibinfo{title}{Principles of Voice Production}}
  (\bibinfo{publisher}{Prentice Hall}),
  \dourl{https://books.google.com/books?id=m48JAQAAMAAJ}.

\bibitem[{Van~den Berg \emph{et~al.}(1957)Van~den Berg, Zantema, and
  Doornenbal~Jr}]{van1957air}
\bibinfo{author}{Van~den Berg, J.}, \bibinfo{author}{Zantema, J.},  and
  \bibinfo{author}{Doornenbal~Jr, P.} (\textbf{\bibinfo{year}{1957}}).
  \enquote{\bibinfo{title}{On the air resistance and the bernoulli effect of
  the human larynx}} \bibinfo{journal}{The Journal of the Acoustical Society of
  America} \textbf{29}(5), \bibinfo{pages}{626--631}.

\bibitem[{Wurzbacher \emph{et~al.}(2006)Wurzbacher, Schwarz, D{\"o}llinger,
  Hoppe, Eysholdt, and Lohscheller}]{wurzbacher2006model}
\bibinfo{author}{Wurzbacher, T.}, \bibinfo{author}{Schwarz, R.},
  \bibinfo{author}{D{\"o}llinger, M.}, \bibinfo{author}{Hoppe, U.},
  \bibinfo{author}{Eysholdt, U.},  and \bibinfo{author}{Lohscheller, J.}
  (\textbf{\bibinfo{year}{2006}}). \enquote{\bibinfo{title}{Model-based
  classification of nonstationary vocal fold vibrations}} \bibinfo{journal}{The
  Journal of the Acoustical Society of America} \textbf{120}(2),
  \bibinfo{pages}{1012--1027}.

\bibitem[{Xue \emph{et~al.}(2012)Xue, Mittal, Zheng, and
  Bielamowicz}]{xue2012computational}
\bibinfo{author}{Xue, Q.}, \bibinfo{author}{Mittal, R.},
  \bibinfo{author}{Zheng, X.},  and \bibinfo{author}{Bielamowicz, S.}
  (\textbf{\bibinfo{year}{2012}}). \enquote{\bibinfo{title}{Computational
  modeling of phonatory dynamics in a tubular three-dimensional model of the
  human larynx}} \bibinfo{journal}{The Journal of the Acoustical Society of
  America} \textbf{132}(3), \bibinfo{pages}{1602--1613}.

\bibitem[{Xue \emph{et~al.}(2011)Xue, Zheng, Bielamowicz, and
  Mittal}]{xue2011sensitivity}
\bibinfo{author}{Xue, Q.}, \bibinfo{author}{Zheng, X.},
  \bibinfo{author}{Bielamowicz, S.},  and \bibinfo{author}{Mittal, R.}
  (\textbf{\bibinfo{year}{2011}}). \enquote{\bibinfo{title}{Sensitivity of
  vocal fold vibratory modes to their three-layer structure: Implications for
  computational modeling of phonation}} \bibinfo{journal}{The Journal of the
  Acoustical Society of America} \textbf{130}(2), \bibinfo{pages}{965--976}.

\bibitem[{Xue \emph{et~al.}(2014)Xue, Zheng, Mittal, and
  Bielamowicz}]{xue2014subject}
\bibinfo{author}{Xue, Q.}, \bibinfo{author}{Zheng, X.},
  \bibinfo{author}{Mittal, R.},  and \bibinfo{author}{Bielamowicz, S.}
  (\textbf{\bibinfo{year}{2014}}). \enquote{\bibinfo{title}{Subject-specific
  computational modeling of human phonation}} \bibinfo{journal}{The Journal of
  the Acoustical Society of America} \textbf{135}(3),
  \bibinfo{pages}{1445--1456}.

\bibitem[{Zanartu \emph{et~al.}(2007)Zanartu, Mongeau, and
  Wodicka}]{zanartu2007influence}
\bibinfo{author}{Zanartu, M.}, \bibinfo{author}{Mongeau, L.},  and
  \bibinfo{author}{Wodicka, G.~R.} (\textbf{\bibinfo{year}{2007}}).
  \enquote{\bibinfo{title}{Influence of acoustic loading on an effective single
  mass model of the vocal folds}} \bibinfo{journal}{The Journal of the
  Acoustical Society of America} \textbf{121}(2), \bibinfo{pages}{1119--1129}.

\bibitem[{Zhang and Yang(2016)}]{zhang2016evaluation}
\bibinfo{author}{Zhang, L.~T.},  and \bibinfo{author}{Yang, J.}
  (\textbf{\bibinfo{year}{2016}}). \enquote{\bibinfo{title}{Evaluation of
  aerodynamic characteristics of a coupled fluid-structure system using
  generalized bernoulli's principle: An application to vocal folds vibration}}
  \bibinfo{journal}{Journal of coupled systems and multiscale dynamics}
  \textbf{4}(4), \bibinfo{pages}{241--250}.

\bibitem[{Zhang and Jiang(2008)}]{zhang2008nonlinear}
\bibinfo{author}{Zhang, Y.},  and \bibinfo{author}{Jiang, J.~J.}
  (\textbf{\bibinfo{year}{2008}}). \enquote{\bibinfo{title}{Nonlinear dynamic
  mechanism of vocal tremor from voice analysis and model simulations}}
  \bibinfo{journal}{Journal of Sound and Vibration} \textbf{316}(1-5),
  \bibinfo{pages}{248--262}.

\bibitem[{Zhang \emph{et~al.}(2020)Zhang, Zheng, and Xue}]{yang2020DNN}
\bibinfo{author}{Zhang, Y.}, \bibinfo{author}{Zheng, X.},  and
  \bibinfo{author}{Xue, Q.} (\textbf{\bibinfo{year}{2020}}).
  \enquote{\bibinfo{title}{A deep neural network based glottal flow model for
  predicting fluid-structure interactions during voice production}}
  \bibinfo{journal}{Applied Sciences} \textbf{10}(2), \bibinfo{pages}{705}.

\bibitem[{Zheng \emph{et~al.}(2010)Zheng, Xue, Mittal, and
  Beilamowicz}]{zheng2010coupled}
\bibinfo{author}{Zheng, X.}, \bibinfo{author}{Xue, Q.},
  \bibinfo{author}{Mittal, R.},  and \bibinfo{author}{Beilamowicz, S.}
  (\textbf{\bibinfo{year}{2010}}). \enquote{\bibinfo{title}{A coupled
  sharp-interface immersed boundary-finite-element method for flow-structure
  interaction with application to human phonation}} \bibinfo{journal}{Journal
  of Biomechanical Engineering} \textbf{132}(11), \bibinfo{pages}{111003}.

\end{thebibliography}

\end{document}